\documentclass[3p]{elsarticle}
\usepackage{amsmath}       
\usepackage{amsfonts}      
\usepackage{mathrsfs}      
                          
\usepackage{hyperref}  
\makeatletter
\providecommand{\doi}[1]{%
  \begingroup
    \let\bibinfo\@secondoftwo
    \urlstyle{rm}%
    \href{http://dx.doi.org/#1}{%
      doi:\discretionary{}{}{}%
      \nolinkurl{#1}%
    }%
  \endgroup
}
\makeatother

\usepackage{bigints}

\usepackage{subcaption}

\usepackage{xifthen}       

\renewcommand{\citet}{\citep}

\newcommand{\GradSymbol}{\operatorname{\mathbf{\nabla}}}
\newcommand{\Grad}{\GradSymbol}
\newcommand{\Div}{\Grad\cdot}
\newcommand{\GradSurf}{\GradSymbol_{\cal C}}
\newcommand{\DivSurf}{\GradSurf\cdot}
\newcommand{\Der}{\partial}
\newcommand{\DerTot}[2][t]
{
  \ifthenelse{\equal{#2}{}}
  {\frac{d #2}{dt}}
  {\frac{d #2}{d #1}}
}
\newcommand{\Diff}{\operatorname{d}\!}

\newcommand{\REAL}{{\mathbb R}}

\newcommand{\ABS}[1]{\ensuremath{\left|#1\right|}}
\newcommand{\NORM}[1]{\ensuremath{\left\| #1 \right\|}}

\newcommand{\scalar}[2]{\prodscal{#1}{#2}}
\newcommand{\prodscal}[2]{\left\langle #1,#2 \right\rangle}

\newcommand{\DET}[1]{\ensuremath{\operatorname{det}(#1)}}

\newcommand{\Point}{\mathbf{P}}
\newcommand{\point}{\mathbf{p}}
\newcommand{\Qpoint}{\mathbf{Q}}
\newcommand{\qpoint}{\mathbf{q}}

\newcommand{\tempo}{t} 
\newcommand{\Tempo}{T} 

\newcommand{\Domain}{\Omega}
\newcommand{\Bnd}{\partial \Omega}
\newcommand{\IT}{I_T} 

\newcommand{\Depth}{\eta}
\newcommand{\density}{\rho}

\newcommand{\heightsymb}{{\cal H}}
\newcommand{\height}[1][]{\ifthenelse{\equal{#1}{}}{\mathbf{\heightsymb}}{\heightsymb_{#1}}}
\newcommand{\FSMsymbol}{{\cal W}}
\newcommand{\BSMsymbol}{{\cal B}}
\newcommand{\LSMsymbol}{{\cal L}}

\newcommand{\FSM}{\heightsymb_{\FSMsymbol}}
\newcommand{\BSM}{\heightsymb_{\BSMsymbol}}

\newcommand{\FreeWaterSurf}{{\cal S}_\FSMsymbol}
\newcommand{\Bottom}{{\cal S}_\BSMsymbol}
\newcommand{\Lateral}{{\cal S}_\LSMsymbol}
\newcommand{\Surf}{{\cal S}}

\newcommand{\mapsymb}{\Psi}
\newcommand{\map}[1][]{\ifthenelse{\equal{#1}{}}{\mathbf{\mapsymb}}{\mapsymb_{#1}}}

\newcommand{\Transsymb}{\Phi}
\newcommand{\Trans}[1][]{\ifthenelse{\equal{#1}{}}{\mathbf{\Transsymb}}{\Transsymb_{#1}}}

\newcommand{\fsymb}{f}
\newcommand{\f}[1][]{\ifthenelse{\equal{#1}{}}{\mathbf{\fsymb}}{\fsymb_{#1}}}

\newcommand{\xvsymb}{x}
\newcommand{\xv}[1][]{\ifthenelse{\equal{#1}{}}{\mathbf{\xvsymb}}{\xvsymb_{#1}}}
\newcommand{\xcg}[1][]{\ifthenelse{\equal{#1}{}}{\xvsymb_1}{\xvsymb_{1}^{(#1)}}}
\newcommand{\ycg}[1][]{\ifthenelse{\equal{#1}{}}{\xvsymb_2}{\xvsymb_{2}^{(#1)}}}
\newcommand{\zcg}[1][]{\ifthenelse{\equal{#1}{}}{\xvsymb_3}{\xvsymb_{3}^{(#1)}}}

\newcommand{\svsymb}{s}
\newcommand{\sv}[1][]{\ifthenelse{\equal{#1}{}}{\mathbf{\svsymb}}{\svsymb}_{#1}}
\newcommand{\svcomp}[1][]{\ifthenelse{\equal{#1}{}}{\svsymb}{\svsymb_{#1}}}
\newcommand{\xcl}[1][]{\ifthenelse{\equal{#1}{}}{\svsymb_1}{\svsymb_{1}^{(#1)}}}
\newcommand{\ycl}[1][]{\ifthenelse{\equal{#1}{}}{\svsymb_2}{\svsymb_{2}^{(#1)}}}
\newcommand{\zcl}[1][]{\ifthenelse{\equal{#1}{}}{\svsymb_3}{\svsymb_{3}^{(#1)}}}

\newcommand{\TanPlane}[2][]
{
  \ifthenelse{\equal{#1}{}}
  {T_{\!\!\point}#2}
  {T_{\!\!#1}#2}
}
\newcommand{\SubsetSymbol}{{\cal A}}
\newcommand{\SubsetU}{\SubsetSymbol}
\newcommand{\SubsetV}{{\cal V}}
\newcommand{\NeighSymbol}{{\cal N}}
\newcommand{\Neigh}[1][]
{
  \ifthenelse{\equal{#1}{}}
  {\NeighSymbol_{\point}}
  {\NeighSymbol_{#1}}
}
\newcommand{\NeighSurf}[1][]
{
  \ifthenelse{\equal{#1}{}}
  {\SubsetSymbol_{\point}}
  {\SubsetSymbol_{#1}}
}
\newcommand{\NormSymb}{\mathbf{n}}
\newcommand{\normalSurf}[1][]
{
  \ifthenelse{\equal{#1}{}}
  {\NormSymb}
  {\NormSymb(#1)}
}

\newcommand{\normalEdge}{\mathbf{n}} 
\newcommand{\basisCC}{t}
\newcommand{\basisGC}{e}
\newcommand{\vecBaseGC}[1][]
{
  \ifthenelse{\equal{#1}{}}
  {\mathbf{\basisGC}}
  {\mathbf{\basisGC}_{#1}}
}
\newcommand{\vecBaseCCcv}[1][]
{
  \ifthenelse{\equal{#1}{}}
  {\mathbf{\basisCC}}
  {\mathbf{\basisCC}_{#1}}
}
\newcommand{\tvecBaseCCcv}[1][]
{
  \ifthenelse{\equal{#1}{}}
  {\tilde{\mathbf{\basisCC}}}
  {\tilde{\mathbf{\basisCC}}_{#1}}
}
\newcommand{\hvecBaseCCcv}[1][]
{
  \ifthenelse{\equal{#1}{}}
  {\hat{\mathbf{\basisCC}}}
  {\hat{\mathbf{\basisCC}}_{#1}}
}
\newcommand{\vecBaseCCctrv}[1][]
{
  \ifthenelse{\equal{#1}{}}
  {\mathbf{\basisCC}}
  {\mathbf{\basisCC}^{#1}}
}

\newcommand{\metricsymbol}{g}
\newcommand{\metrTensCv}[1]{\metricsymbol_{#1}}
\newcommand{\metrTensCtrv}[1]{\metricsymbol^{#1}}
\newcommand{\metrcoefsymbol}{h}
\newcommand{\metrcoef}[1]{\metrcoefsymbol_{#1}}

\newcommand{\defVecBaseCCcv}[1]
{
  \left(
    \dfrac{\Der\xcg[]}{\Der\svcomp[#1]},
    \dfrac{\Der\ycg[]}{\Der\svcomp[#1]}, 
    \dfrac{\Der\zcg[]}{\Der\svcomp[#1]} 
  \right)
}
\newcommand{\First}[1][]{\ifthenelse{\equal{#1}{}}{\ensuremath{\operatorname{\cal G}}}{\ensuremath{\operatorname{\cal G}_{#1}}}}
\newcommand{\first}[1]{%
  \IfEqCase{#1}{%
    {1}{\operatorname{E}}
    {2}{\operatorname{F}}
    {3}{\operatorname{G}}
  }%
  [\PackageError{first}{Undefined option to first: #1}{}]%
}%

\newcommand{\second}[1]{
  \IfEqCase{#1}{
    {1}{\operatorname{e}}
    {2}{\operatorname{f}}
    {3}{\operatorname{g}}
  }
  [\PackageError{first}{Undefined option to first: #1}{}]%
}

\newcommand{\ChristSymb}[2]{\Gamma_{#1}^{#2}}
\newcommand{\DerFunct}[2][]{\ifthenelse{\equal{#1}{}}{\f_{#2}}{{#1}_{#2}}}

\newcommand{\velSymbol}{u}
\newcommand{\vectvel}[1][]
{
   \ifthenelse{\equal{#1}{}}
   {\mathbf{\velSymbol}}
   {\mathbf{\velSymbol}(#1)}
}
\newcommand{\velcomp}[2][i]
{
   \ifthenelse{\equal{#2}{}}
   {\velSymbol_{#1}}
   {\velSymbol_{#1}(#2)}}
\newcommand{\VelSymbol}{\bar{u}}
\newcommand{\vectVel}[1][]
{
   \ifthenelse{\equal{#1}{}}
   {\mathbf{\VelSymbol}}
   {\mathbf{\VelSymbol}(#1)}
}
\newcommand{\Velcomp}[2][i]
{
   \ifthenelse{\equal{#2}{}}
   {\VelSymbol_{#1}}
   {\VelSymbol_{#1}(#2)}
}
\newcommand{\VprimoSymbol}{\tilde{u}}
\newcommand{\Vprimo}[1][]
{
   \ifthenelse{\equal{#1}{}}
   {\mathbf{\VprimoSymbol}}
   {\mathbf{\VprimoSymbol}(#1)}
}
\newcommand{\VprimoComp}[2][i]
{
   \ifthenelse{\equal{#2}{}}
   {\VprimoSymbol_{#1}}
   {\VprimoSymbol_{#1}(#2)}
}
\newcommand{\QSymbol}{q}
\newcommand{\Qdisch}[1][]
{
   \ifthenelse{\equal{#1}{}}
   {\mathbf{\QSymbol}}
   {\mathbf{\QSymbol}(#1)}
}
\newcommand{\Qcomp}[2][i]
{
   \ifthenelse{\equal{#2}{}}
   {\QSymbol_{#1}}
   {\QSymbol_{#1}(#2)}
}
\newcommand{\FricSymbol}{f}
\newcommand{\vectFric}[1][]
{
   \ifthenelse{\equal{#1}{}}
   {\mathbf{\FricSymbol}}
   {\mathbf{\FricSymbol}(#1)}
}
\newcommand{\Friccomp}[2][i]
{
   \ifthenelse{\equal{#2}{}}
   {\FricSymbol_{#1}}
   {\FricSymbol_{#1}(#2)}}
\newcommand{\BFsymbol}{\tau}
\newcommand{\BottomFriction}[1][]{\ifthenelse{\equal{#1}{}}{\BFsymbol_{b}}{\BFsymbol_{b_#1}}}
\newcommand{\IDSymbol}{\mathbb{I}}
\newcommand{\IDtens}[1][]{\ifthenelse{\equal{#1}{}}{\IDSymbol}{\IDSymbol(#1)}}

\newcommand{\tensSymbol}{\mathbb{T}}
\newcommand{\tenscompSymbol}{\tau}
\newcommand{\tens}[1][]{\ifthenelse{\equal{#1}{}}{\tensSymbol}{\tensSymbol(#1)}}
\newcommand{\tenscomp}[2][ij]
{
  \ifthenelse{\equal{#2}{}}
  {\tenscompSymbol_{#1}}
  {\tenscompSymbol_{#1}(#2)}
}
\newcommand{\tensrow}[2][i]
{
  \ifthenelse{\equal{#2}{}}
  {\tensSymbol_{#1}}
  {\tensSymbol_{#1}(#2)}
}

\newcommand{\MCxl}[1][]{\ifthenelse{\equal{#1}{}}{h_1}{h_{1,#1}}} 
\newcommand{\MCyl}[1][]{\ifthenelse{\equal{#1}{}}{h_2}{h_{2,#1}}} 
\newcommand{\MCzl}[1][]{\ifthenelse{\equal{#1}{}}{h_3}{h_{3,#1}}}

\newcommand{\press}{p}
\newcommand{\grav}{g}
\newcommand{\vectgrav}{\mathbf{g}}

\newcommand{\Cinf}[1][]{
  \ifthenelse{\equal{#1}{}}
  {\mathscr{C}^{\infty}}
  {\mathscr{C}^{\infty}(#1)}
}

\newcommand{\curveparam}{\lambda}
\newcommand{\meshparam}{\ell}
\newcommand{\Triang}{{\cal T}_\meshparam}
\newcommand{\Edge}{\sigma}
\newcommand{\Cell}[1][]{\ifthenelse{\equal{#1}{}}{T}{T_{#1}}}
\newcommand{\GeodCurve}{\mathbf{c}}
\newcommand{\CellArea}[1][]
{
  \ifthenelse{\equal{#1}{}}
    {\ABS{\Cell}}
    {\ABS{\Cell[#1]}}
}
\newcommand{\arcLength}{\ABS{\Edge}} 
\newcommand{\edgeLength}{\ABS{\Edge_{ij}}}
\newcommand{\subVolume}[2]{V_{#1}^{#2}}

\newcommand{\Source}{\mathbf{S}}
\newcommand{\SourceTermsMC}[1]{S^{\mbox{\tiny{M}}}_{#1}}
\newcommand{\SourceTermsForces}[1]{S^{\mbox{\tiny{F}}}_{#1}}

\newcommand{\ConservVar}[2]{\mathbf{U}_{#1}^{#2}}
\newcommand{\tConservVar}[2]{\tilde{\mathbf{U}}_{#1}^{#2}}
\newcommand{\Flux}{\underline{\underline{F}}}
\newcommand{\FluxEdge}{\mathbf{F}_{ij}}
\newcommand{\jacobMatrix}{\hat{A}_{ij}}

\newcommand{\Sect}[1]{S_{#1}}

\newcommand\Rey{\mbox{\textit{Re}}}

\journal{Advances in Water Resources}
\biboptions{sort&compress} 
\begin{document}

\begin{frontmatter}
  \title{Modeling Shallow Water Flows on General Terrains}

  \author[KUL]{Ilaria Fent\corref{cor1}}
  \ead{ilaria.fent@uclouvain.be}
  \cortext[cor1]{Corresponding author}
  \address[KUL]{ Institute of Mechanics, Materials and Civil
    Engineering, Universit\'e Catholique de Louvain, Belgium}
  
  \author[MAT]{Mario Putti}
  \ead{mario.putti@unipd.it}
  \address[MAT]{Department of Mathematics, University of Padua, Italy}

  \author[TESAF]{Carlo Gregoretti}
  \ead{carlo.gregoretti@unipd.it}
  \address[TESAF]{Department of Land, Environment, Agriculture and
    Forestry, University of Padua, Italy}
  
  \author[ICEA]{Stefano Lanzoni}
  \ead{stefano.lanzoni@unipd.it}
  \address[ICEA]{Department of Civil, Environmental and Architectural
    Engineering, University of Padua, Italy}
  
  \begin{abstract}
    A formulation of the two-dimensional shallow water equations
    adapted to general and complex terrains is proposed.
    Its derivation starts from the observation that the typical
    approach of depth integrating the Navier-Stokes equations along
    the direction of gravity forces is not exact in the general case
    of a tilted curved bottom.
    We claim that an integration path that better adapts to the
    shallow water hypotheses follows the ``cross-flow'' surface, i.e., a
    surface that is normal to the velocity field at any point of the
    domain.
    Because of the implicitness of this definition, we approximate
    this ``cross-flow'' path by performing depth integration along a
    local direction normal to the bottom surface, and propose a
    rigorous derivation of this approximation and its numerical
    solution as an essential step for the future development of the
    full ``cross-flow'' integration procedure.
    We start by defining a local coordinate system, anchored on the
    bottom surface to derive a covariant form of the Navier-Stokes
    equations.
    Depth integration along the local normals yields a covariant
    version of the shallow water equations, which is characterized by
    flux functions and source terms that vary in space because of the
    surface metric coefficients and related derivatives.
    The proposed model is numerically discretized with a first order
    FORCE-type Godunov Finite Volume scheme that allows straight
    forward implementation of spatially variable fluxes.
    We investigate the validity of our SW model and the effects of the
    geometrical characteristics of the bottom surface by means of
    three synthetic test cases that exhibit non negligible slopes and
    surface curvatures.
    The results show the importance of taking into consideration
    bottom geometry even for relatively mild and slowly varying
    curvatures.
    By comparison with the numerical solution of vertically integrated
    models, we observe differences of almost 20\%, in particular for
    the peak values and the shape of the hydrographs calculated at
    given sections of the fluid domain.
\end{abstract}

\end{frontmatter}


\section{Introduction}

Shallow Water Equations (SWE) are classically used as models of
environmental fluid dynamics when the flow field has one component
that is negligible with respect to the other two, e.g., the vertical
velocity component is small with respect to the horizontal
(longitudinal and lateral) components.  This is the so called Shallow
Water (SW) hypothesis, and is considered true when the process has a
dominant characteristic dimension.  Applications of SWE range from
large-scale ocean modeling~\citep{art:Higdon2006} to atmospheric
circulation~\citep{book:Holton2004}, from river
morphodynamics~\citep{art:Zolezzi2001,art:Lanzoni2006} to dam break
and granular
flows~\citep{art:Fraccarollo2002,art:RosattiBegnudelli2013,art:IversonGeorge2014,art:George:2014},
to avalanches~\citep{art:Gray-et-al1999}.  The common derivation of
this hyperbolic system of balance laws is based on the integration of
the Navier-Stokes (NS) equations over the fluid depth in combination
with an asymptotic analysis implementing the SW
assumption~\citep{art:Decoene-et-al2009}.  For slowly varying bottom
topographies fluid depth is evaluated along the vertical direction as
an approximation of the bottom normal. This approach is generally used
in modeling large scale ocean dynamics or atmospheric flows, where the
bottom boundary is the geo-sphere and the normal direction coincides
with the direction of gravitational forces~\citep{book:Pedlosky1979}.
Also at smaller scales, typical of models of river hydraulics or
granular and snow avalanches, the vertical direction is ordinarily
utilized~\citep{book:Jansen1979}.  However, this approximation is
valid only for relatively small angles, commonly estimated at about
six degrees with respect to horizontal~\citep{book:Chow1959}, and for
negligible curvatures of the bottom relief.  To improve accuracy, ad
hoc pressure corrections are often devised to take into account
deviations of the vertical pressure profile from the hydrostatic
behavior due to bathimetry variability~\citep{art:Higdon2006}.  More
recent attempts look at extensions of the Boussinesq scaling approach
to evaluate these corrective terms~\citep{art:Donahue2015}, employing
sufficiently low order Green-Naghdi polynomial expansions of the
vertical pressure profile to combine accuracy and computational
efficiency of the resulting model~\citep{art:Zhang2013}. Another
recent non-hydrostatic pressure solver for the nonlinear shallow water 
equations proposed in~\citet{art:Arico2016}. 

For bed shapes with more general geometries, vertical integration is
inaccurate and the normal to the bottom profile is preferred.  Studies
that attempt to quantify the accuracy of the vertical integration
approximation are scarce.  Perturbation approaches have been used to
derive the SW equations on general topography with resulting models
usually valid for low enough Reynolds numbers (\Rey). For
example,~\citet{art:RuyerQuil-Manneville-1998,art:RuyerQuil-Manneville-2000}
model thin film flows in lubrication theory ($\Rey\sim$ 300) to
simulate instabilities that develop at long wavelengths using
approaches resting on truncated gradient expansions.  However, for
larger values of \Rey\ typical of geophysical applications, even when
considering steady flow conditions, the leading order solution poses
the problem of matching together the velocity profiles in the inertial
and the viscous layers~\citep{book:TennekesLumley1972}.  In addition,
not less important is the consideration of the spatial variability of
the bed roughness that always characterizes geophysical flows and the
ensuing uncertainty (see, e.g.,~\citet{art:Nikora-et-al-2001,art:Butler2015}).

Within our context of environmental fluid dynamics, the early work
of~\citet{art:SavageHutter1989,art:SavageHutter1991} posed the
foundation for studies of non planar beds by developing a formulation
of the SW model in local curvilinear coordinates based on depth
integration along the normal to the topography. This approach is valid
only for small and essentially one-dimensional bottom curvatures, and,
in practice, it assumes that the fluid surface is parallel to the
bottom.  More recently, \citet{art:Bouchout2004} extended the
methodology to more general bottom surfaces, but its theoretical and
practical limitations have not been extensively studied, as well as
its numerical solution
  , although a number of applications that use this model have been
  published~\citep[e.g.,][]{art:FernandezNieto-et-al2008,art:Moretti-et-al2015}.
Using a different approach, \citet{art:Rossmanith-et-al2004}
considered SW equations on manifolds, developing an original a
modification of the wave propagation algorithm described
in~\citet{art:Bale2002} to work on non-autonomous fluxes arising from
geometrical information, and apply it to SWE on the sphere.

  In this paper we derive a governing system that
  resembles in many ways the model of~\cite{art:Bouchout2004}, but our
  derivation allows the direct identification of the neglected/retained
  terms during the enforcement of the SW assumption.  This entails a
  better understanding of the limitations and assumptions intrinsic to
  the final governing PDE, include the accurate identification of the
  actual hypothesis on the bed geometry.  Thus, the equations are
  written in coordinates instead of intrinsic operators with the
  intent of isolating the terms leading to these limitations, in order
  to carefully identify the correct mathematical assumptions that lead
  to the proposed model in covariant form.  

Our developments start from the observation that, in flows over
general terrains, streamlines deviate substantially from a rectilinear
behavior and may assume generally curved shapes that are independent
of the bed configuration. In this case, the SW hypothesis of small
vertical velocity needs to be adapted to the curvilinear path of the
streamlines. This adaptation can be intuitively explained as follows.
We first note that the SW assumption is identically satisfied on lines
that are orthogonal to the velocity vector at each point. In other
words, the velocity components that are tangential to these so-called
``cross-flow'' lines, and that play the role of the vertical
components, are always zero. If we now define a local curvilinear
reference system anchored on the bottom surface and with the third
direction following the cross-flow lines, we can proceed to integrate
along these paths to arrive at the SW system, once the profile of the
velocity normal to the cross-flow direction is assumed.  A similar
concept is already contained in~\citet{art:Boutounet-et-al2008}.
However, the definition of cross-flow path is implicit and thus
impractical as it requires the knowledge of the NS velocity field, an
unknown of the problem.  To solve this difficulty,
\citet{art:Bresh-Noble-2011} and~\citet{art:Noble-Vila2013} propose a
discretization of the cross-flow path by means of a discrete Fourier
transformation.  The resulting model approaches the complexity of a
full three-dimensional simulation.  Alternatively, we may think of
approximating the cross-flow lines by discretization using piecewise
linear polynomials. This leads to a a multi-layer system where the
cross-flow lines can be approximated with the straight direction
locally normal to each layer bottom.  This strategy is affected by an
error that depends on the thickness of the layers and its full
definition contains still some outstanding issues.  In this paper we
address one of them, namely the derivation of an appropriate
curvilinear reference system and the definition of the SW model
structure for a single layer by performing depth integration along the
local normals.

The completion of the depth integration procedure requires the
knowledge of the velocity variation along the integration path.  In
general, the shape of this profile varies with the type of fluid and
flow regime.  Linear, parabolic, or logarithmic approximations have
been proposed with reference to the specific characteristics of the
fluid (e.g., water, granular mixtures) and of the flow (e.g., laminar,
turbulent, grain-inertial, quasi static).  In our derivation, the
departure of this profile from the depth averaged value is accounted
for by adding multiplicative coefficients to the advective fluxes that
take into consideration the vertical non-uniformity of cross-flow
velocities. These are often called differential advection
terms~\citep{book:vreugdenhil1994}, and, in the geophysical
literature, are alternatively thought of as residual dispersive
stresses~\citep{book:nakagawa1993,art:Kim2009,art:Kim2011} and are
mostly neglected~\citep{book:vreugdenhil1994,art:Iverson2001}.  
For all these approaches, empirical formulations accounting for bottom
stresses are often employed. As shown
by~\citet{art:Decoene-et-al2009}, this strategy yields solutions that
are very similar to more sophisticated and rigorous developments, so
that their use is strictly justified, although a-posteriori.

  Note that our work is different from the so called
  boundary-fitted numerical schemes, typically used in atmospheric
  flows, where a curvilinear coordinate system follow mesh edges that
  are built on the boundary-following surfaces. This approach has been
  proposed recently for the SW equation on general topography
  by~\citet{art:Gallerano2011} and~\citet{art:Gallerano2012}.  In this
  approach the authors rewrite the standard SW equation in the
  curvilinear coordinate system following cell boundaries. As such,
  the SWE do not embody all the geometric information arising from the
  bottom surface.  

  An accurate solution of the SW equations requires the
  analysis of the well-balance problem.  Typically, corrective factors
  that maintain consistency of the scheme are introduced in these
  situations (see the booklet by~\citet{book:Bouchout2004}).
  Determination of these corrections is still an open question in the
  case of general surfaces (see~\cite{art:Dziuk2013}), and ensuring
  consistency is not trivial.  Our choice is then to perform ''exact''
  analytical calculations as much as possible and to resort to
  numerical approximations that are consistent independently of the
  fact that we take into account geometrical effects or not.
  Indeed, all our test cases are designed in such a way that
  well-balance errors are negligible, i.e., the fluid is always moving
  with enough speed.

In this work we present the development of the proposed SW model and
its numerical discretization by finite volumes. By comparison on
simple test cases, numerical results show that neglecting bottom
curvature information leads to significant differences.  We start by
describing the NS equations written in covariant form with respect to
a well-defined local reference system and proceed by performing depth
integration along the normal defined at each point of the bottom
surface.  We then 
%
  formulate a first order FORCE-type
  Godunov Finite Volume scheme for the numerical solution of the
  resulting equations.  We borrow our formulation from the work of
  ~\citet{art:Canestrelli2010,art:Canestrelli2012} who propose a
  second-order FORCE-based SW solver with wetting-and-drying and
  well-balance properties. This approach has been tested on several
  standard problems and a number of real-world applications.  The
  rationale for the use of this approach is that the FORCE scheme is
  based on a central flux approximation that does not require the
  definition of a Riemann problem. 
%
Simulations performed on synthetic but realistic test cases defined on
smooth curved domains show that the influence of the bottom geometry
is important even in the presence of small curvatures.

\section{Equations of motion}

We start this section by first writing the Navier-Stokes equations
using a global Cartesian (orthogonal) reference frame.  The equations
are assumed to be defined in a three-dimensional domain bounded by
smooth surfaces.  Next, we switch our description to a local
curvilinear reference system positioned on the surface defining the
topography of the bottom.  Then, all the developments, including depth
integration, will be carried out with respect to this local reference
system.  A rigorous definition of this curvilinear reference system is
fundamental to understand, at least qualitatively, the limitations
introduced by our approximations.

A remark on notation: in the following we never use contravariant
quantities but only covariant or physical components, and hence we do
not use Einstein summation convention and do not indicate
contravariant or covariant vectors using superscripts or subscripts,
as typical of classical tensor analysis.  Instead, we always report
the complete expressions and definitions of all the operators and
quantities used in the developments, including the summation indices
and their bounds explicitly, and pointing at specialized literature if
necessary.

\subsection{Equations of motion in global  Cartesian coordinates}
  
The equations of motion for an incompressible fluid using a Global
Cartesian Coordinate System (GCS) can be written
as~\citep{book:Batchelor2000}:
\begin{align}
  & \frac{\Der\vectvel}{\Der \tempo}+\Div(\vectvel\otimes\vectvel)=
    -\frac{1}{\density}\Div\left(\press \IDtens +\tens \right)+\vectgrav,
    \nonumber \\[-1em]
  \label{eq:momentCons3D}\\[-1em]
  & \Div\vectvel = 0, \nonumber
\end{align}
where $\vectvel$ is the three-dimensional velocity vector,
$\tempo\in\IT=[0,\Tempo]$ is time, $\Grad$ indicates the gradient of a
function in Cartesian coordinates, $\Div$ is the divergence of a
vector or a tensor (in this latter case $\Div$ operates on each row of
the tensor), $\otimes$ is the tensor product of two vectors
($\vectvel\otimes\vectvel=\vectvel\vectvel^T$), $\density$ is the
density of the fluid, $\press$ is the dynamic fluid pressure,
$\IDtens$ is the identity tensor, $\tens$ is the deviatoric component
of the turbulent stress tensor, and $\vectgrav$ is the gravitational
acceleration vector.

The NS equations are defined for each time $t$ on a domain $\Domain$,
a subset of $\REAL^3$ with boundary $\Bnd$ formed by the union of
three material surfaces identifying the bottom surface ($\Bottom$),
the fluid free surface ($\FreeWaterSurf$), and the lateral surface
($\Lateral$).  We denote with $\xv[]$ the position vector of a generic
point of the domain $\Domain$ with respect to a global Cartesian
(orthogonal) coordinate system spanned by the canonical basis vectors
$(\vecBaseGC[1],\vecBaseGC[2],\vecBaseGC[3])$:
\begin{equation*}
  \xv[]=(\xcg[],\ycg[],\zcg[]) 
          =\xcg[]\vecBaseGC[1]+
           \ycg[]\vecBaseGC[2]+
           \zcg[]\vecBaseGC[3].
\end{equation*}
We assume that the bottom and free surfaces are
regular~\citep{book:AbateTovena2012}.  Mathematically speaking, a
regular surface is a connected subset $\Surf\subset\REAL^3$ such that
for every $\Point\in\Surf$ there exists a $\Cinf$ map
$\map:\SubsetU\mapsto\REAL^3$, with $\SubsetU\subset\REAL^2$, having
the ensuing properties: i) there exists an open neighborhood
$\SubsetV\subset\REAL^3$ of $\Point$ such that
$\map(\SubsetU)=\SubsetV\cap\Surf$; ii) $\map$ is a homeomorphism of
its image; iii) the differential of $\map$ in $\Point$, represented by
the $3\times 2$ Jacobian matrix
$\Diff\map_{\Point}=\{\Der \map[i]/\Der \xv[j]\}$, is injective (it has
maximum rank).  This definition of a surface focuses on the local
properties of the map $\map(\SubsetU)$ that contains all the
information needed for the definition of the local coordinate system.
Essentially, $\Surf\subset\REAL^3$ is a regular surface if it contains
no self-intersections, every point of $\Surf$ can be described by a
continuous, differentiable, and invertible local parametrization,
i.e., the map $\map$, and at each point $\Point\in\Surf$ the tangent
plane $\TanPlane[\point]{\Surf}$ is well defined, or, in other words,
it does not reduce to a line or a point~\citep{book:DoCarmo1976}.

\begin{figure}
  \centerline{
    \includegraphics[width=\textwidth]{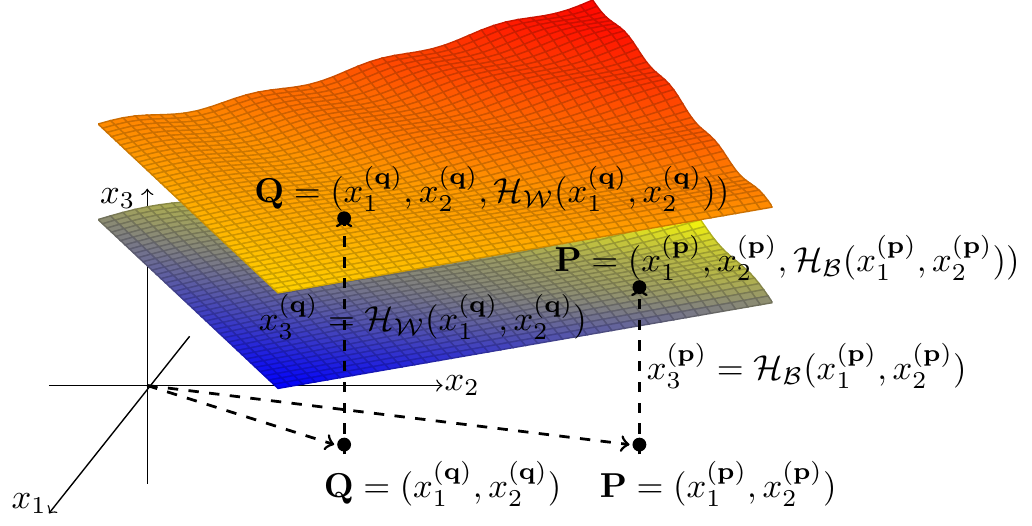}
  }
  \caption{Example of the use of Monge parametrization to define
    points $\Point_{\BSM}$ and $\Point_{\FSM}$ on the bottom and free
    surfaces, respectively.}
  \label{fig:monge}
\end{figure}

We will identify these surfaces as the zero level set of an implicit
function, and will give more precise definitions when needed:
\begin{align*}
  \FreeWaterSurf:&\REAL^3\times\REAL\rightarrow\REAL  
  &
    \FreeWaterSurf&=\{\xv\in\REAL^3: \FreeWaterSurf(\xv,t)=0\}; \\  
  \Bottom:&\REAL^3\times\REAL\rightarrow\REAL 
  & 
    \Bottom&=\{\xv\in\REAL^3: \Bottom(\xv,t)=0\}; \\  
  \Lateral:&\REAL^3\times\REAL\rightarrow\REAL  
  & 
    \Lateral&=\{\xv\in\REAL^3: \Lateral(\xv,t)=0\}.
\end{align*}
To simplify the notation, we use the same symbols to identify the
subset (of $\REAL^3$) of the points belonging to the surface, and the
function defining the surface itself.

Since all regular surfaces of $\REAL^3$ are locally a graph of a
$\Cinf$ function, our surfaces can be described using the so called
Monge parametrization: for each point $\Point\in\Surf$ there exists a
smooth function $\map$ mapping points from an open set
$\SubsetU\subseteq\REAL^2$ to $\Surf\subset\REAL^3$ such that:
\begin{equation*}
  \begin{aligned}
    \map : \SubsetU\subseteq\REAL^2 &\mapsto \REAL^3;\\
    \xv=(\xcg,\ycg) &\mapsto 
        \sv=(\xcl(\xcg,\ycg),\ycl(\xcg,\ycg),\zcl(\xcg,\ycg)).
  \end{aligned}
\end{equation*}
Monge parametrization is often expressed by means of a height function
$\height(\xcg,\ycg)$ as:
\begin{equation}\label{eq:MongeParamBottom}
  \map: \left\{
  \begin{array}{l}
    \xcl=\xcg, \\  \ycl=\ycg, \\  \zcl=\height(\xcg,\ycg).
  \end{array}
  \right.
\end{equation}
Figure~\ref{fig:monge} shows an example of the use of Monge
parametrization to describe a point $\Point$ on the bottom surface
$\Point=(\xcg[\point],\ycg[\point],\BSM(\xcg[\point],\ycg[\point]))$
and a point $\Qpoint$ on the the free surface
$\Qpoint=(\xcg[\qpoint],\ycg[\qpoint],\FSM(\xcg[\qpoint],\ycg[\qpoint]))$.

\subsection{Local curvilinear coordinate system}\label{sec:LCS}

In view of the depth integration procedure to be carried out later, we
need to define a local system of curvilinear coordinates, i.e., a
coordinate system spanning the fluid domain in a neighborhood
$\Neigh[\Point]$ of a point $\Point$ belonging to the bottom surface,
and with origin in $\Point$.  To this end, we need a local coordinate
frame, i.e., a triplet of basis vectors attached to each point
$\Point\in\Bottom$ that can be used to describe all other points in
$\Neigh[\Point]$.  This reference system will be called the ``Local
Curvilinear coordinate System'' (LCS)~\citep{thesis:Balsemin2015}.

Our construction of the LCS is based on the assumption that there
exists a neighborhood $\Neigh$ of $\Point$ such that the
transformation $\Trans_{\Point}$ of each point $\Point\in\Neigh$ from
the global to the local coordinate is a differentiable map whose
inverse is differentiable (it is a diffeomorphism).  We identify these
transformations as:
\begin{equation}\label{eq:transfcoord}
  \Trans_{\Point}:\REAL^3\mapsto\REAL^3; \quad\xv[\Point]\mapsto\sv[\Point], 
  \qquad\qquad
  \Trans_{\Point}^{-1}:\REAL^3\mapsto\REAL^3; \quad\sv[\Point]\mapsto\xv[\Point],
\end{equation}
where $\xv[\Point]=(\xcg[\point],\ycg[\point],\zcg[\point])$ are the
coordinates with respect to the GCS and
$\sv[\Point]=(\xcl[\point],\ycl[\point],\zcl[\point])$ are the
corresponding coordinates with respect to the LCS. 
We start the actual construction by defining a reference frame for the
tangent plane $\TanPlane[\point]{\Bottom}$ at every point
$\Point\in\Bottom$ on the bottom surface, that is a pair of linearly
independent vectors tangent to $\Bottom$ in $\Point$,
$(\vecBaseCCcv[1,\point],\vecBaseCCcv[2,\point])\in\TanPlane[\point]{\Bottom}$.
To simplify the notation, we will often drop the subscript $\Point$,
but all the quantities will refer to the general point
$\Point\in\Bottom$.  The local frame is then completed by choosing the
unit vector orthogonal to $\TanPlane[\point]{\Bottom}$.  The two
tangent vectors at $\Point$ can be calculated as the differential of
the transformation applied to the canonical basis of the GCS, or
equivalently, as the derivatives of the transformation with respect to
the local coordinates.  We can write then:
\begin{equation*}
  \tvecBaseCCcv[i,\point] = \Diff\Trans_{\Point}(\vecBaseGC[i,\point])
     =\defVecBaseCCcv{i}, \qquad i=1,2,
\end{equation*}
where $\Diff\Trans_{\Point}$ is the Jacobian of the coordinate
transformation~\eqref{eq:transfcoord}.  For a regular surface, these
two tangent vectors are guaranteed to exist and be linearly
independent.  However, since their direction depends on the curvatures
of $\Bottom$ at $\Point$, they may become approximately parallel,
potentially leading to numerical ill-conditioning.  For this reason,
vector $\tvecBaseCCcv[2]$ is orthogonalized with respect to
$\tvecBaseCCcv[1]$ via Gram-Schmidt, yielding $\vecBaseCCcv[1]$ and
$\vecBaseCCcv[2]$ such that
$\scalar{\vecBaseCCcv[1]}{\vecBaseCCcv[2]}=0$.  Note that, to make
sure that our LCS is properly defined, normalization of these two
basis vectors cannot be done.  In fact, a local orthonormal coordinate
frame cannot exist as this would amount to assume a zero local
Gaussian curvature of $\Bottom$~\citep{book:AbateTovena2012}, which
would imply $\FreeWaterSurf$ parallel to $\Bottom$, i.e., the same
assumption implicitly contained
in~\citet{art:SavageHutter1991} and~\cite{art:Hutter-et-al2005}. Finally, the
frame-completing vector $\vecBaseCCcv[3]$ can be chosen to be unitary
and orthogonal to $\vecBaseCCcv[1]$ and $\vecBaseCCcv[2]$.

\begin{figure}
  \centerline{
    \includegraphics[width=0.5\textwidth]{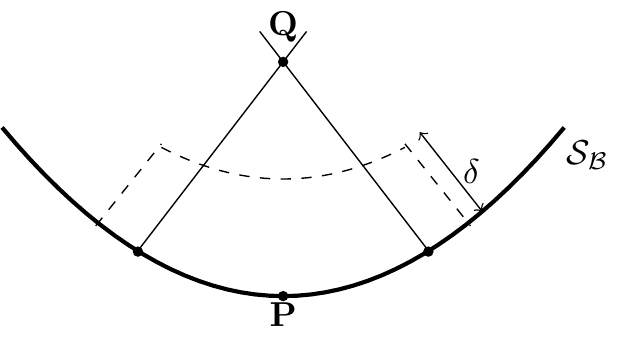}
  }
  \caption{Example of singular point $\Qpoint\in\Neigh$. The figure
    represents a vertical section of a local neighborhood
    $\Neigh[\point,\delta]$ of a parabolic point $\Point\in\Bottom$.}
  \label{fig:NormalAxesIntersect}
\end{figure}

In the case of the Monge parametrization, indicating with subscripts
partial differentiation with respect to the same variable, e.g.,
$\DerFunct[\BSM]{\xcl}={\Der\BSM}/{\Der\xcl}$, we can write explicitly
the expressions for the frame vectors. After Gram-Schmidt
orthogonalization and appropriate normalization, they read:
\begin{align*}
  \vecBaseCCcv[1,\point]
  &=
    \left[1; 0 ; \DerFunct[\BSM]{\xcl} \right],
  \\ 
  \vecBaseCCcv[2,\point]
  &= 
    \left[
    -\frac
    {
    \DerFunct[\BSM]{\xcl}\DerFunct[\BSM]{\ycl}
    }
    {
    1+\DerFunct[\BSM]{\xcl}^2
    } ;
    1;
    \frac
    {
    \DerFunct[\BSM]{\ycl}
    }
    {
    1+\DerFunct[\BSM]{\xcl}^2
    }
    \right],
  \\
  \vecBaseCCcv[3,\point]
  &
    = \normalSurf[\Point]
    = 
    \frac
    {
    \tvecBaseCCcv[1,\point]\wedge\tvecBaseCCcv[2,\point]
    }
    {
    \NORM{\tvecBaseCCcv[1,\point]}\NORM{\tvecBaseCCcv[2,\point]}
    }
    =
    \frac{1}{\sqrt{\DET{\First[\Point]}}}
    \left[
    -\DerFunct[\BSM]{\xcl},-\DerFunct[\BSM]{\ycl},1
    \right].
\end{align*}
The first fundamental form $\First[\Point]$ of the surface at
$\Point\in\Bottom$, or metric tensor, is derived directly from natural
$\REAL^3$ scalar product and is given explicitly by:
\begin{equation}\label{eq:FirstFondamForm}
  \First[\Point]= \left[\metrTensCv{ij}\right]
  =
  \begin{bmatrix}
    \NORM{\vecBaseCCcv[1]}^2 & 0  \\
    0 & \NORM{\vecBaseCCcv[2]}^2
  \end{bmatrix}
  =
  \begin{bmatrix}
    1+\DerFunct[\BSM]{\xcl}^2 & 0  \\
    0 & 
      \dfrac
      {
        1+\DerFunct[\BSM]{\xcl}^2+\DerFunct[\BSM]{\ycl}^2
      }{
        1+\DerFunct[\BSM]{\xcl}^2
      }
  \end{bmatrix}.
\end{equation}
Given a neighborhood $\Neigh$ of $\Point$, every point
$\Qpoint\in\Neigh$ can be expressed in the LCS as follows.  Let
$\Qpoint\in\Neigh$ be given in the GCS by
$\Qpoint=(\xcg[\qpoint],\ycg[\qpoint],\zcg[\qpoint])$.  Consider the
line passing through $\Qpoint$ and parallel to $\normalSurf[\Point]$,
which can be given the following parametric form:
\begin{equation*}
  \gamma(\curveparam):\curveparam\mapsto 
    (\xcg[\qpoint],\ycg[\qpoint],\zcg[\qpoint]) - 
        \normalSurf[\Point] \curveparam,
\end{equation*}
with $r=\gamma(\bar{\curveparam})$ the intersection between the
coordinate line $\gamma$ and $\Bottom$.  Hence, the direct and inverse
transformations of $\Qpoint$ are explicitly given by:
\begin{align*}
  \Trans_{\Qpoint}:
  & 
    \Big(\xcl[\qpoint],\ycl[\qpoint],\zcl[\qpoint]\Big):=
    \Big(\xcg[r],\ycg[r],\bar{\curveparam}\Big), \\
  \Trans^{-1}_{\Qpoint}:
  & 
    \Big(\xcg[\qpoint],\ycg[\qpoint],\zcg[\qpoint]\Big):= 
    \bigg(\xcl[\qpoint],\ycl[\qpoint],
    \BSM\Big(\xcl[\qpoint],\ycl[\qpoint]\Big)
    +\normalSurf[\Point]\zcl[\qpoint]\bigg). 
\end{align*}
The LCS thus defined is not a global bijection.  In fact, singular
points may arise, for example, at the intersection of normal vectors
leaving $\Bottom$ at points belonging to $\Neigh$, as exemplified in
figure~\ref{fig:NormalAxesIntersect} where a neighborhood of a
parabolic point is shown.  However, it can be proved that this LCS is
a diffeomorphism in a $\delta$-neighborhood of $\Point$.  In other
words, there exists a positive real number $\delta$ such that, given a
neighborhood of $\Point$ on the bottom surface,
$\NeighSurf\subset\Bottom$, we can define
$\Neigh[\point,\delta]=\NeighSurf\times [0,\delta]$ where
$\Trans_{\Qpoint}$ and $\Trans^{-1}_{\Qpoint}$ exist for each point
$\Qpoint$ in this neighborhood and are continuous, i.e., the
transformation of coordinates is a diffeomorphism.  Thus, the normal
depth of the fluid domain must in general be smaller than $\delta$, a
condition that is satisfied if $\delta$ is chosen small enough.  As
exemplified in figure~\ref{fig:NormalAxesIntersect}, the intuitive
explanation of this condition is that the flow depth must be smaller
than the minimum radius of curvature.  This would be a rigorous
statement when using the intrinsic coordinates, i.e., coordinates
defined along geodesic (minimum length) curves on the surface. However
calculation of geodesic curves is a difficult numerical and analytical
task, and this is the reason why intrinsic coordinates are not used in
this work.  This introduces limitations on the curvature of $\Bottom$
at $\Point$~\citep{book:Folland1995}.  Note that if we use the exact
cross-flow paths to define a curvilinear $\zcl$ and
$\vecBaseCCcv[3,\point]$ the ensuing LCS will always be a
diffeormphism with no limitations on the curvature.  Finally, it can
be shown that the frame vector fields of this LCS commute, i.e., their
Lie Bracket vanishes~\citep{book:AbateTovena2012}, a necessary and
sufficient condition for the proper definition of the coordinate
system~\citep{book:Boothby2003}.  This last property implies that
every point on the bottom surface can be reached by following
coordinate curves independently of their order.

We use the above defined LCS to express every point of
$\Neigh[\point,\delta]$ and to write the NS equations in local
curvilinear coordinates.  Since our intention is to depth-average
these equations by integration along the normal direction and not the
cross-flow paths, we can assume that the coordinate curve along $\zcl$
is rectilinear.  Hence, the expression for the elements of the
diagonal three-dimensional metric tensor in $\Neigh[\point,\delta]$
can be written as:
\begin{equation*}
  \metrcoef{i} = \sqrt{\metrTensCv{ii}}, \qquad i=1,2.
\end{equation*}
The derivatives of these functions can also be calculated explicitly,
yielding the following properties of the metric tensor: 
\begin{equation}\label{eq:der_metrcoef}
  \metrcoef{3}=1,
  \qquad
  \frac{\partial \metrcoef{3}}{\partial \sv[i]}=0, \quad i=1,2,3;
  \qquad 
  \frac{\partial \metrcoef{i}}{\partial \zcl}=0, \quad i=1,2,3.
\end{equation}
The affine connection (or Christoffel symbols) of $\Bottom$ at
$\Point$ can be written as: 
\begin{equation*}
  \ChristSymb{ij}{k}=
  \sum_{m=1}^3
  \frac12 \metrTensCtrv{mk}
  \left(
    \frac{\Der\metrTensCv{mi}}{\Der \xv[j]} + 
    \frac{\Der\metrTensCv{mj}}{\Der \xv[i]} -
    \frac{\Der\metrTensCv{ij}}{\Der \xv[m]}
  \right),
\end{equation*}
where $\metrTensCtrv{mk}$ are the elements of the inverse
metric tensor defined in~\eqref{eq:FirstFondamForm}, so that
$\sum_{m}\metrTensCtrv{im}\metrTensCv{mj}=\delta_{ij}$.

\subsubsection{Operators in curvilinear coordinates}

In this section we write out the explicit formulas for the operators
needed to express the NS equation in covariant form
where~\eqref{eq:der_metrcoef} is enforced.  The covariant gradient of
a scalar function $f$ is:
\begin{equation*}
  \GradSurf f = 
  \left( 
    \frac{1}{\metrcoef{1}} \frac{\Der f}{\Der\xcl},\;
    \frac{1}{\metrcoef{2}} \frac{\Der f}{\Der\ycl},\;
                  \frac{\Der f}{\Der\zcl}
  \right).
\end{equation*}
For example, given the gravitational potential $\zcg$, the gravity
acceleration vector can be transformed as:
\begin{align*}
  \vectgrav=-\grav\Grad\zcg = 
  0\;\vecBaseGC[1]
  +0\;\vecBaseGC[2]
  -\grav\;\vecBaseGC[3]
  &
    = -\grav\GradSurf\zcg\\
  & = 
    -\grav\frac{1}{\metrcoef{1}}\frac{\Der\zcg}{\Der\xcl}\;\vecBaseCCcv[1]
    -\grav\frac{1}{\metrcoef{2}}\frac{\Der\zcg}{\Der\ycl}\;\vecBaseCCcv[2]
    -\grav\frac{\Der\zcg}{\Der\zcl}\;\vecBaseCCcv[3].
\end{align*}
The covariant divergence of a vector
$\vectvel=\velcomp[1]{}\vecBaseCCcv[1]+\velcomp[2]{}\vecBaseCCcv[2]+\velcomp[3]{}\vecBaseCCcv[3]$
takes on the form:
\begin{equation*}
  \DivSurf \vectvel = 
     \frac{1}{\metrcoef{1}\metrcoef{2}} 
        \left(
           \frac{\Der\left(\velcomp[1]{}\metrcoef{2}\right)}{\Der \xcl} +
           \frac{\Der\left(\velcomp[2]{}\metrcoef{1}\right)}{\Der \ycl} +
           \frac{\Der\left(\velcomp[3]{}\metrcoef{1}\metrcoef{2}\right)}
                {\Der\zcl}
       \right).
\end{equation*}
The covariant divergence $\DivSurf\tens$ of a tensor
$\tens=\left\{\tenscomp{}\right\}$, written in physical
components, is the vector whose $j-th$ element is given by:
\begin{gather*}
  \left(\DivSurf\tens\right)_{j} = 
     \metrcoef{j}\DivSurf\tens_{(j)}
     +
     \frac{1}{\metrcoef{1}\metrcoef{j}}
     \left(
       2 \tenscomp[1j]{}\frac{\Der\metrcoef{j}}{\Der\xcl}
       - \tenscomp[11]{}\frac{\Der\metrcoef{1}}{\Der\sv[j]}
     \right) \\
     +
     \frac{1}{\metrcoef{2}\metrcoef{j}}
     \left(
       2 \tenscomp[2j]{}\frac{\Der\metrcoef{j}}{\Der\ycl}
       - \tenscomp[22]{}\frac{\Der\metrcoef{2}}{\Der\sv[j]}
     \right),
\end{gather*}
where $\DivSurf\tens_{(j)}$ is the covariant divergence of the vector
$\tens_{(j)}=[\tenscomp[j1]{}, \tenscomp[j2]{},
\tenscomp[j3]{}]/\metrcoef{j}$.

\subsection{Equations of motion in the local curvilinear system}

Using the previous expressions, we can write the covariant form of
Navier-Stokes equations for a viscous incompressible fluid. 
The compact form takes the expression:
\begin{subequations} \label{eq:MassConsMomentCons_Surf} 
  \begin{align}
    & \frac{\Der\vectvel}{\Der t}+\DivSurf(\vectvel\otimes\vectvel)=
      -\frac{1}{\density}\DivSurf\left(\press \IDtens +\tens \right)+\vectgrav,
      \label{eq:momentCons3D_Surf}\\
    & \DivSurf \vectvel = 0, \label{eq:massCons_Surf}
  \end{align}
\end{subequations}
where all the differential operators have been defined in the previous
paragraph.  We would like to note that, although the form of these
equations is equal to the standard compact form of the NS
equations~\eqref{eq:momentCons3D}, the subscript in the gradient
symbol $\GradSurf$ indicates that they are written in the LCS and
contain the information (metric) of the bottom surface.  These
equations are used to perform the depth integration over $\zcl$, i.e.,
along the direction normal to the bed, using as integration limits the
bottom topography and the water surface.

\section{Integration along the normal depth}
\label{sec:NANS}

\begin{figure}
  \centerline{
    \includegraphics[width=0.6\textwidth]{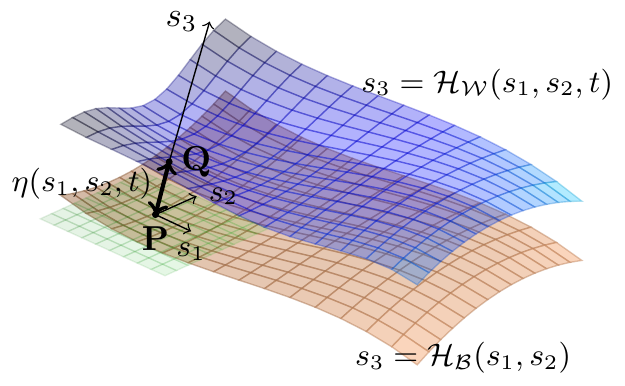}
    \includegraphics[width=0.4\textwidth]{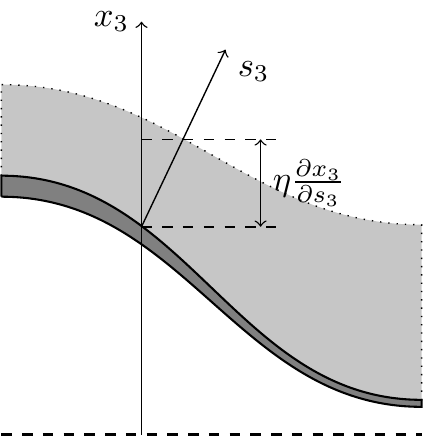}
  }
  \caption{Definition of the water depth $\Depth(\xcl,\ycl,\tempo)$ on
    the local Monge patch at point $\Point\in\Bottom$ as the distance
    between $\Bottom$ and $\FreeWaterSurf$ evaluated along the normal
    direction starting from $\Point$ (left panel).  
    Pressure
    distribution along the normal direction $\zcl$ (right panel).
  }
  \label{fig:Monge_Depth}
\end{figure}

For each point $\Point\in\Bottom$, depth integration of the NS
equations~\eqref{eq:MassConsMomentCons_Surf} is carried out along the
normal direction $\zcl$.  To this aim, we identify the bottom and the
fluid free surfaces as:
\begin{align*}
  \Bottom(\xcl,\ycl,\zcl)
  &=\left\{\left(\xcl,\ycl,\zcl\right)\in\REAL^3:
           \zcl=\BSM(\xcl,\ycl)=0\right\},\\ 
  \FreeWaterSurf(\xcl,\ycl,\zcl,\tempo)
  &=\left\{\left(\xcl,\ycl,\zcl\right)\in\REAL^3:\right.  \\
  & \qquad \qquad \left.
    \zcl=\FSM(\xcl,\ycl,\tempo)=\Depth(\xcl,\ycl,\tempo)
           \mbox{ for each }\tempo\in [0,T]\right\},
\end{align*}
where
$\Depth(\xcl,\ycl,\tempo)=\FSM(\xcl,\ycl,\tempo)-\BSM(\xcl,\ycl)$
identifies the fluid depth (figure~\ref{fig:Monge_Depth}, left panel).
Note that we assume that the bottom is not eroding and thus maintains
a fixed geometry, while the fluid surface is a function of time.  For
this reason, we have included the time-dependence in the functional
forms of both $\FreeWaterSurf$ and $\Depth$ but not in $\Bottom$.  The
kinematic condition at the fluid-air boundary postulates that the free
surface is a material interface that moves with the fluid.  The bottom
boundary is assumed to be impermeable.  Thus we have:
\begin{align} 
  \DerTot{\FreeWaterSurf}\ =&\ \
  \frac{\Der\FreeWaterSurf}{\Der t} +
  \vectvel_{\FSM} \cdot \Grad \FreeWaterSurf = 0, \qquad \zcl=\Depth,
  \label{eq:KinematicCondWatSurf} \\
  \DerTot{\Bottom}\ =&\
  \vectvel_{\BSM} \cdot \Grad \Bottom = 0,\qquad \zcl=0;
  \label{eq:KinematicCondBott} 
\end{align}
where the surface velocities coincide with the three-dimensional fluid
velocity, i.e., $\vectvel_{\BSM}=\vectvel|_{\BSM}$ and
$\vectvel_{\FSM}=\vectvel|_{\FSM}$.  Using the definition of
$\FreeWaterSurf$ and $\Depth$, we have immediately:
\begin{equation}
  \frac{\Der\Depth}{\Der \tempo} +
  \vectvel \cdot \GradSurf \Depth = 0, \qquad \zcl=\Depth.
  \label{eq:KinematicCondWatDepth} 
\end{equation}
Assuming that external actions on the free surface are negligible, the
dynamic condition at the fluid-air interface translates into a
zero-stress boundary equation:
\begin{equation}\label{eq:DynamicCondSurf}
  \tens_{_{\FSM}}\cdot\normalSurf_{_{\FSM}}=0\ ,
  \qquad
  \normalSurf_{_{\FSM}}=\frac{\Grad \FSM}{\ABS{\ABS{\Grad \FSM}}}
\end{equation}
where $\normalSurf_{_{\FSM}}$ is the normal vector to the free water
surface $\FSM$.  The bed boundary condition imposes the value of the
shear stress:
\begin{equation}\label{eq:DynamicCondBot}
  \tens_{_{\BSM}}\cdot\normalSurf=\vectFric_{\BSM}=
  \press_{\BSM}\normalSurf + \BottomFriction[1]{}\vecBaseCCcv[1] +
  \BottomFriction[2]{}\vecBaseCCcv[2], 
\end{equation}
where $\press_{\BSM}$ is the pressure at the bed surface and
$\BottomFriction[i]{}$ denotes bottom friction stresses.

The velocity vector is split into a depth average value, $\vectVel$,
plus a departure from the average, $\Vprimo$:
\begin{equation}\label{eq:fluctuation}
  \vectvel=\vectVel+\Vprimo, \qquad
  \vectVel(\xcl,\ycl,\tempo)=
      \frac{1}{\Depth} 
        \int_{0}^{\Depth}
        \vectvel(\sv,\tempo)\;\Diff\zcl, \qquad
   \int_{0}^{\Depth}
        \Vprimo(\sv,\tempo)\;\Diff\zcl=0.
\end{equation}
We will often use the following relationship derived directly from
Leibnitz rule:
\begin{multline}
    \int_{0}^{\Depth}
    \frac{\Der\velcomp[{i}]{}(\sv,\tempo)}{\Der\tempo}\;\Diff\zcl
    = 
    \frac{\Der}{\Der\tempo}
    \left( 
      \int_{0}^{\Depth}\velcomp[{i}]{}(\sv,\tempo)\;\Diff\zcl
    \right)
    - 
    \left[\velcomp[{i}]{}\frac{\Der\FSM}{\Der\tempo}\right]_{\zcl=\Depth}
    \label{eq:Leibnitz}
    \\
    = \frac{\Der\Depth\Velcomp[{i}]{}}{\Der\tempo}
    - 
    \left[\velcomp[{i}]{}\frac{\Der\FSM}{\Der\tempo}\right]_{\zcl=\Depth}.
\end{multline}
Depth integration starts with the mass continuity
equation~\eqref{eq:massCons_Surf}. We have:
\begin{equation*}
  \int_{0}^{\Depth}
  \left[
    \frac{\Der\left(\velcomp[1]{}\metrcoef{2}\right)}{\Der\xcl}
    + \frac{\Der\left(\velcomp[2]{}\metrcoef{1}\right)}{\Der\ycl}
    + \frac{\Der\left(\velcomp[3]{}\metrcoef{1}\metrcoef{2}\right)}{\Der\zcl}
  \right]\; \Diff\zcl = 0.
\end{equation*}
Application of Leibnitz Rule~\eqref{eq:Leibnitz} yields:
\begin{equation*}
  \frac{\Der}{\Der\xcl}
  \int_0^{\Depth}\velcomp[1]{}\metrcoef{2} \, \text{d}\zcl
  + \frac{\Der}{\Der\ycl}
  \int_0^{\Depth}\velcomp[2]{} \metrcoef{1} \, \text{d}\zcl
-\left[\vectvel\cdot\GradSurf\Depth\right]_{\zcl=\Depth}
 =0.
\end{equation*}
From the definition of the average velocity $\vectVel$, using
eq.~\eqref{eq:KinematicCondWatDepth} we obtain:
\begin{align*}
  \int_{0}^{\Depth}\velcomp[i]{}\metrcoef{j}\Diff\zcl &=
  \int_{0}^{\Depth}(\Velcomp[i]{}+\VprimoComp[i]{})\metrcoef{j}\Diff\zcl
  \\
  &=
  \Velcomp[i]{}\metrcoef{j}\int_{0}^{\Depth}\Diff\zcl +
  \metrcoef{j}\int_{0}^{\Depth}\VprimoComp[i]{}\Diff\zcl
  =
  \Depth\;\Velcomp[i]{}\metrcoef{j}, \qquad i,j=1,2. 
\end{align*}
where we have used the fact that $\metrcoef{j}$ does not depend on
$\zcl$.  Setting $\Qcomp[{i}]{}=\Depth\;\Velcomp[{i}]{}$ 
and using the chain rule
we finally obtain:
\begin{multline*}
    \frac{\Der\Depth}{\Der\tempo}
    + \frac{\Der}{\Der\xcl}
    \left(
      \frac{\Qcomp[1]{}}{\metrcoef{1}}
    \right)
    + \frac{\Der}{\Der\ycl}
    \left(
      \frac{\Qcomp[2]{}}{\metrcoef{2}}
    \right)\\
    + \left(
      \frac{1}{\metrcoef{2}}
      \frac{\Der \metrcoef{2}}{\Der\xcl}
      + \frac1{\metrcoef{1}}
      \frac{\Der\metrcoef{1}}{\Der\xcl}
    \right)
    \frac{\Qcomp[1]{}}{\metrcoef{1}}
    +
    \left(
      \frac{1}{\metrcoef{2}}
      \frac{\Der\metrcoef{2}}{\Der\ycl}
      + \frac1{\metrcoef{1}}\frac{\Der\metrcoef{1}}{\Der\ycl}
    \right)
    \frac{\Qcomp[2]{}}{\metrcoef{2}}
    =0,
\end{multline*}
%
  Note that the use of Leibnitz theorem and the related chain rule of
  differentiation implicitly assumes continuity of the solution and its
  derivatives (both water depth $\Depth$ and velocity $\vectvel$) thus
  in principle ruling out the formation of shocks.  Only recently,
  theoretical studies on the use of the chain rule for BV (Bounded
  Variation) functions have been
  proposed~\citep{art:Crasta2011,art:Ambrosio2012}, but their practical
  applications in numerical solvers seems to be still far reaching.

Now we turn our attention to the third ($\zcl$) equation
of~\eqref{eq:momentCons3D_Surf}.  It is worthwhile writing this
equation on a fully curvilinear system (i.e., $\zcl$ is curvilinear
and not straight as in our LCS) so that the effects of following the
cross-flow are rendered explicitly:
\begin{align*}
    \frac{\Der\velcomp[3]{}}{\Der\tempo}&
      + \frac{1}{\metrcoef{1}\metrcoef{2}\metrcoef{3}}
        \left(
    \frac{\Der\left(\velcomp[1]{}\velcomp[3]{}\metrcoef{2}\metrcoef{3}\right)}{\Der\xcl}
    +\frac{\Der\left(\velcomp[2]{}\velcomp[3]{}\metrcoef{1}\metrcoef{3}\right)}{\Der\ycl}
    +\frac{\Der\left(\velcomp[3]{}^2\metrcoef{1}\metrcoef{2}\right)}{\Der\zcl}
        \right) \\
    &\qquad
    -\frac{\velcomp[1]{}^2}{\metrcoef{1}\metrcoef{3}}\frac{\Der\metrcoef{1}}{\Der\zcl}
    +\frac{\velcomp[1]{}\velcomp[3]{}}{\metrcoef{1}\metrcoef{3}}\frac{\Der\metrcoef{3}}{\Der\xcl}
    -\frac{\velcomp[2]{}^2}{\metrcoef{2}\metrcoef{3}}\frac{\Der\metrcoef{2}}{\Der\zcl}
    +\frac{\velcomp[2]{}\velcomp[3]{}}{\metrcoef{2}\metrcoef{3}}\frac{\Der\metrcoef{3}}{\Der\ycl}
    = \\
    &= 
    -\frac{1}{\density}\frac{1}{\metrcoef{3}}\frac{\Der\press}{\Der\zcl}
    -\frac{1}{\metrcoef{3}}\frac{\Der\zcg}{\Der\zcl}\grav
    -\frac{1}{\density}\frac{1}{\metrcoef{1}\metrcoef{2}\metrcoef{3}}
       \Big(
    \frac{\Der\left(\tenscomp[31]{}\metrcoef{2}\metrcoef{3}\right)}{\Der\xcl} \\
    &\qquad\qquad\qquad
      +\frac{\Der\left(\tenscomp[32]{}\metrcoef{1}\metrcoef{3}\right)}{\Der\ycl}
    +\frac{\Der\left(\tenscomp[33]{}\metrcoef{1}\metrcoef{2}\right)}{\Der\zcl}
       \Big) \\
    &\qquad 
    +\frac{1}{\density}
    \left( 
      \frac{\tenscomp[11]{}}{\metrcoef{1}\metrcoef{3}}\frac{\Der\metrcoef{1}}{\Der\zcl}
      -\frac{\tenscomp[13]{}}{\metrcoef{1}\metrcoef{3}}\frac{\Der\metrcoef{3}}{\Der\xcl}
      +\frac{\tenscomp[22]{}}{\metrcoef{2}\metrcoef{3}}\frac{\Der\metrcoef{2}}{\Der\zcl}
      -\frac{\tenscomp[23]{}}{\metrcoef{2}\metrcoef{3}}\frac{\Der\metrcoef{3}}{\Der\ycl}
    \right).
\end{align*}
Note the presence of terms containing $\metrcoef{j}$ and its
derivatives. In the case of a fully curvilinear LCS these terms may be
important, depending on the curvatures of the cross-flow. In our case
they are zero because of~\eqref{eq:der_metrcoef}.  The SW
approximation in our curvilinear system is equivalent to assume that
the $\zcl$-component of the velocity is negligible.  Hence, in the
previous equation we drop all the terms that contain $\velcomp[3]{}$,
including $\tenscomp[3i]{}=\tenscomp[i3]{}$, $i=1,2,3$, to obtain:
\begin{equation*}
  \int_{0}^{\Depth}\frac{1}{\density}\frac{\Der\press}{\Der\zcl}\;\text{d}\zcl
  +\int_{0}^{\Depth}\frac{\Der\zcg}{\Der\zcl}\grav\;\text{d}\zcl\approx
  0,
\end{equation*}
which states that the pressure 
  varies proportionally to $\Der\zcg/\Der\zcl$ along the
  direction normal to the bed.
Note that if we followed the cross-flow paths
during depth integration, since along these surface $\velcomp[3]{}=0$,
the above equation would be exact, but with extra terms involving
$\metrcoef{3}$ and $\Der\metrcoef{3}/\Der\svcomp[i]{}$.  The
expression for the pressure at the bottom surface then reads (see
figure~\ref{fig:Monge_Depth}, right panel):
\begin{equation}\label{eq:hydrostatic_pressure}
  \press_{\BSM}=\density\grav\Depth\frac{\Der\zcg}{\Der\zcl}.
\end{equation}
This expression will be used during depth-integration of the $\xcl$
and $\ycl$ momentum conservation equations to evaluate the pressure
terms as a function of the depth $\Depth$.  We would like to remark
that $\zcg=\zcg(\xcl,\ycl,\zcl)$ and its derivative are evaluated
through the scalar product:
\begin{equation*}
  \frac{\Der\zcg}{\Der\zcl}=
  \frac{\vecBaseGC[3]\cdot\normalSurf}
       {\NORM{\vecBaseGC[3]}\NORM{\normalSurf}}=
  \frac1{\sqrt{1+\DerFunct[\BSM]{\xcl}^2+\DerFunct[\BSM]{\ycl}^2}}.
\end{equation*}

Next, we focus on the $\xcl$-momentum conservation equation.  Applying
depth-integration, using Leibnitz rule together with the kinematic
equations~\eqref{eq:KinematicCondWatDepth}
and~\eqref{eq:KinematicCondBott}, and the dynamic
conditions~\eqref{eq:DynamicCondSurf} and~\eqref{eq:DynamicCondBot},
and incorporating the hydrostatic pressure
condition~\eqref{eq:hydrostatic_pressure}, we obtain:
\begin{equation*}
  \begin{split}
    \frac{\Der\Qcomp[1]{}}{\Der\tempo}
    &
    + \frac{\Der}{\Der\xcl}
    \left(
      \frac{\alpha_{11}}{\metrcoef{1}}
      \frac{\Qcomp[1]{}^2}{\Depth}
      +\grav\frac{\Depth^2}{2\metrcoef{1}}
      \frac{\Der\zcg}{\Der\zcl}
    \right)
    + \frac{\Der}{\Der\ycl}
    \left(
      \frac{\alpha_{12}}{\metrcoef{2}}
      \frac{\Qcomp[1]{}\Qcomp[2]{}}{\Depth}
    \right) \\
    & +
    \left(
      \frac{1}{\metrcoef{1}}
      \frac{\Der \metrcoef{1}}{\Der\xcl}
      + \frac{1}{\metrcoef{2}}\frac{\Der\metrcoef{2}}{\Der\xcl}
    \right)
    \frac{\alpha_{11}}{\metrcoef{1}}
    \frac{\Qcomp[1]{}^2}{\Depth} \\
    & + 
    \left(
      \frac{2}{\metrcoef{1}}\frac{\Der\metrcoef{1}}{\Der\ycl}
      +
      \frac{1}{\metrcoef{2}}\frac{\Der\metrcoef{2}}{\Der\ycl}
    \right)
    \frac{\alpha_{12}}{\metrcoef{2}}
    \frac{\Qcomp[1]{}\Qcomp[2]{}}{\Depth}
    - 
    \left(
      \frac{1}{\metrcoef{1}}
      \frac{\Der\metrcoef{2}}{\Der\xcl}
    \right)
    \frac{\alpha_{22}}{\metrcoef{2}}
    \frac{\Qcomp[2]{}^2}{\Depth} \\
  = & -\grav\Depth\frac1{\metrcoef{1}}
    \frac{\Der\zcg}{\Der\xcl}
    -\frac{1}{2\metrcoef{1}}\grav\Depth^2
    \frac{\Der}{\Der\xcl}
    \left(
      \frac{\Der\zcg}{\Der\zcl}
    \right)
    -\frac{1}{2\metrcoef{1}^2}\grav\Depth^2
    \frac{\Der\metrcoef{1}}{\Der\xcl}
    \frac{\Der\zcg}{\Der\zcl}
    -\frac{1}{\density}\tensrow[\BSM,1]{}\cdot\normalSurf \\
    &
    -\frac{1}{\density}
    \left[
      \frac{\Der}{\Der\xcl}
      \left(
        \frac{1}{\metrcoef{1}}
        \int_0^{\Depth}\tenscomp[11]{}\Diff\zcl
      \right)
      +
      \frac{\Der}{\Der\ycl}
      \left(
        \frac{1}{\metrcoef{2}}
        \int_0^{\Depth}\tenscomp[12]{}\Diff\zcl
      \right)\right. \\
      & \qquad + 
      \left(
        \frac{1}{\metrcoef{1}}
        \frac{\Der\metrcoef{1}}{\Der\xcl} +
        \frac{1}{\metrcoef{2}}
        \frac{\Der\metrcoef{2}}{\Der\xcl}
      \right)
      \frac{1}{\metrcoef{1}}
      \int_0^{\Depth}\tenscomp[11]{}\Diff\zcl \\
      & \qquad +
      \left(
        \frac{2}{\metrcoef{1}}
        \frac{\Der\metrcoef{1}}{\Der\ycl} 
        +
        \frac{1}{\metrcoef{2}}
        \frac{\Der\metrcoef{2}}{\Der\ycl}
      \right)
        \frac{1}{\metrcoef{2}}
      \int_0^{\Depth}\tenscomp[12]{}\Diff\zcl \\
      & \qquad\left. -
      \frac{1}{\metrcoef{1}}
      \frac{\Der\metrcoef{2}}{\Der\xcl}      
      \frac{1}{\metrcoef{2}}
      \int_0^{\Depth}\tenscomp[22]{}\Diff\zcl
    \right],
  \end{split}
\end{equation*}
where:
\begin{equation*}
  \alpha_{ij}=\frac{1}{\eta}\int_0^\eta 
  \left( 
    1+\frac{\VprimoComp[i]{}\VprimoComp[j]{}}{\Velcomp[i]{}\Velcomp[j]{}}
  \right) \Diff\zcl
\end{equation*}
with obvious analogue for the $\ycl$ counterpart.  The coefficients
$\alpha_{ij}$ incorporate deviations from the the average of the
velocity profile along the local normal direction.  Effectively, they
can be considered as a contribution to the lateral momentum exchanges
and thus added to the stresses $\tenscomp[ij]{}$ (e.g., viscous and
turbulent).  Taking origin from the
decomposition~\eqref{eq:fluctuation}, these terms are often called
differential advection terms~\citep{book:vreugdenhil1994}, or,
considering their stress-like nature, as residual dispersive
stresses~\citep{book:nakagawa1993,art:Kim2009,art:Kim2011}.  In
principle, their evaluation requires the knowledge of the fully
three-dimensional structure of the flow field.  However, unlike
typical turbulent and viscous stresses, they cannot be modeled by a
simple diffusive approach~\citep{book:vreugdenhil1994}.  In some
special cases, quasi-three dimensional models have then been used to
model their
effects~\citep{art:Zolezzi2001,book:SchlichtingGersten2017,art:Frascati2013}.
More often, they are disregarded in conventional two-dimensional
models of geophysical flows~\citep{art:Iverson2001} and even in
Boussinesq-type models that assumes a depth-varying turbulence
averaged velocity~\citep{art:Kim2009}.  For these reasons, as a first
approximation we neglect these terms as well as the depth averaged
contributions of $\tenscomp[ij]{}$, thus effectively incorporating their
effects into the uncertainty of the empirical friction coefficients.

In summary, the final covariant form of the SWE written in the local
orthogonal curvilinear coordinate system for $\alpha_{ij}=1$ is given
by:
\begin{subequations} \label{eq:sistSaintVenantCoordCurv}
  \begin{align}  
    & \frac{\Der\Depth}{\Der\tempo}
      + \frac{\Der}{\Der\xcl}
      \left(
      \frac{\Qcomp[1]{}}{\metrcoef{1}}
      \right)
      + \frac{\Der}{\Der\ycl}
      \left(
      \frac{\Qcomp[2]{}}{\metrcoef{2}}
      \right)
      + \SourceTermsMC{1} =0,
    \\[1.2em]
    \begin{split}
      & \frac{\Der\Qcomp[1]{}}{\Der\tempo}
      + \frac{\Der}{\Der\xcl}
      \left(
        \frac{1}{\metrcoef{1}}
        \frac{\Qcomp[1]{}^2}{\Depth}
        + \grav\frac{\Depth^2}{2\metrcoef{1}}
        \frac{\Der\zcg}{\Der \zcl}
      \right)
      + \frac{\Der}{\Der \ycl}
      \left(
        \frac{1}{\metrcoef{2}}
        \frac{\Qcomp[1]{}\Qcomp[2]{}}{\Depth}
      \right)
      + \SourceTermsForces{2}
      + \SourceTermsMC{2}  =0,
    \end{split}
    \\[1.2em]
    \begin{split}
      & \frac{\Der\Qcomp[2]{}}{\Der\tempo}
      + \frac{\Der}{\Der\xcl}
      \left(
        \frac{1}{\metrcoef{1}}
        \frac{\Qcomp[1]{}\Qcomp[2]{}}{\Depth}
      \right)
      + \frac{\Der}{\Der\ycl}
      \left(
        \frac{1}{\metrcoef{2}}\frac{\Qcomp[2]{}^2}{\Depth}
        + \grav\frac{\Depth^2}{2\metrcoef{2}}\frac{\Der\zcg}{\Der\zcl}
      \right)
      + \SourceTermsForces{3}
      + \SourceTermsMC{3} =0.
    \end{split}
  \end{align}
\end{subequations}
The terms $\SourceTermsForces{k}$ and $\SourceTermsMC{k}$ are the
$k$-th components of the vectors containing gravity and friction
forces, and derivatives of the metric coefficients, as expressed by:
\begin{equation*} 
  \mathbf{\SourceTermsForces{}}=
  \begin{bmatrix}
    0 \\ 
    \begin{split}
      \BottomFriction[1]{}
      + \grav\frac{\Depth}{\metrcoef{1}}\frac{\Der\zcg}{\Der \xcl}
    \end{split}
    \\
    \begin{split}
      \BottomFriction[2]{}
      + \grav\frac{\Depth}{\metrcoef{2}}\frac{\Der\zcg}{\Der\ycl},
    \end{split}
  \end{bmatrix}
\end{equation*}
and
\begin{equation*} 
  \mathbf{\SourceTermsMC{}}=
  \begin{bmatrix}
    \begin{split}
      \left(
      \frac{1}{\metrcoef{2}}
      \frac{\Der \metrcoef{2}}{\Der\xcl}
      +\frac1{\metrcoef{1}}
      \frac{\Der\metrcoef{1}}{\Der\xcl}
      \right)
      \frac{\Qcomp[1]{}}{\metrcoef{1}}
      +
      \left(
      \frac{1}{\metrcoef{2}}
      \frac{\Der\metrcoef{2}}{\Der\ycl}
      +\frac1{\metrcoef{1}}\frac{\Der\metrcoef{1}}{\Der\ycl}
      \right)
      \frac{\Qcomp[2]{}}{\metrcoef{2}}
    \end{split}
    \\[2em]
    \begin{split}
      \frac{\Der\metrcoef{1}}{\Der\xcl}
      \frac{1}{\metrcoef{1}^2}
      \left(
        \frac{\Qcomp[1]{}^2}{\Depth}
        + g\frac{\Depth^2}{2}
        \frac{\Der\zcg}{\Der\zcl}
      \right)
    & +
      \frac{\Der\metrcoef{1}}{\Der\ycl}
      \frac{2}{\metrcoef{1}\metrcoef{2}}
      \frac{\Qcomp[1]{}\Qcomp[2]{}}{\Depth}
      +
      \frac{\Der\metrcoef{2}}{\Der\xcl}
      \frac{1}{\metrcoef{1}\metrcoef{2}}
      \left(
        \frac{\Qcomp[1]{}^2}{\Depth}
        -\frac{\Qcomp[2]{}^2}{\Depth}
      \right) \\
    &  +
      \frac{\Der\metrcoef{2}}{\Der\ycl}
      \frac{1}{\metrcoef{2}^2}
      \frac{\Qcomp[1]{}\Qcomp[2]{}}{\Depth}
      +
      g\frac{\Depth^2}{2\metrcoef{1}}
      \frac{\Der}{\Der\xcl}
      \left(
        \frac{\Der\zcg}{\Der\zcl}
      \right)
    \end{split}
    \\[3em]
    \begin{split}
      \frac{\Der\metrcoef{1}}{\Der\xcl}
      \frac{1}{\metrcoef{1}^2}
      \frac{\Qcomp[1]{}\Qcomp[2]{}}{\Depth}
      +
      \frac{\Der\metrcoef{1}}{\Der\ycl}
      \frac{1}{\metrcoef{1}\metrcoef{2}}
    &
      \left(
        \frac{\Qcomp[2]{}^2}{\Depth}
        -\frac{\Qcomp[1]{}^2}{\Depth}
      \right)
      +
      \frac{\Der\metrcoef{2}}{\Der\xcl}
      \frac{2}{\metrcoef{1}\metrcoef{2}}
      \frac{\Qcomp[1]{}\Qcomp[2]{}}{\Depth} \\
    & +
      \frac{\Der\metrcoef{2}}{\Der\ycl}
      \frac{1}{\metrcoef{2}^2}
      \left(
        \frac{\Qcomp[2]{}^2}{\Depth}
        +
        g\frac{\Depth^2}{2}
        \frac{\Der\zcg}{\Der\zcl}
      \right)
      +
      g\frac{\Depth^2}{2\metrcoef{2}}
      \frac{\Der}{\Der\ycl}
      \left(
        \frac{\Der\zcg}{\Der\zcl}
      \right).
    \end{split}
  \end{bmatrix}
\end{equation*}
Equation~\eqref{eq:sistSaintVenantCoordCurv} is finally written in
vector form as:
\begin{equation} \label{eq:vectorialSWeq}
  \frac{\Der \ConservVar{}{}}{\Der \tempo} + \Div \Flux + \Source = 0,
\end{equation}
where
\begin{equation*}
  \ConservVar{}{}=
     \left[
       \begin{array}{c}
         \Depth \\ \Qcomp[1]{}\\\Qcomp[2]{}
       \end{array}
     \right],
  \qquad
  \Flux=
     \left[
       \begin{array}{cc}
         \frac{\Qcomp[1]{}}{\metrcoef{1}}
       & 
         \frac{\Qcomp[2]{}}{\metrcoef{2}} \\
          \frac{1}{\metrcoef{1}}
          \frac{\Qcomp[1]{}^2}{\Depth}
          + \grav\frac{\Depth^2}{2\metrcoef{1}}
          \frac{\Der\zcg}{\Der \zcl}
       &
          \frac{1}{\metrcoef{2}}
          \frac{\Qcomp[1]{}\Qcomp[2]{}}{\Depth} \\
          \frac{1}{\metrcoef{1}}
          \frac{\Qcomp[1]{}\Qcomp[2]{}}{\Depth}
       &
          \frac{1}{\metrcoef{2}}\frac{\Qcomp[2]{}^2}{\Depth}
           + \grav\frac{\Depth^2}{2\metrcoef{2}}\frac{\Der\zcg}{\Der\zcl}
       \end{array}
     \right],
     \qquad
     \Source=\mathbf{\SourceTermsForces{}}+\mathbf{\SourceTermsMC{}}.
\end{equation*}

\section{The Numerical Model}
\label{sec:numerics}

The numerical solution of equation~\eqref{eq:vectorialSWeq} is
obtained by means of a Finite Volume (FV) scheme based on a first
order accurate Godunov method~\citep{book:Leveque2002} combined with a
centered flux approximation as provided by the FORCE
scheme~\citep{art:Toro2009}.  The reason for the choice of a low order
discretization is that we want to highlight the importance of the
correct geometrical formulation of the SWE system rather than focus on
their numerical discretization.  Thus, we adopt a simple but robust
solver, favoring resiliency over accuracy.  In fact, robustness and
accuracy seldom coexist and higher order schemes often lead to
erroneous results (see e.g. \citet{art:Mazzia2011}) because of
potential ill-conditioning intrinsic to the discretization
method~\citep{art:Kershaw1981}.  Hence, we adopt a robust
central-upwind Godunov-type scheme as proposed
by~\citet{art:Kurganov2001,art:CastroDiaz2012}.  The numerical fluxes
are evaluated using the FORCE approximation that does not require
ad-hoc Riemann solvers~\citep{art:Toro2009}.
%
  In our case of non-autonomous fluxes, a precise
  definition of the Riemann problem is not available. A few attempts
  have been proposed in the
  literature~\citep{art:Bale2002,art:Andreianov2011}, but more work
  along this direction is needed. The use of such an approach would
  introduce uncontrolled errors in the wave speeds and characteristic
  curves. Hence we decided to use a "non-geometric" solver based on a
  central flux approximation that does not require the definition of a
  Riemann problem.

Application of the FV method requires the discretization of the bottom
surface. For the reasons stated above, i.e., to focus exclusively on
the geometrical characteristics of the proposed SW equations, we
decide to define the bottom topography with a known analytical
function, so that arc lengths, surface areas, and integrals on the FV
cells can be evaluated as accurately as possible. 
This is achieved by first triangulating the planar domain formed by
the union of the two-dimensional Monge patches
$\bigcup\SubsetU\subset\REAL^2$ (i.e., the projection of the bottom
surface on the $(\xcg,\ycg)$ coordinate plane) and then raising the
vertical coordinate of the resulting mesh nodes using the Monge
function. 
Thus, the global coordinates of mesh node $i$ on the bottom
surface are given by $\left(\xcg[i],\ycg[i],\BSM(\xcg[i],\ycg[i])\right)$.
Potentially badly shaped surface triangles are corrected a posteriori
by manual intervention or by local mesh refinement.
The relevant geometrical characteristics, such as metric coefficients,
are defined analytically on each triangle, which is thus
identified with the local Monge patch. 
For practical applications, the description of bottom topography is
generally based on observed data and the above procedure no longer
applies. 
Starting from a measured DEM (Digital Elevation Model),
metric coefficients will need 
to be evaluated numerically, with consequential loss of accuracy.
The issue of bottom surface definition from observations and the
numerical approximation of the needed geometrical quantities is
beyond the purpose of the current paper and will be addressed in
future work. 
For the same reason, problems related to well-balance and
fluid-at-rest, as well as wetting-and-drying algorithms are not
addressed specifically. 

\subsection{Description of the surface triangulation}

\begin{figure}
  \centerline{
    \includegraphics[width=0.5\textwidth]{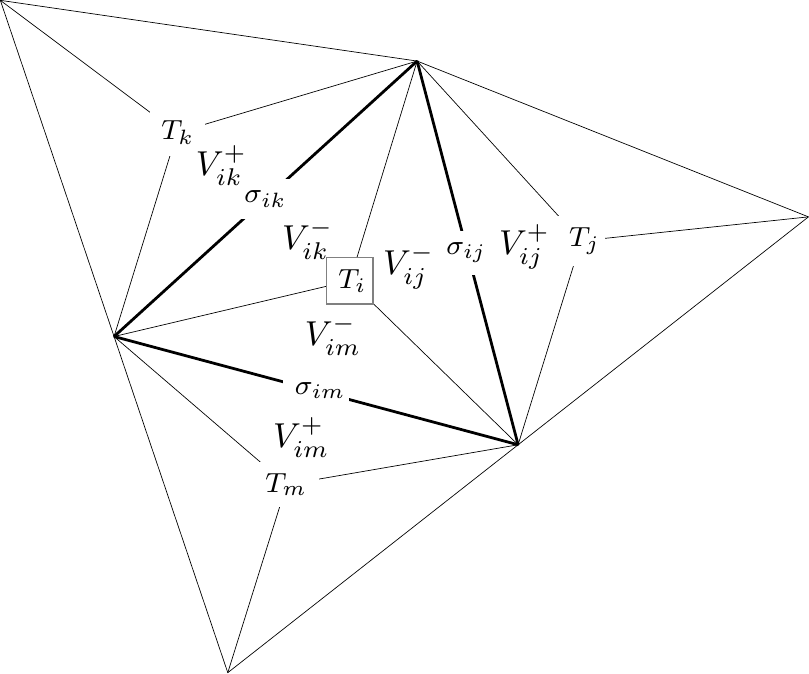}
  }
  \caption{Notation for a general configuration on an unstructured
    triangular mesh.}
  \label{fig:triangulation}
\end{figure}

Let the domain be discretized by a surface triangulation $\Triang$,
which is assumed to be regular and identified by the maximum arc
length $\meshparam$. The domain of definition of~\eqref{eq:vectorialSWeq}
$\Domain$ is assumed to be a polygonally shaped closed subset of
$\BSM$ and is formed by the union of non-intersecting triangular cells
$\left(\Domain=\Triang=\bigcup_{i}\Cell[i]\right)$.  We denote the
surface area of $\Cell[i]$ with the symbol $\CellArea[i]$, while
$\Edge_{ij}$ represents the edge between cell $\Cell[i]$ and
$\Cell[j]$, $\Edge_{ij}=\Cell[i] \cap \Cell[j]$, whose arc length is
given by $\edgeLength$.  The unit normal vector is indicated with
$\normalEdge_{ij}$ and is directed from $\Cell[i]$ towards $\Cell[j]$
(see figure~\ref{fig:triangulation}).  
This process is adopted as it ensures that discrete
values of the metric coefficients converge to the continuous
counterparts when the mesh is refined. This convergence may not be
warranted for meshes made up of quadrilateral or triangular
elements with center of gravity defined on the
surface~\citep{book:Morvan2008}. 

The computation of $\CellArea[i]$ and of $\edgeLength$ are performed
exploiting the functional form of the surface itself defined through
the LCS.
Accordingly, we have
\begin{equation*}
  \CellArea[i] = 
     \int_{\Cell[i]} 
        \metrcoef{1}(\xcl,\ycl)\metrcoef{2}(\xcl,\ycl)\ 
     \Diff\xcl\,\Diff\ycl.
\end{equation*}
The integral above is computed using a 7-point Gaussian quadrature
rule.
The edge arc length is approximated as follows. Define the curve that
connects the edge nodes $\Point_{\alpha}$ and $\Point_{\beta}$ in
parametric form as:
\begin{equation}\label{eq:paramEdge}
  \Edge(\curveparam)=
  \left\{
  \begin{array}{l}
    \xcl(\curveparam)=\left(\xcl[\alpha]-\xcl[\beta]\right)\curveparam
    + \xcl[\alpha], \\
    \ycl(\curveparam)=\left(\ycl[\alpha]-\ycl[\beta]\right)\curveparam
    + \ycl[\alpha], \\
    \zcl(\curveparam)=\BSM\left({\xcl(\curveparam),\ycl(\curveparam)}\right),
  \end{array}
  \right.
\end{equation}
where $\Point_{\alpha}$ and $\Point_{\beta}$ are the two limit points
of the arc, $\xcl$ and $\ycl$ are curvilinear coordinates written as
function of a parameter $\curveparam$, and $\zcl$ is given by Monge
parametrization~\eqref{eq:MongeParamBottom}. 
The arc length of each edge is thus the integral 
of curve~\eqref{eq:paramEdge}:
\begin{equation*}
  \begin{aligned}
    \arcLength &= \int_{\Point_{\alpha}}^{\Point_{\beta}} d\Edge \\
    &=\bigintss_{0}^{1}
    \bigg(
          \left(\xcl[\alpha]-\xcl[\beta]\right)^2 +
          \left(\ycl[\alpha]-\ycl[\beta]\right)^2 + 
     \\
    &\qquad\qquad
          \left(
            \frac{\Der\BSM(\curveparam)}{\Der\xcl}
            \left(\xcl[\alpha] - \xcl[\beta]\right) +
            \frac{\Der\BSM(\curveparam)}{\Der\ycl}
            \left(\ycl[\alpha] - \ycl[\beta]\right)
          \right)^2
          \bigg)^{\frac{1}{2}}
    \Diff\curveparam,
  \end{aligned}
\end{equation*}
where we have written for simplicity 
$\BSM(\curveparam)=\BSM(\xcl(\curveparam),\ycl(\curveparam))$.
This integral is evaluated numerically by means of a 2-point Gaussian
quadrature rule.

\subsection{The finite volume  scheme}

Integrating eq.~\eqref{eq:vectorialSWeq} over $\Domain$, using the
fact that $\int_{\Domain}=\sum_{i=1}^N\int_{\Cell}$, exchanging the
time-derivative with the integral sign, and applying 
Gauss' Divergence Theorem we obtain:
\begin{equation*} 
  \frac{\Der}{\Der \tempo}
  \int_{\Cell}\ConservVar{}{} =
  -\int_{\partial \Cell} \Flux \cdot \normalEdge + \int_{\Cell} \Source,
  \qquad\qquad \forall\Cell\in\Triang.
\end{equation*}
Following Godunov's approach, we define the cell-averaged quantities
over the triangulated surface as:
\begin{equation*} 
  \ConservVar{i}{}=
  \frac{1}{\CellArea[i]}\int_{\Cell[i]}\ConservVar{}{}\ ;
  \quad
  \Source_i=
  \frac{1}{\CellArea[i]}  \int_{\Cell[i]} \Source{}{},
\end{equation*}
and the edge-averaged numerical flux as:
\begin{equation*}
  \FluxEdge=
  \frac{1}{\edgeLength}\int_{\Edge_{ij}}\Flux\cdot\normalEdge
  = \scalar{\Flux}{\normalEdge_{ij}}_{\Edge_{ij}}.
\end{equation*}
Using a time-splitting approach to handle the source term together
with a forward Euler time-discretization, we obtain:
%
\begin{align*} 
  \tConservVar{i}{k+1}&=
  \ConservVar{i}{k} -
  \frac{\Delta \tempo}{\CellArea[i]} 
   \sum_{j=1}^{N_s}\edgeLength \FluxEdge^{FORCE}, \\
  \ConservVar{i}{k+1} & =
     \tConservVar{i}{k+1}-\Delta\tempo\;\;\Source\!\left(\tConservVar{i}{k+1}\right),
\end{align*}
%
where the {\it FORCE} flux is defined
as the average of the Lax-Friedrichs and the Lax-Wendroff
fluxes:
\begin{equation*} 
  \FluxEdge^{FORCE}=
  \frac12\left( \FluxEdge^{LF} +   \FluxEdge^{LW} \right)
\end{equation*}
and these fluxes adapted to two-dimensional triangulations are given
by~\citep{art:Toro2009}: 
\begin{equation*} 
  \FluxEdge^{LF}=
  \frac{
    \subVolume{j}{-}\FluxEdge\left(\ConservVar{j}{k}\right)
  + \subVolume{j}{+}\FluxEdge\left(\ConservVar{i}{k}\right)
  }{
    \subVolume{j}{-} + \subVolume{j}{+}
  }
  -
  \frac{\subVolume{j}{-}\subVolume{j}{+}}{\subVolume{j}{-}+\subVolume{j}{+}}
  \frac{2}{\Delta\tempo\edgeLength}
    \left(\ConservVar{j}{k}-\ConservVar{i}{k}\right),
\end{equation*}
and
\begin{equation*} 
  \FluxEdge^{LW}=
  \frac
  {
    \subVolume{j}{+}\FluxEdge\left(\ConservVar{j}{k}\right)
    + \subVolume{j}{-}\FluxEdge\left(\ConservVar{i}{k}\right)
  }
  {
    \subVolume{j}{-}+\subVolume{j}{+}
  }
  -  \frac12\frac{\edgeLength\Delta\tempo}{\subVolume{j}{-}+\subVolume{j}{+}}
   \jacobMatrix\left(\FluxEdge\left(\ConservVar{j}{k}\right)
  -\FluxEdge\left(\ConservVar{i}{k}\right)\right).
\end{equation*}
In the above equations we have used the symbols $\subVolume{j}{+}$ and
$\subVolume{j}{-}$ to indicate the one-sided diamond cells
(see figure~\ref{fig:triangulation}), while matrix $\jacobMatrix$ is the
Jacobian of the flux
matrix~\citep{art:DalMaso-et-al1995,art:Canestrelli2010}.  
The fluxes are evaluated at edge midpoints using the exact metric
tensor~\eqref{eq:FirstFondamForm} of the surface, and the integrals of
the source terms are evaluated using the midpoint rule with exact
surface triangular area.

Finally, in order to ensure the stability of the scheme, the standard
CFL (Courant-Friedrichs-Lewy) condition must be satisfied.
Note that the wetting-and-drying algorithm presented
in~\citet{art:Canestrelli2012} has been implemented in the current
version of the model but has not been specifically adapted for the
actual covariant form of the SWE.
Thus we expect potential oscillatory behavior ahead of wave fronts.
For this reason, in this first step of our research, we will design
testing examples that consider only bed drying, thus avoiding
the difficulties connected with wetting and drying processes.

\subsection{Parallel transport}

The above FV formulation requires the evaluation of differences between
flux and velocity vectors calculated at the cell interfaces. 
Since these vectors are defined on each cell, and since each cell lies
on a different tangent plane, these differences must be performed
after parallel transport of the vector quantities, given the
relevant parametrization.  
Parallel transport of a vector $\vectvel\in\BSM$ is obtained as
the solution of the following system of ordinary differential
equations defined on the surface curve connecting geodetically the
gravity centers of two adjacent cells:
\begin{equation}
  \label{parallel-transport}
    \Grad\velcomp[i]{}\cdot\frac{d\GeodCurve}{d\curveparam}
      + \vectvel\cdot\left\{\ChristSymb{kl}{i}\right\}
        \frac{d\GeodCurve}{d\curveparam}
    =0,
\end{equation}
where $\GeodCurve(\curveparam)$ is the parametric expression of the
geodesic curve joining the two gravity centers and
$\{\ChristSymb{kl}{i}\}$ is the matrix of the $i$-th Christoffel
symbol.  We implement parallel transport following the same
simplifications made by~\citet{art:Rossmanith-et-al2004}.  Thus, the
ODE system~\eqref{parallel-transport} is solved numerically using a
first order explicit Euler method along the curve parameter
$\curveparam$.  Given two triangles $\Cell[i]$ and $\Cell[j]$ adjacent
through edge $\Edge_{ij}$, we denote with $\sv[i]$ the LCS coordinates
of the gravity center of $\Cell[i]$ and with $\sv[ij]$ the LCS
coordinates of the midpoint of $\Edge_{ij}$.  The index (subscript or
superscript) $(i)$ relates quantities calculated on cell $\Cell[i]$
using the related LCS.  We also assume that the geodesic curve is
always along the $\xcl$ direction, which is then assumed to coincide
with the direction of the principal curvature.  With all these
assumptions, a generic vector $\vectvel\in\BSM$ is parallel
transported using the following formulas:
\begin{align*}
  \velcomp[1]{}^{(i)}(\sv[ij])
     & = \velcomp[1]{\sv[i]} - (\svcomp[1,ij]-\svcomp[1,i])
     \left(
         \ChristSymb{11}{1,(i)}\velcomp[1]{\sv[i]} +
         \ChristSymb{21}{1,(i)}\velcomp[2]{\sv[i]}
     \right), \\
  \velcomp[2]{}^{(i)}(\sv[ij])
     & = \velcomp[2]{\sv[i]} - (\svcomp[1,ij]-\svcomp[1,i])
     \left(
         \ChristSymb{11}{2,(i)}\velcomp[1]{\sv[i]} +
         \ChristSymb{21}{2,(i)}\velcomp[2]{\sv[i]}
     \right), \\[1.5em]
  \velcomp[1]{}^{(j)}(\sv[ij])
     & = \velcomp[1]{\sv[j]} - (\svcomp[1,ij]-\svcomp[1,j])
     \left(
         \ChristSymb{11}{1,(j)}\velcomp[1]{\sv[j]} +
         \ChristSymb{21}{1,(j)}\velcomp[2]{\sv[j]}
     \right), \\
  \velcomp[2]{}^{(j)}(\sv[ij])
     & = \velcomp[2]{\sv[j]} - (\svcomp[1,ij]-\svcomp[1,j])
     \left(
         \ChristSymb{11}{2,(j)}\velcomp[1]{\sv[j]} +
         \ChristSymb{21}{2,(j)}\velcomp[2]{\sv[j]}
     \right).
\end{align*}
%

\section{Simulations and Numerical Results}
\label{sec:results}

The present section describes three synthetic test cases designed to
analyze the influence of the bottom geometry (slope and curvature) on
the solution of the SWE.
The design needs to be aware of the assumptions leading to our
formulation, and most importantly of the hypothesis of slowly varying
topography with small curvatures. 
Moreover, we want to minimize the influence of critical, but not
relevant to geometrical effects, numerical algorithms, such as
wetting-drying treatment or water-at-rest enforcement.
At the same time boundary and initial conditions should have minimal
effects on the accuracy of the numerical solution.

  Our experiments must then have always a nonzero downstream water
  depth to avoid wetting phenomena. Moreover, the fluid velocity must
  always be large enough so that well-balance errors are
  negligible. Then we are forced to use initial conditions that
  are in non-equilibrium states, a situation that is difficult to
  to reproduce in laboratory experiments. 
  Moreover, there are no explicit solutions of our mathematical model
  and thus we cannot evaluate discretization errors. 
  As a consequence, the only possible and significative comparison is
  versus the solution of the same model without geometric information,
  i.e., setting $\metrcoef{i}=1$. 

\begin{figure}
  \centerline{
    \includegraphics[width=0.5\textwidth]{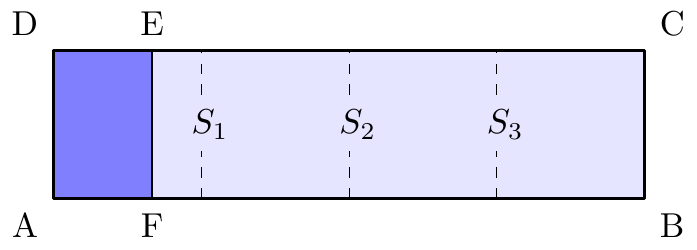}
  }
  \caption{Sketch of the two-dimensional Monge patch
    $\SubsetU\subset\REAL^2$ defining the domain of all the test
    cases. The water depth initial conditions are shown in blue shades
    (dark identifies deeper upstream depth) together with the location
    of the three cross-sections where discharges are evaluated.}
  \label{fig:DomainSketch}
\end{figure}

To achieve these goals the domain is defined via Monge parametrization
starting from a regular triangulation of a rectangular subset
$\SubsetU\subset\REAL^2$ (Figure~\ref{fig:DomainSketch}) and using a
sufficiently smooth height function. 
All the test cases implement a simulation where a gravity wave moves
downward starting with initial conditions involving water depths that
are sufficiently large to minimize the drying of the bottom and exclude
re-wetting.  
We avoid topographies with valleys or depressions so that
water-at-rest issues are not relevant. 

Figure~\ref{fig:DomainSketch} shows the generic rectangular Monge
domain used in all test cases. To implement our simulations, we employ
the following boundary conditions: edges $\overline{\text{AB}}$,
$\overline{\text{DC}}$ and $\overline{\text{DA}}$ are impermeable
walls, while $\overline{\text{CB}}$ represents the open outlet.  In
order to avoid the reflection of outgoing waves and discontinuous
profiles at this latter boundary, a smooth transition of the bottom
geometry is enforced at the outlet.  The edge $\overline{\text{EF}}$
describes the position of the initial water depth discontinuity, with
upstream (shown with darker color in figure~\ref{fig:DomainSketch})
and downstream depths of 2~m and 1~m, respectively.  The same figure
displays three uniformly distributed cross-sections $\Sect{i}, i=1,2,3$
orthogonal to the lateral walls.  These sections are used to evaluate
streamflows.  All the test cases consider the Gauckler-Strickler
formula to describe the flow resistance at the bottom:
\begin{equation*}
  \BottomFriction[i]{}=\frac{1}{k_s^2}\frac{\velcomp[i]{} }{R_h^{4/3}} 
\end{equation*}
with the resistance coefficient $k_s=40$~m$^{1/3}$/s.

The first test case considers sloping planes with variable
inclinations to study the effects of small and large sloping angles on
the dynamics of the flow.
The second test case introduces simple one-dimensional curvatures on a
sloping domain, and the last examines a fully three-dimensional bottom
surface. 

\subsection{Test case 1: sloping planes.}

\begin{figure}
    \centering
    \begin{subfigure}[b]{0.45\textwidth}
        \includegraphics[width=\textwidth]{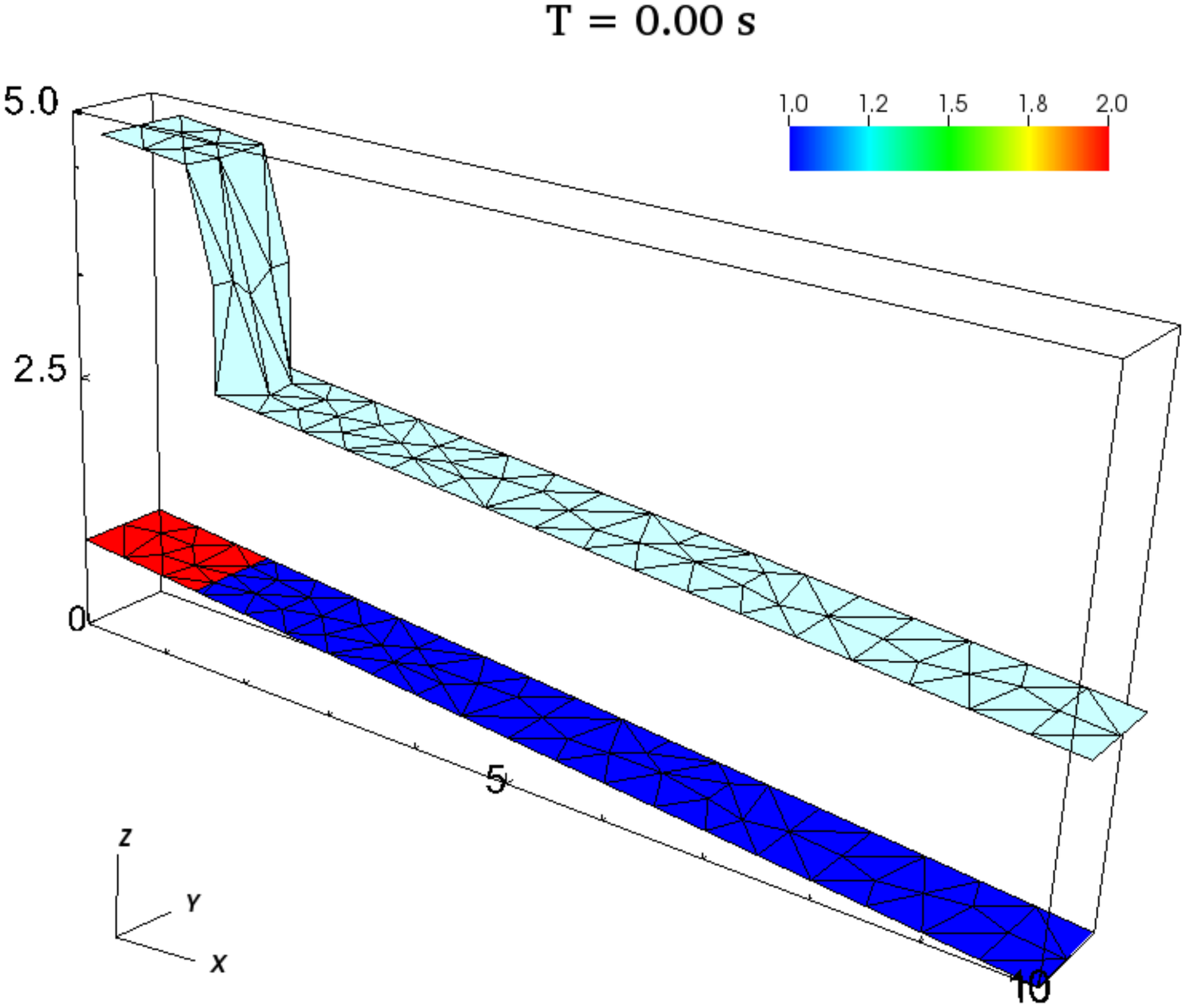}
        \caption{}
        \label{subfig:TC1_5deg_t=000s}
    \end{subfigure}
    \begin{subfigure}[b]{0.45\textwidth}
        \includegraphics[width=\textwidth]{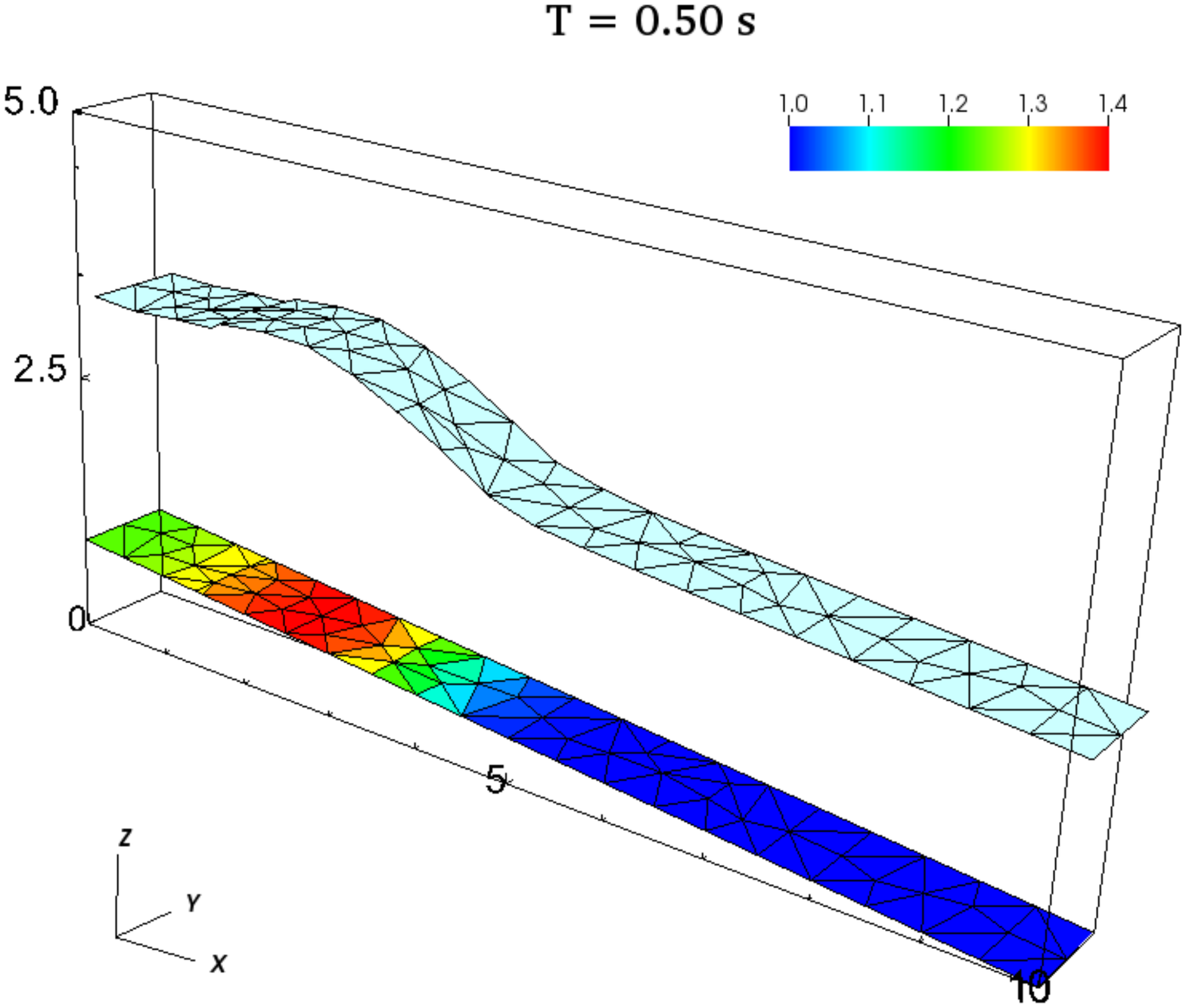}
        \caption{}
        \label{subfig:TC1_5deg_t=050s}
    \end{subfigure}
    \\
    \begin{subfigure}[b]{0.45\textwidth}
        \includegraphics[width=\textwidth]{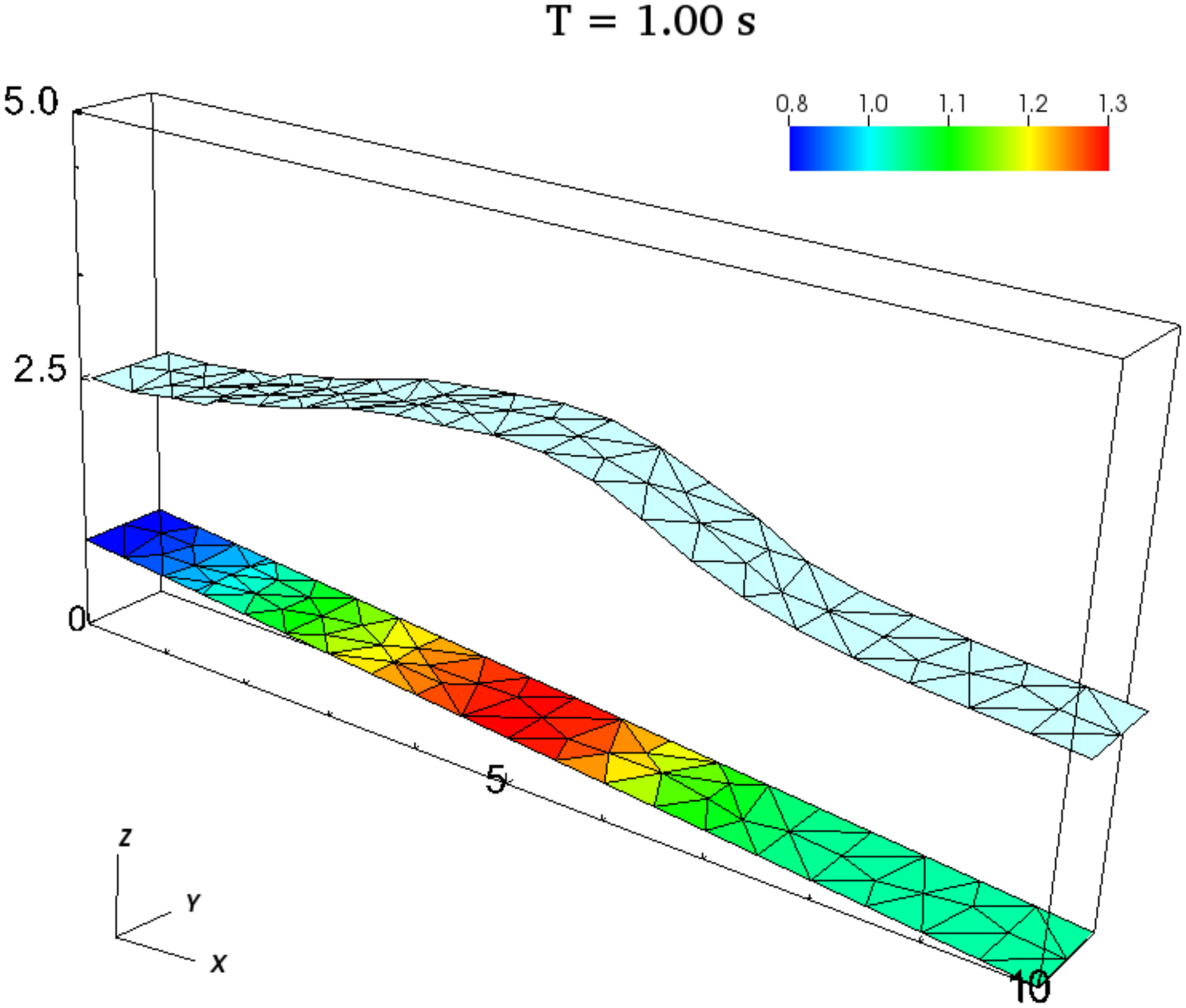}
        \caption{}
        \label{subfig:TC1_5deg_t=100s}
    \end{subfigure}
    \begin{subfigure}[b]{0.45\textwidth}
        \includegraphics[width=\textwidth]{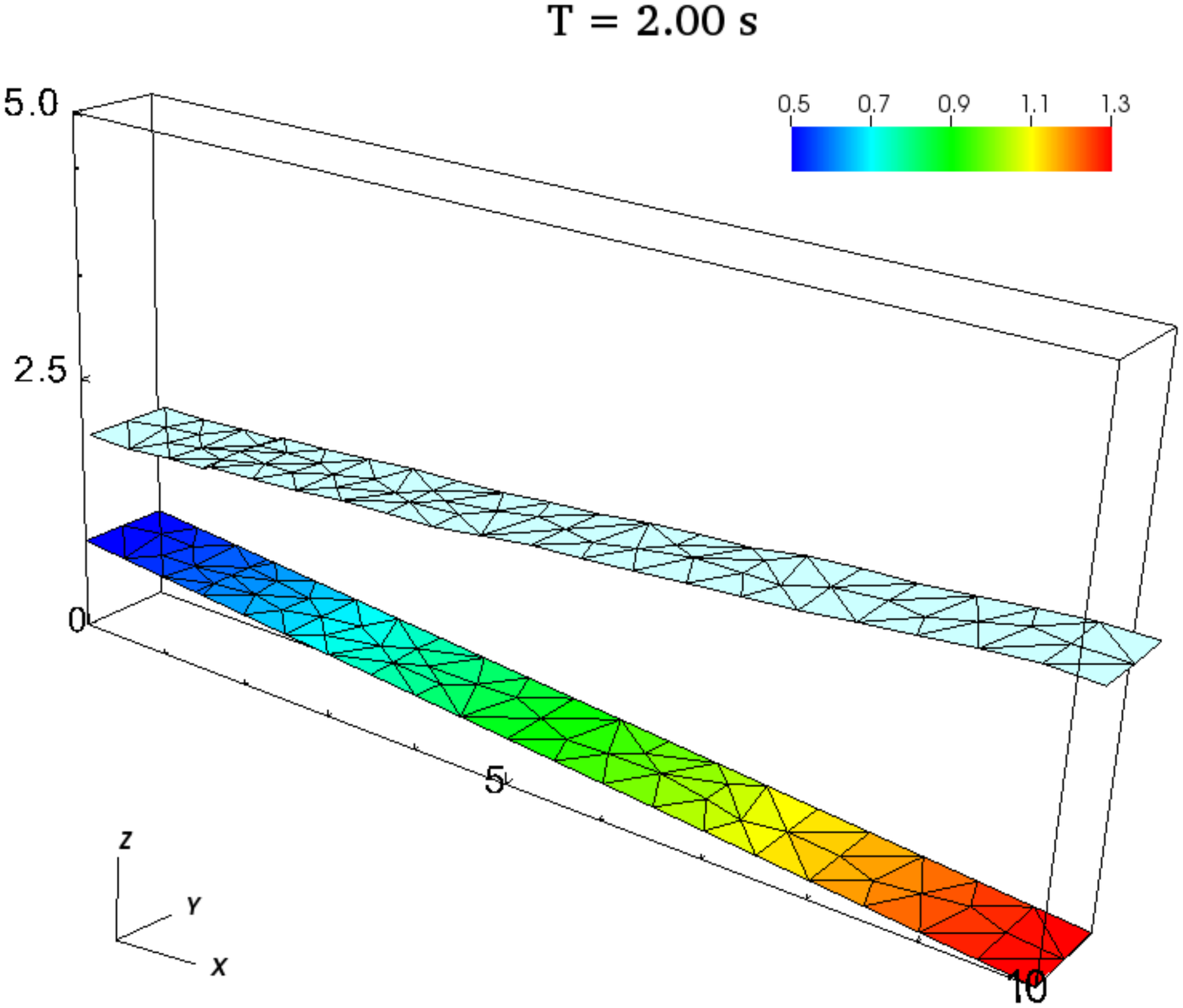}
        \caption{}
        \label{subfig:TC1_5deg_t=200s}
    \end{subfigure}
    \caption{Test case 1: uniformly sloping bottom. Water depth [m]
      evolution of the simulation for the sub-case with 5$^{\circ}$
      shown both as color codes and depth elevation, the latter with a
      magnification factor of 2.  }
    \label{fig:TC1_5deg_results}
\end{figure}

\begin{figure}
    \centering
    \begin{subfigure}[b]{0.45\textwidth}
        \includegraphics[width=\textwidth]{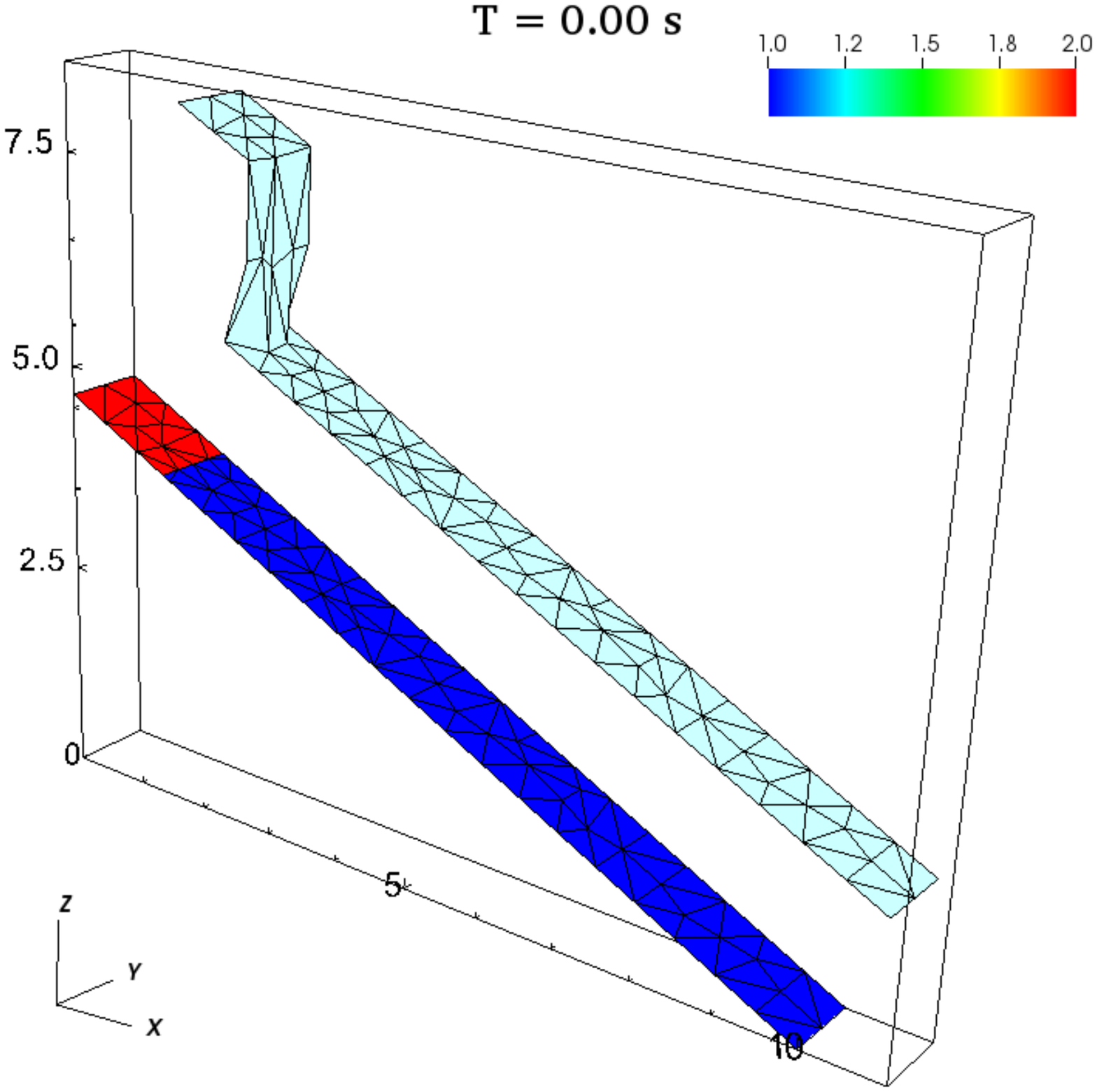}
        \caption{}
        \label{subfig:TC1_25deg_t=000s}
    \end{subfigure}
    \begin{subfigure}[b]{0.45\textwidth}
        \includegraphics[width=\textwidth]{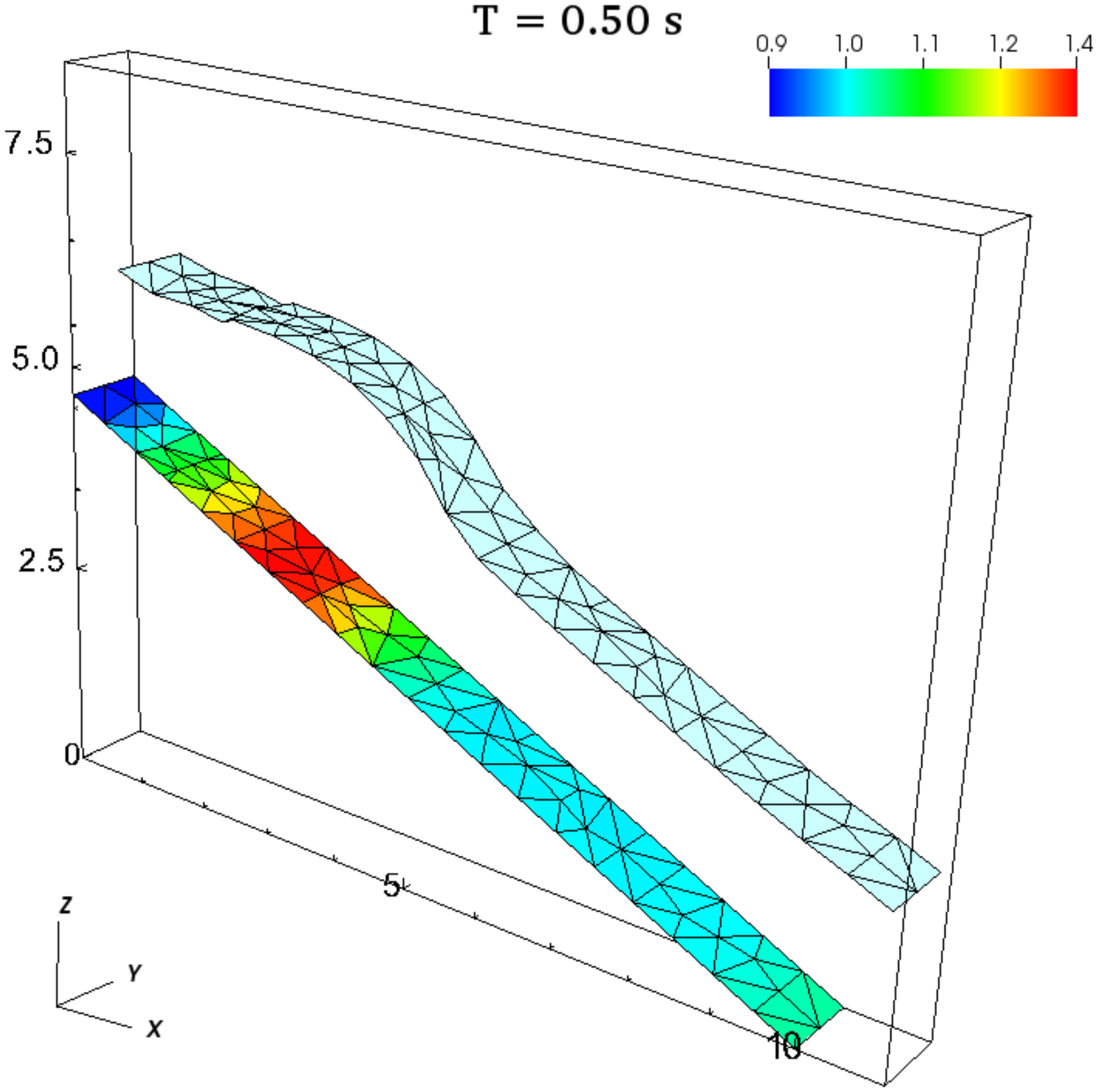}
        \caption{}
        \label{subfig:TC1_25deg_t=050s}
    \end{subfigure}
    \\
    \begin{subfigure}[b]{0.45\textwidth}
        \includegraphics[width=\textwidth]{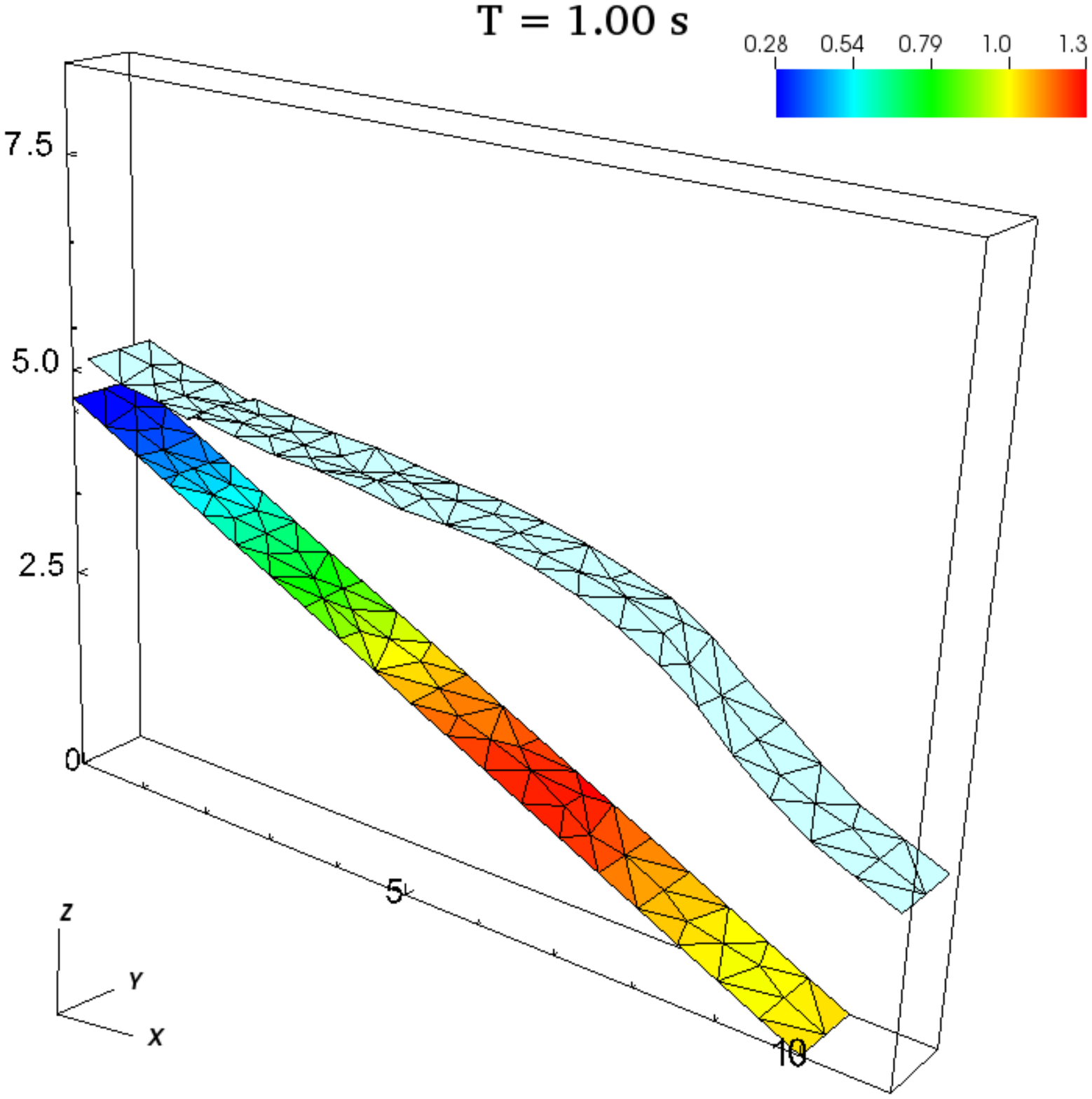}
        \caption{}
        \label{subfig:TC1_25deg_t=100s}
    \end{subfigure}
    \begin{subfigure}[b]{0.45\textwidth}
        \includegraphics[width=\textwidth]{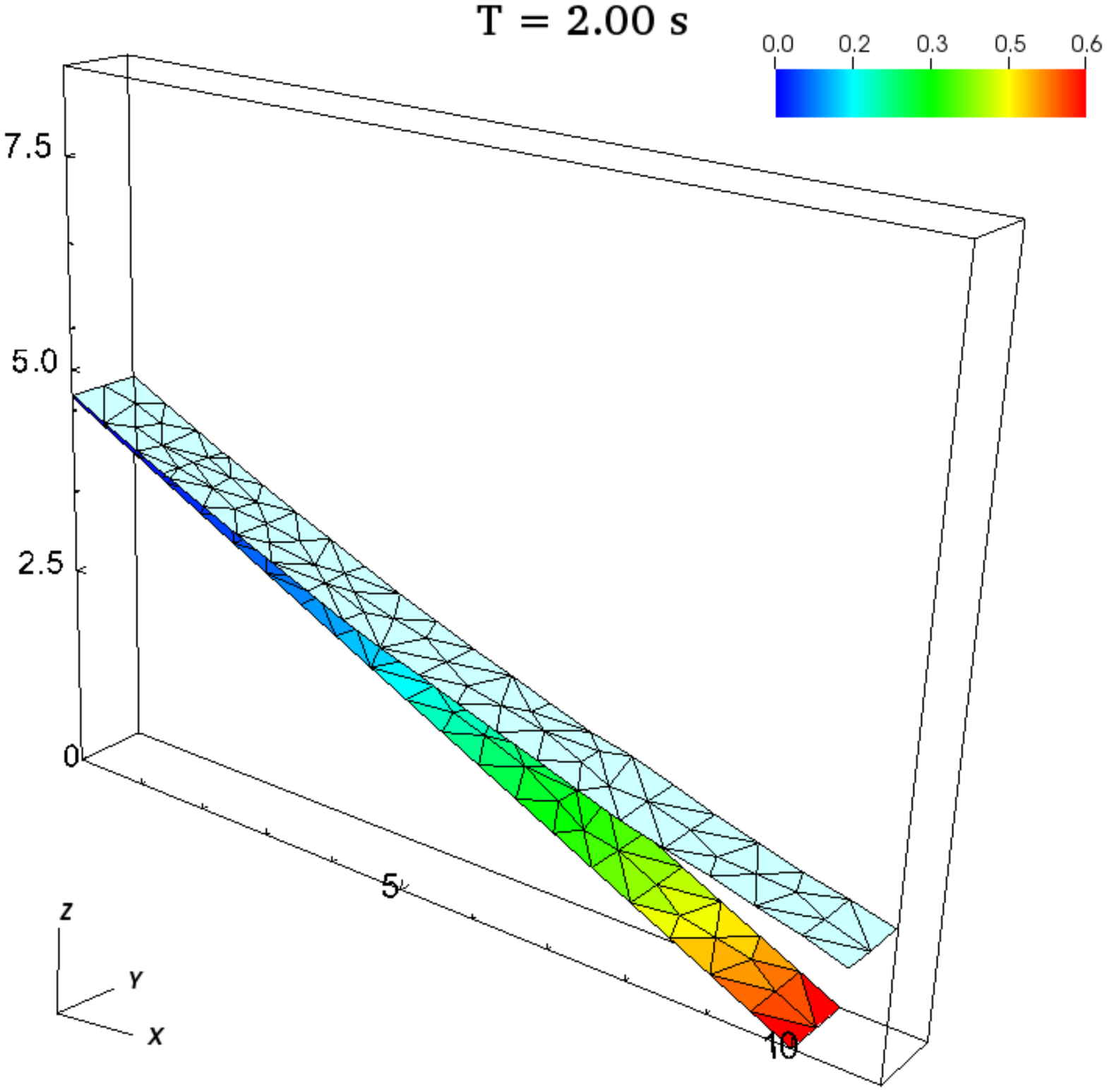}
        \caption{}
        \label{subfig:TC1_25deg_t=200s}
    \end{subfigure}
    \caption{Test case 1: uniformly sloping bottom. Water depth [m]
      evolution of the simulation for the sub-case with 25$^{\circ}$
      shown both as color codes and depth elevation, the latter with a
      magnification factor of 2.  }
    \label{fig:TC1_25deg_results}
\end{figure}

This test case considers a rectangular Monge subset $\SubsetU$ with the
following dimensions: $\overline{\text{AB}}$=10~m,
$\overline{\text{AD}}=1$~m, $\overline{\text{AF}}=1.40$~m. 
Two sub-cases are defined by considering bottom surfaces with constant
slopes of 5$^{\circ}$ and 25$^{\circ}$, respectively.
The discretization of $\SubsetU$ is obtained with a Delaunay
triangulation with average mesh parameter $\meshparam$=0.5~m, yielding
a total of 114 FV surface cells.
Figures~\ref{fig:TC1_5deg_results} and~\ref{fig:TC1_25deg_results}
show the numerically evaluated evolution 
of the gravity wave in terms of water depth $\Depth$~[m] at
$\tempo$~=~0.0~s, 0.4~s, 0.8~s, and 1.2~s for the sub-case of $5^\circ$
and $25^\circ$, respectively.  
The results show a non-oscillatory solution for both sub-cases with a
strong numerical viscosity, typical of first order numerical schemes,
that extends for more than three cells upstream and downstream the
wave front.
At the last time and for the largest slope sub-case,
(figure~\ref{subfig:TC1_25deg_t=200s}), 
the upstream portion of the domain reaches a dry state, without displaying
numerical inconsistencies. Moreover, at the downstream boundary no
backward waves are forming. 

\begin{figure}
    \centering
    \begin{subfigure}[b]{0.45\textwidth}
        \includegraphics[width=1.2\textwidth]{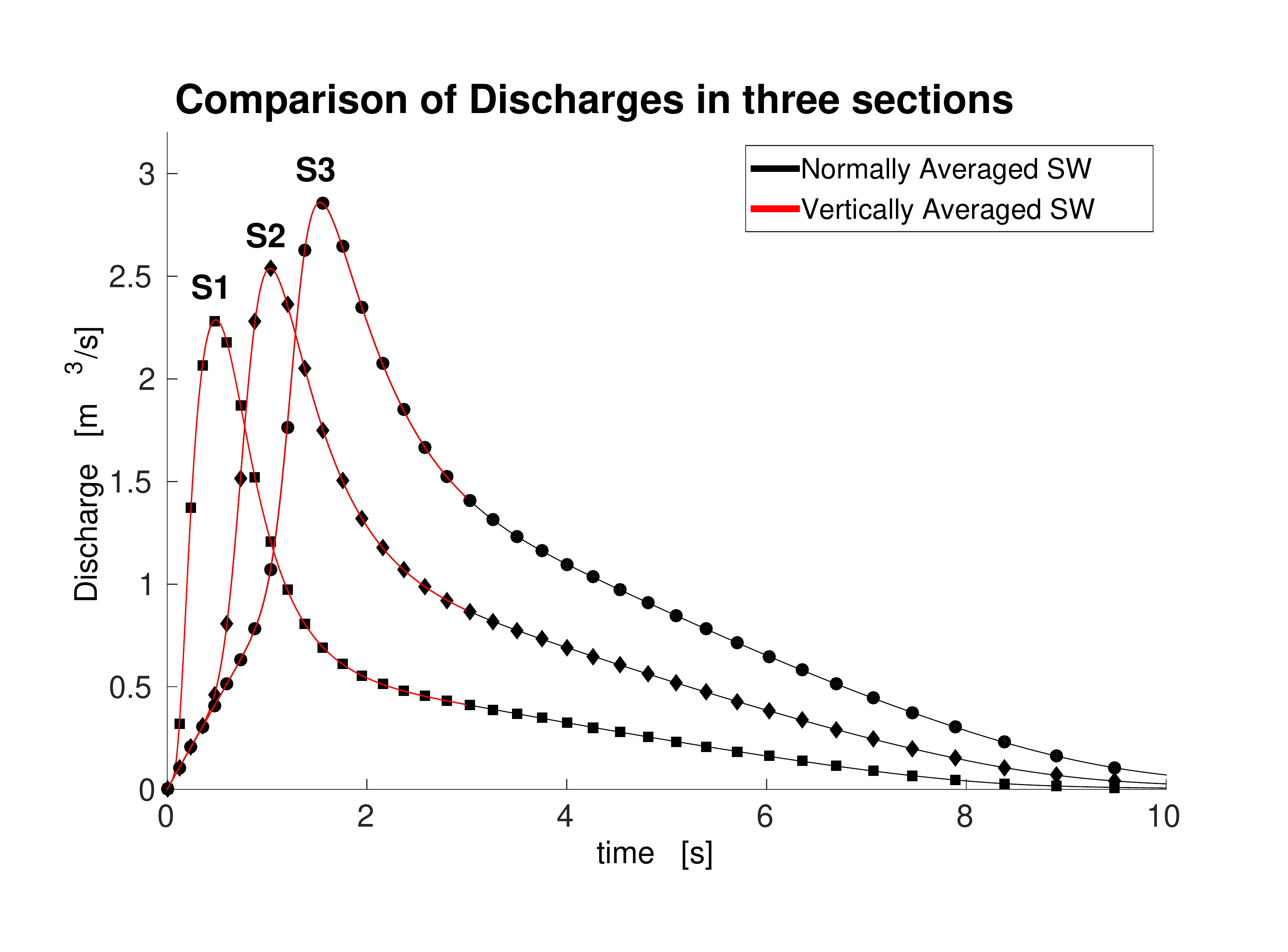}
        \caption{5$^\circ$ bottom slope}
        \label{subfig:DischSect_TC1_05deg}
    \end{subfigure}
    \begin{subfigure}[b]{0.45\textwidth}
        \includegraphics[width=1.2\textwidth]{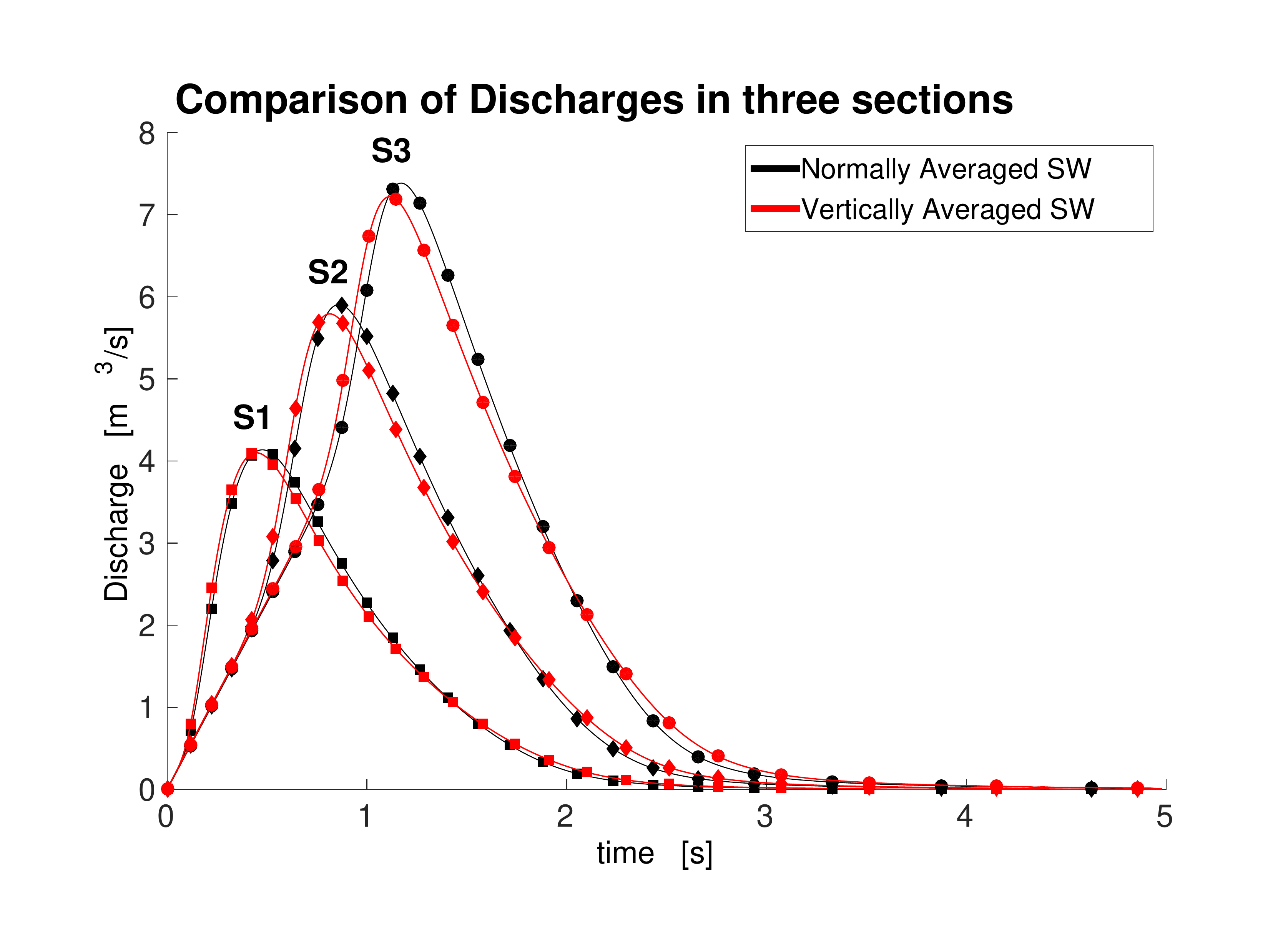}
        \caption{25$^\circ$ bottom slope}
        \label{subfig:DischSect_TC1_25deg}
    \end{subfigure}
    \caption{Test case 1: uniformly sloping bottom. Comparison of
      cross-sectional discharges computed in Sections $\Sect{1}$,
      $\Sect{2}$, and $\Sect{3}$ of figure~\ref{fig:DomainSketch} for
      different sloping angles and using the normally and vertically
      averaged SWE.}
    \label{fig:TC1_discharge}
\end{figure}

The quantitative differences between the results obtained with the
normally averaged SWE and those obtained with the vertically averaged
SWE approach are evaluated by looking at cross-sectional discharges
passing through the three sections of figure~\ref{fig:DomainSketch}. 
The results, reported in figure~\ref{fig:TC1_discharge}, show significant
differences only for the largest slope sub-case
(figure~\ref{subfig:DischSect_TC1_25deg}), with increased peak discharge
and a later time of arrival for the streamflows obtained from the
normally averaged SWE.
The total volumes, evaluated as the integrals under the curves, are
the same for both vertically and normally averaged approaches,
confirming that our FV method is conservative.

\subsection{Test case 2: double parabola.}

\begin{figure}
    \centering
    \begin{subfigure}[b]{0.45\textwidth}
        \includegraphics[width=\textwidth]{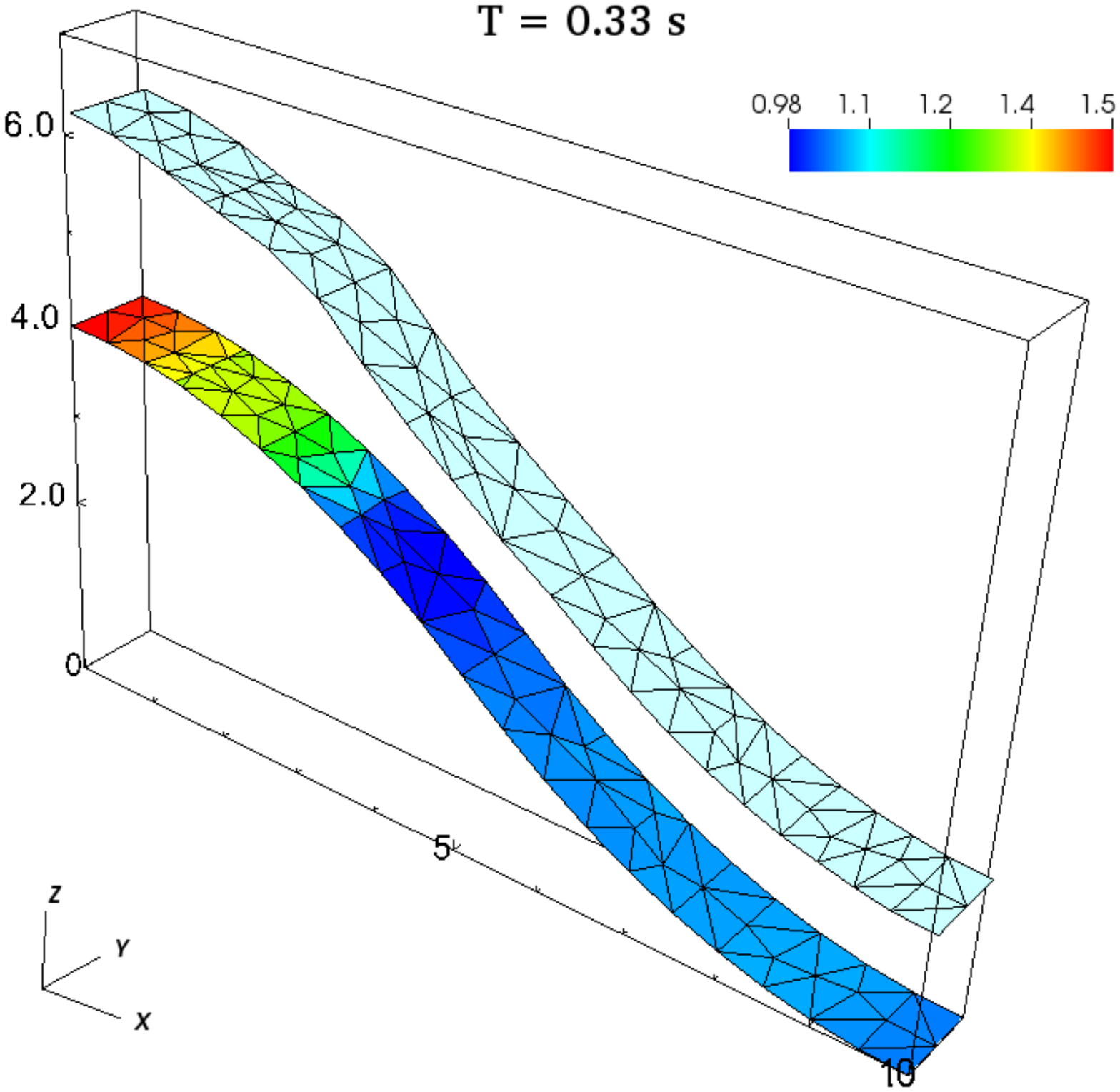}
        \caption{}
        \label{subfig:TC2t=033s}
    \end{subfigure}
    \begin{subfigure}[b]{0.45\textwidth}
        \includegraphics[width=\textwidth]{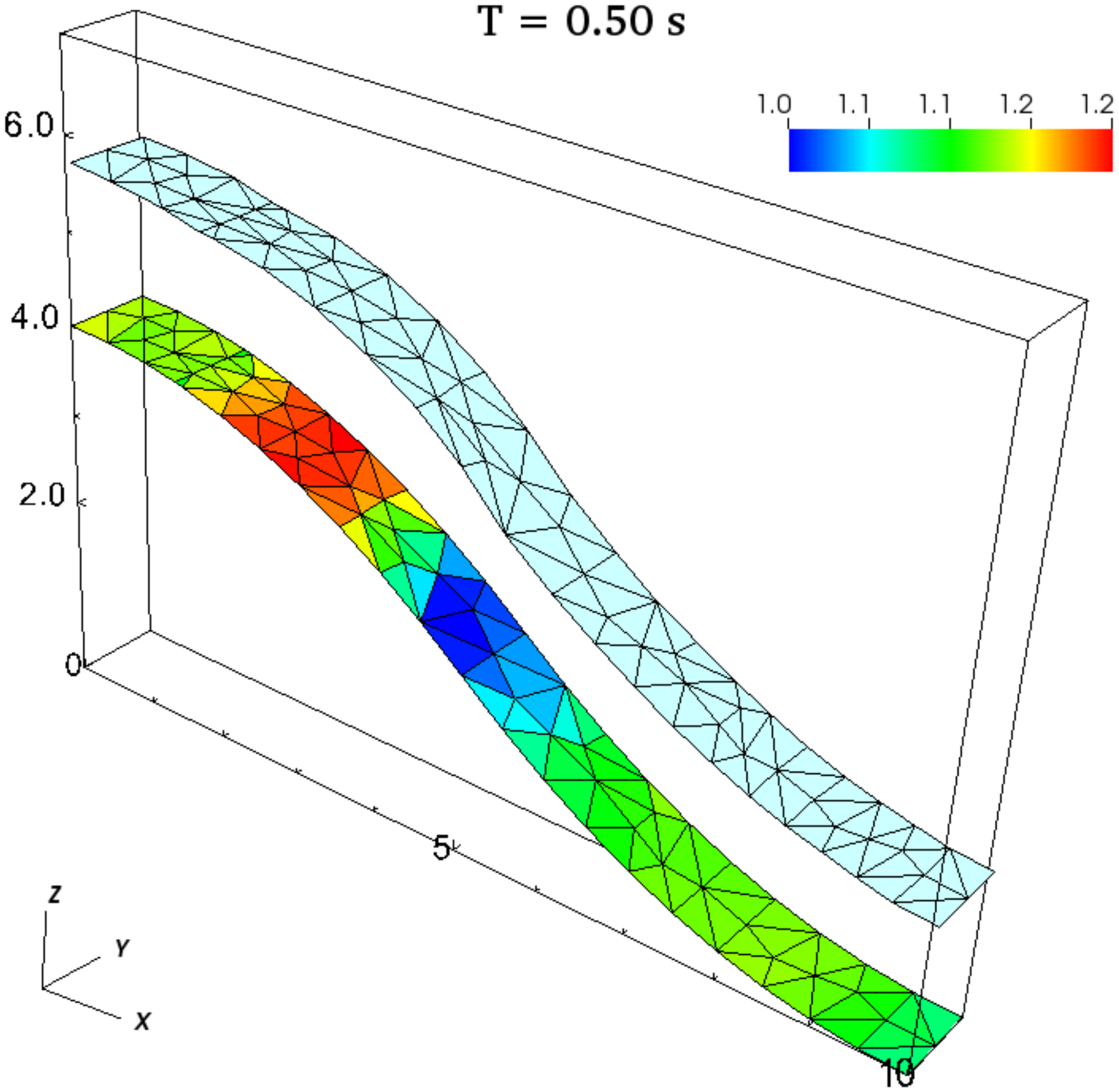}
        \caption{}
        \label{subfig:TC2t=050s}
    \end{subfigure}
    \\
    \begin{subfigure}[b]{0.45\textwidth}
        \includegraphics[width=\textwidth]{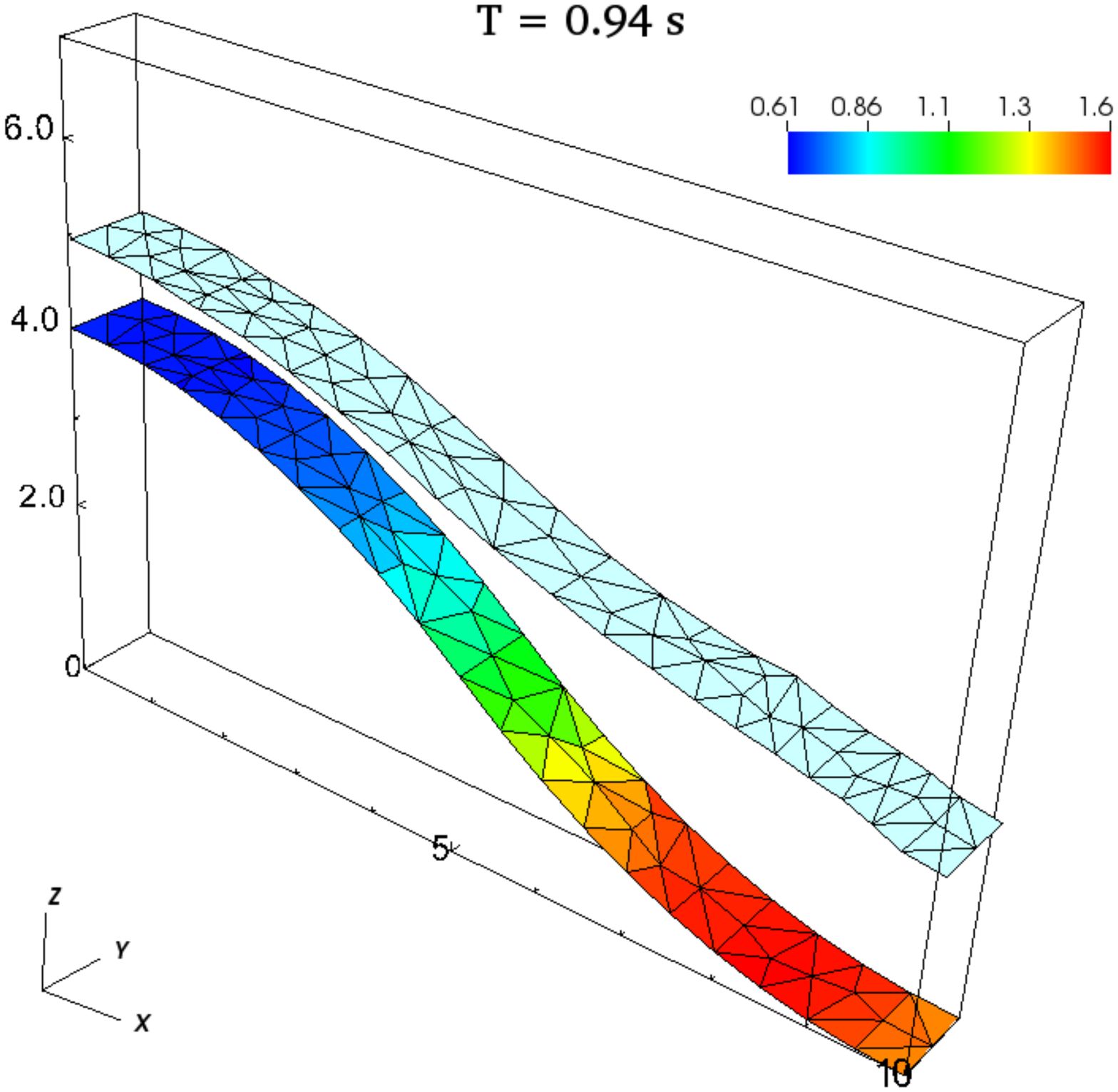}
        \caption{}
        \label{subfig:TC2t=094s}
    \end{subfigure}
    \begin{subfigure}[b]{0.45\textwidth}
        \includegraphics[width=\textwidth]{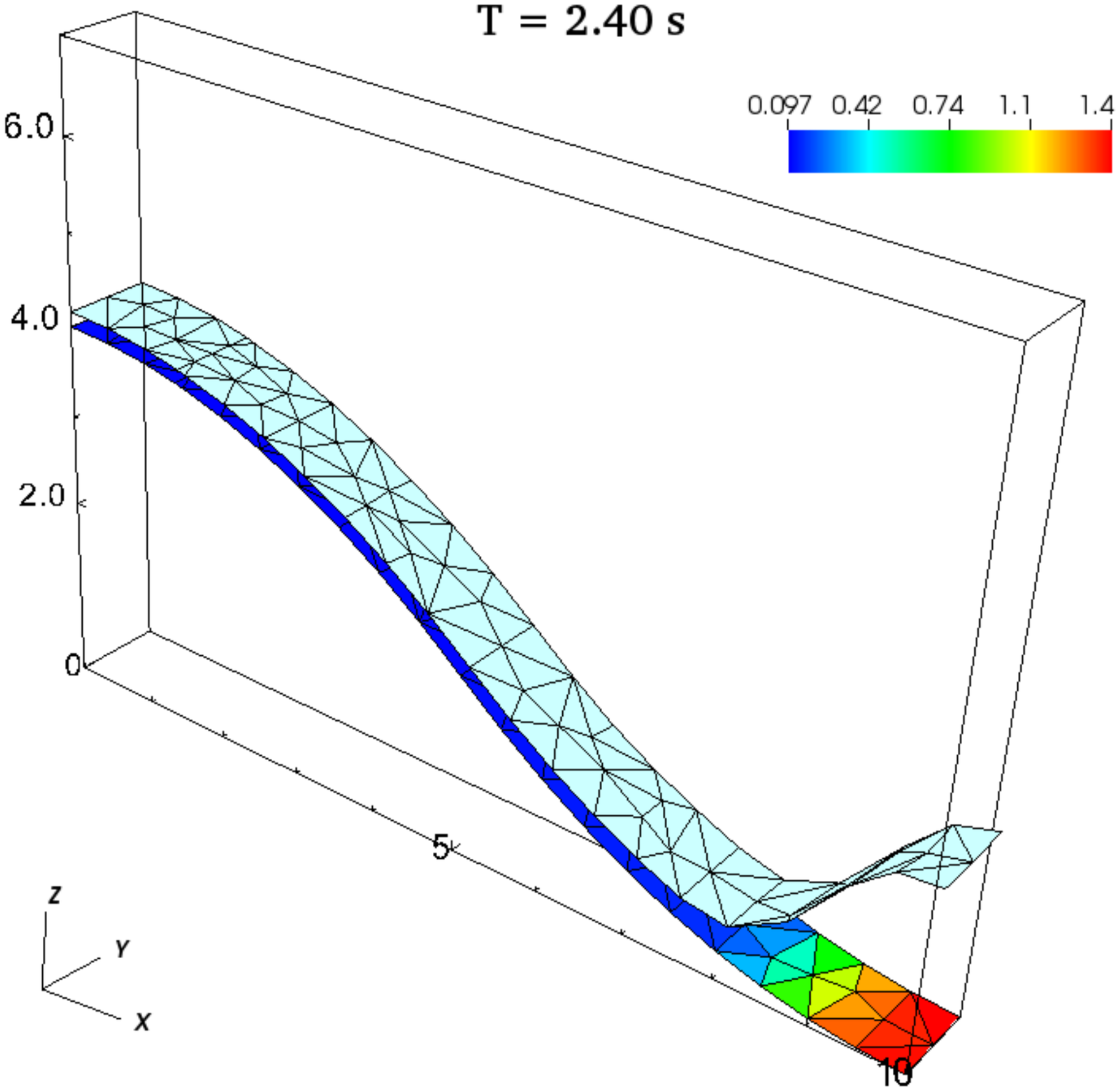}
        \caption{}
        \label{subfig:TC2t=240s}
    \end{subfigure}
    \caption{Test case 2: one-dimensional smooth curved
      bottom. Evolution of the gravity wave, shown both as color codes
      and depth elevation, the latter with a magnification factor of
      1.5.  }
    \label{fig:TC2_results}
\end{figure}

This test case aims at verifying the effects of curvatures for a
simple one-dimensional propagation of the flow.
Thus we consider a rectangular Monge subset $\SubsetU$ with the
same dimensions as in the previous test case.
The Monge height function that describes the bottom surface is:
\begin{equation*}
  \BSM(\xcg,\ycg)=
  \begin{cases}
    -\dfrac{2}{25}\xcg^2+4 & \mbox{ if } \xcg\le 5, \\
    \dfrac{2}{25}(\xcg-10)^2 & \mbox{ otherwise. }
  \end{cases}
\end{equation*}
The discretization of $\SubsetU$ is the same Delaunay triangulation
with average mesh parameter $\meshparam$~=~0.5~m, yielding a total of
114 FV surface cells. Initial conditions are the same as in test case 1.

Figure~\ref{fig:TC2_results} reports the numerically evaluated evolution
of the gravity wave in terms of water depth $\Depth$~[m] at
$\tempo$~=~0.33~s, 0.50~s, 0.94~s, and 2.4~s.
The wave moves slowly in the upper portion of the channel
(figure~\ref{subfig:TC2t=033s}) and increases its speed approaching the
steeper central portion (figure~\ref{subfig:TC2t=050s}). At around 0.33~s a
new gravity wave forms in the domain center because of the acceleration
induced by the change of bottom slope. Figure~\ref{subfig:TC2t=094s} shows that
before $\tempo\approx 1$~s the initial wave collapses on the tail of
the later gravity wave, which is slowed down by the downstream
decreasing bottom slope.
This behavior is quantifiable by noticing that at $\tempo$~=~0.33~s the
wave front is located at around 2~m from the upstream boundary,
while at $\tempo$~=~0.50~s the same wave has traveled for around 2~m. 
On the other hand, in the next half a second ($t=0.94$~s), the
initial wave has traveled almost to the outlet.
At the end of the simulation (figure~\ref{subfig:TC2t=240s}) most
of the water volume has left the domain. 
Note that this occurs in a relatively long time, more than 1.5~s,
because of the flatness of the downstream bottom topography.

\begin{figure}
  \centering
  \begin{subfigure}[b]{0.5\textwidth}
    \includegraphics[width=1.3\textwidth]{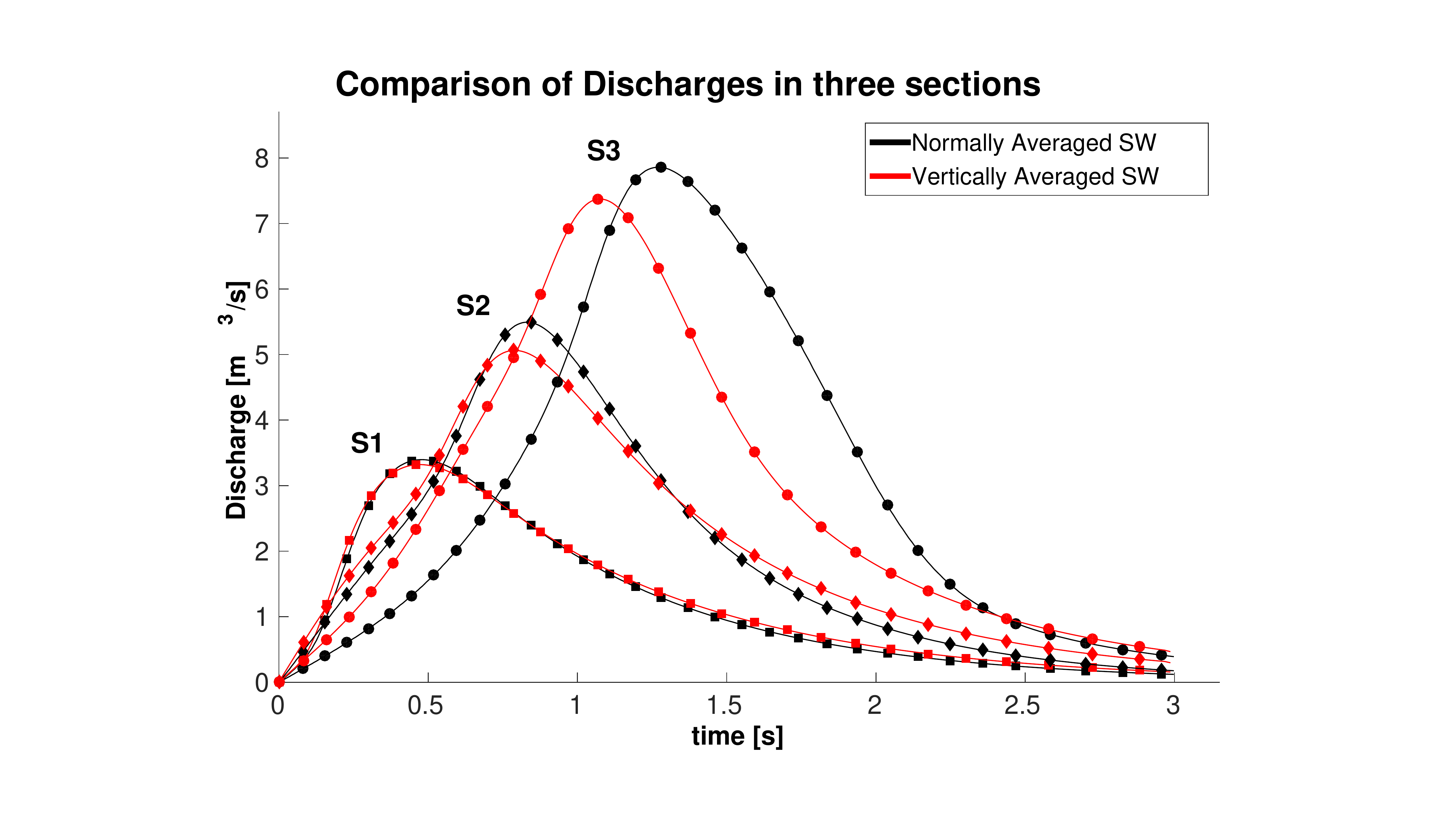}
    \caption{Discharge}
    \label{subfig:TC2_discharge}
  \end{subfigure}
  \begin{subfigure}[b]{0.45\textwidth}
    \includegraphics[width=1.3\textwidth]{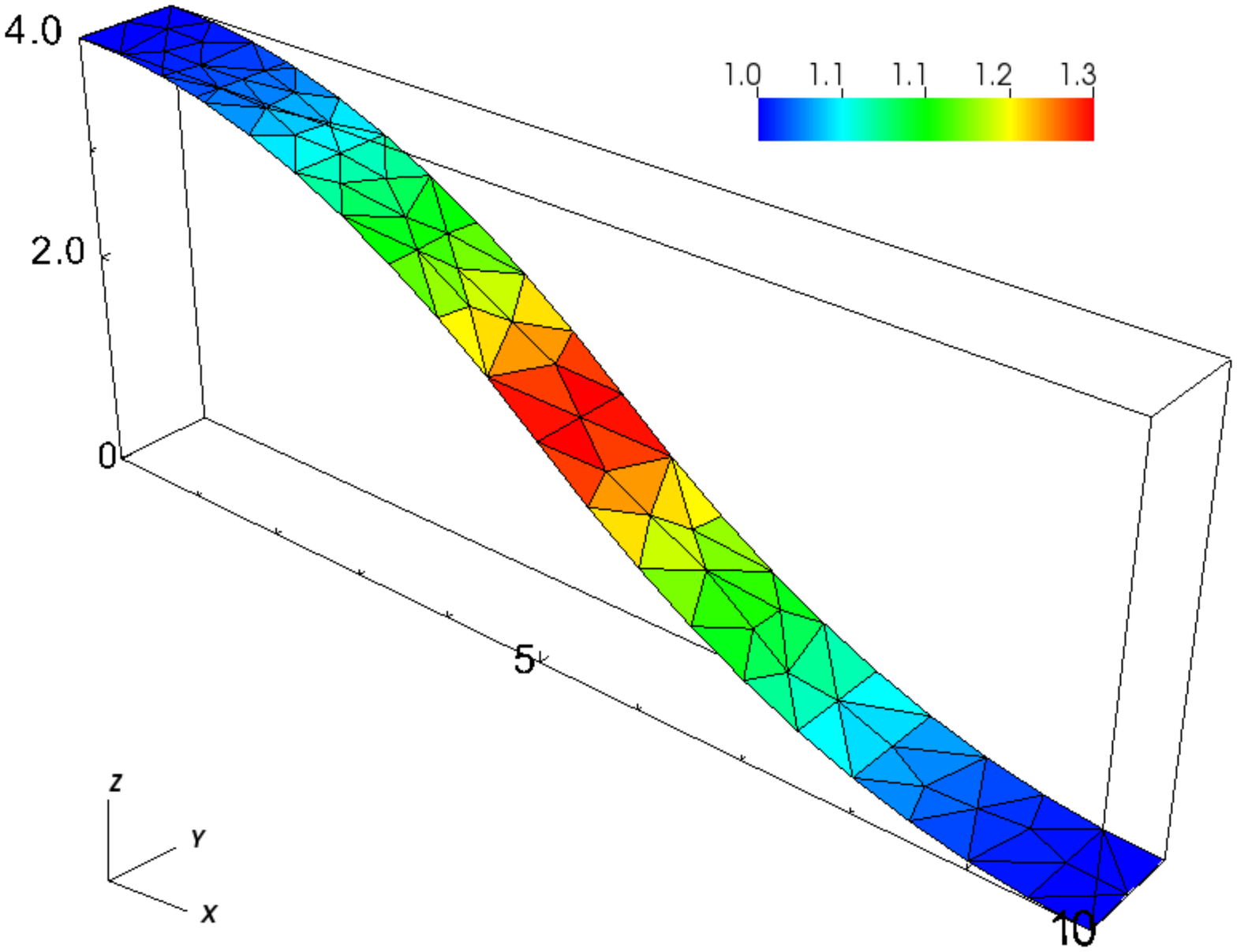}
    \caption{Metric coefficient $\metrcoef{1}$}
    \label{subfig:TC2metrCoeff1}
  \end{subfigure}
  \caption{Test case 2: one-dimensional smooth curved
      bottom. Panel (a): comparison of discharges computed in the
      cross-sections $\Sect{1}$, $\Sect{2}$, $\Sect{3}$, obtained
    from the normally and vertically averaged SWE. Panel (b):
    spatial distribution of the metric coefficient $\metrcoef{1}$}
  \label{fig:TC2_discharge_metrcoef}
\end{figure}

Better appreciation of the influence of bottom geometry is achieved by
comparing simulated discharges at the three channel cross sections,
obtained by using normally and vertically averaged SWE.
The results of this comparison are reported in
figure~\ref{subfig:TC2_discharge}.
The differences in the first section are rather small, a consequence
of an almost flat topography upstream of $\Sect{1}$.
Much larger discrepancies are visible for the results of the other two
sections, where higher discharge peaks and longer arrival times
characterize the streamflow evaluated using the normally averaged
model.
This behavior can be attributed to the effects of the curvatures that
cause lower discharge during the rising limb and a consequential later
peak increase, so that mass conservation is satisfied.
The effects of the bottom curvatures can be appreciated also by looking at the
spatial distribution of the metric coefficient $\metrcoef{1}$,
reported in figure~\ref{subfig:TC2metrCoeff1}.
The largest values are localized in correspondence of section
$\Sect{2}$, which is where the concavity of the bottom topography
changes. 
The presence of $\metrcoef{1}$ in the denominator of the terms 
of equation~\eqref{eq:sistSaintVenantCoordCurv} contained in the
divergence operator is responsible for local reductions of the
conserved fluxes. 
These reductions are concentrated in the central portion
of the domain, where the metric coefficient is largest.

\begin{table}
  \centerline{
  \begin{tabular}{|c|c|c|c|c|}
    \hline
    $\meshparam$ 
    & $\NORM{\Depth_{\meshparam}-\tilde{\Depth}}_{2}$
    & ratio
    & $\NORM{\Depth_{\meshparam}-\tilde{\Depth}}_{\infty}$
    & ratio \\
      \hline
    1/10 & 0.2560 & -     & 0.1662 & -      \\
    1/20 & 0.1951 & 1.312 & 0.1244 & 1.335  \\
    1/40 & 0.09766& 1.998 & 0.07881& 1.579 \\
    \hline\hline
    $\meshparam$ 
    & {\small 
      $\Big\|\ABS{\Qdisch[]_{\meshparam}}-\ABS{\tilde{\Qdisch[]_{\meshparam}}}\Big\|_{2}$
      }
    & ratio
    & {\small
      $\Big\|\ABS{\Qdisch[]_{\meshparam}}-\ABS{\tilde{\Qdisch[]_{\meshparam}}}\Big\|_{\infty}$
      }
    & ratio \\
      \hline
    1/10 & 1.684 & -     & 0.9546 & -      \\
    1/20 & 1.237 & 1.361 & 0.7982 & 1.196  \\
    1/40 & 0.7667& 1.613 & 0.4042 & 1.975 \\
    \hline
  \end{tabular}
  }
  \caption{Test case 2: convergence of the proposed FV scheme. The
    errors are estimated by using the fine-grid ($\meshparam=1/80$)
    solution $\tilde{\Depth}$ and $\ABS{\tilde{\Qdisch[]{}}}$ as
    reference solution. The word ``ratio'' indicates the
    ratio between successive error norm estimates.}
  \label{tab:convergence}
\end{table}

In this test case we also evaluate the experimental convergence of the
proposed FV scheme to verify its theoretical properties. To this aim, 
we look at norms of the differences between solutions at successively
refined meshes and the solution obtained at the finest mesh.
We use a rectangular domain of dimensionless length $1\times0.1$, 
and the same ``double-parabola'' surface of Test case 2, scaled on
this dimensionless domain.
Then a sequence of uniform triangulations $\Triang$ is 
obtained by subdividing the domain in
square cells, each one subdivided again into two triangles. The mesh
levels are characterized by $\ell=1/10, 1/20, 1/40, 1/80$. 
The solution obtained at $\ell=1/80$ is used as reference solution,
and is indicated with a tilde. 
Table~\ref{tab:convergence} shows the $L^2$ and $L^{\infty}$ norms of
the errors evaluated on $\Depth$ and $\ABS{\Qdisch[]{}}$.
The error norms decrease at a ratio that approaches the value 2,
confirming that the proposed FV scheme is first order accurate.

\subsection{Test case 3: three dimensional topography.}

\begin{figure}
  \centering
  \begin{subfigure}[b]{0.45\textwidth}
    \includegraphics[width=\textwidth]{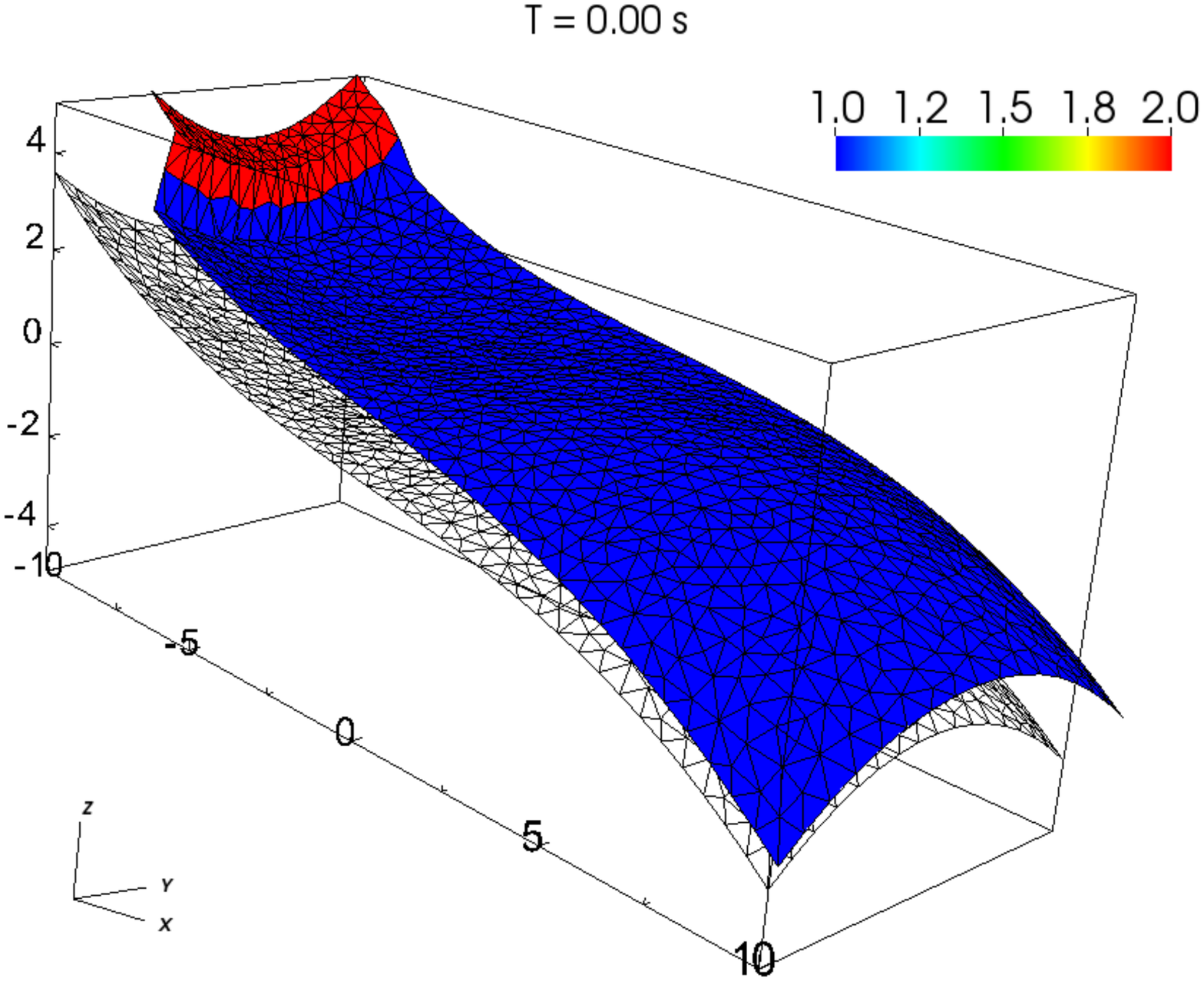}
    \caption{Initial conditions}
    \label{subfig:TC3t=00s}
  \end{subfigure}
  \begin{subfigure}[b]{0.45\textwidth}
    \includegraphics[width=\textwidth]{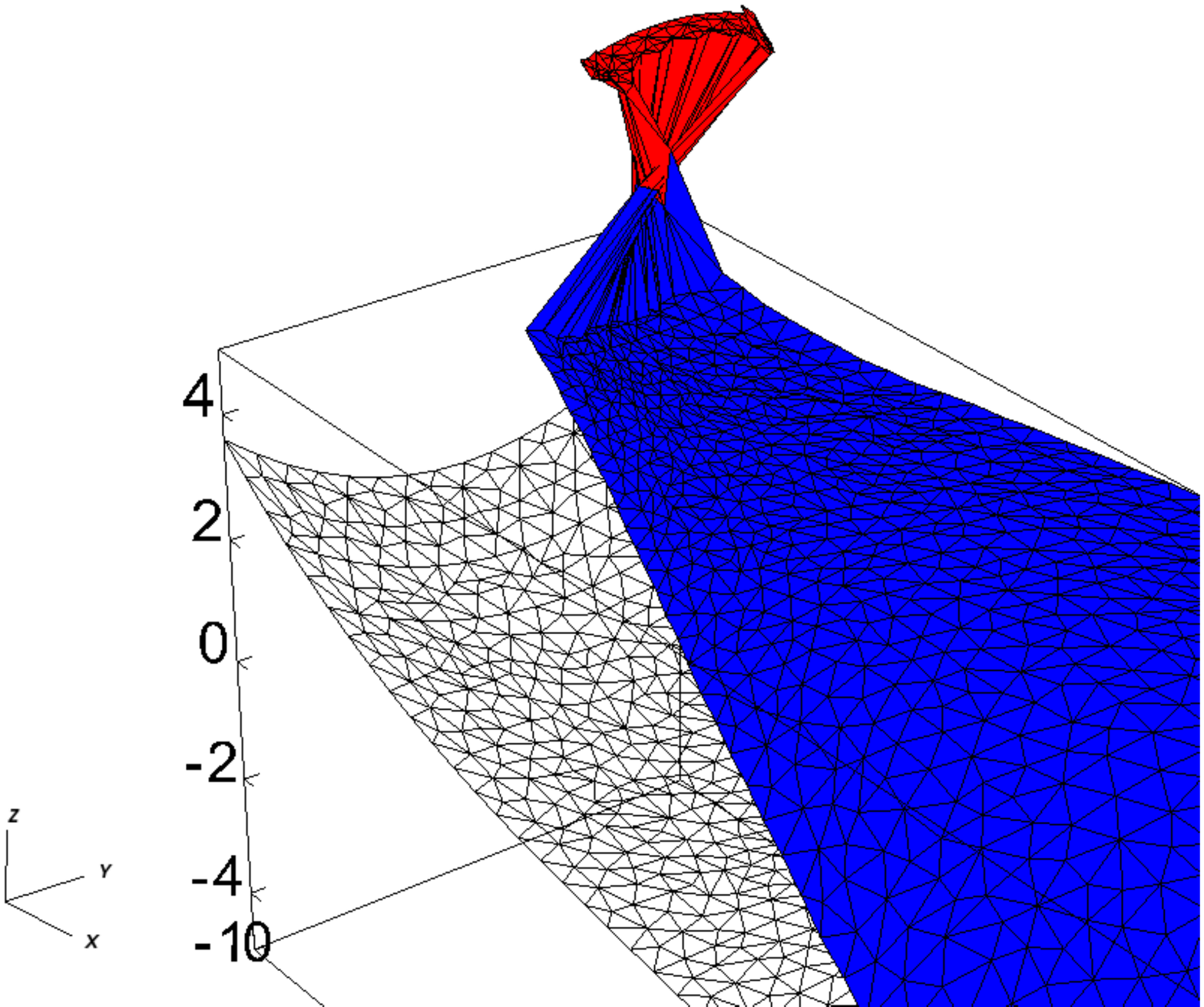}
    \caption{Crossing bottom normals.}
    \label{subfig:TC3singularPoints}
  \end{subfigure}
  \caption{Test case 3: three-dimensional smooth curved bottom. Mesh
    and initial conditions (Panel~\ref{subfig:TC3t=00s}) for the test
    case problem defined on a fully three-dimensional surface.
    Panel~\ref{subfig:TC3singularPoints} shows an example of the
    occurrence of singular points in the coordinate transformation, in
    the case case of initial condition of 10~m upstream of the initial
    wave discontinuity.  }
  \label{fig:TC3_mesh_IC}
\end{figure}

The final test case considers a fully three-dimensional surface,
built starting from a rectangular Monge subset $\SubsetU$ with the
dimensions: $\overline{\text{AB}}=20$~m, $\overline{\text{AD}}$=8~m,
$\overline{\text{AF}}=1.50$~m.  
The Monge height function is given by:
\begin{equation*}
  \BSM(\xcg,\ycg)=-\dfrac{1}{500}\xcg^3-\dfrac{1}{100}\xcg\ycg^2.
\end{equation*}
The final triangulation, shown in figure~\ref{subfig:TC3t=00s}, is
characterized by an average mesh parameter $\meshparam$=0.5~m, and a
total of 1924 FV surface cells.
The initial conditions (figure~\ref{subfig:TC3t=00s}) again consider a
uniform water depth of 2~m upstream of $\xcg$=-8.5~m, and 1~m
downstream. We would like to note that the choice of initial
conditions of a 2~m deep reservoir avoids the singularities of the
coordinate transformation by ensuring that water depth is sufficiently
shallow so that bottom normals do not intersect within the fluid domain,
as described in Section~\ref{sec:LCS}. For example,
figure~\ref{subfig:TC3singularPoints} 
shows what could happen had we chosen a water depth of 10~m in the
reservoir. 
In this case the LCS cannot be used to perform the depth integration.

\begin{figure}
    \centering
    \begin{subfigure}[b]{0.45\textwidth}
        \includegraphics[width=0.9\textwidth]{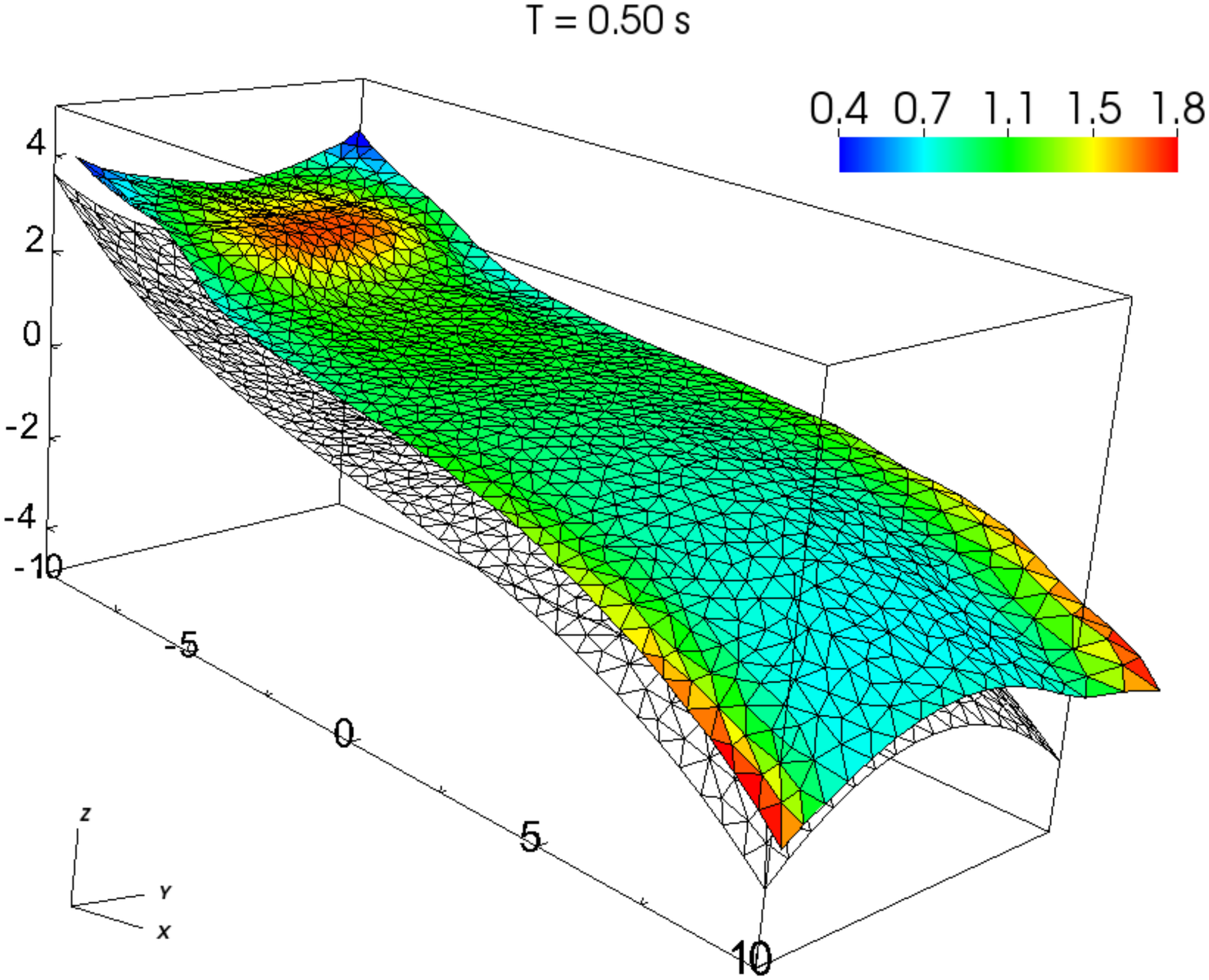}
        \caption{}
        \label{subfig:TC3t=050s}
    \end{subfigure}
    \begin{subfigure}[b]{0.45\textwidth}
        \includegraphics[width=0.9\textwidth]{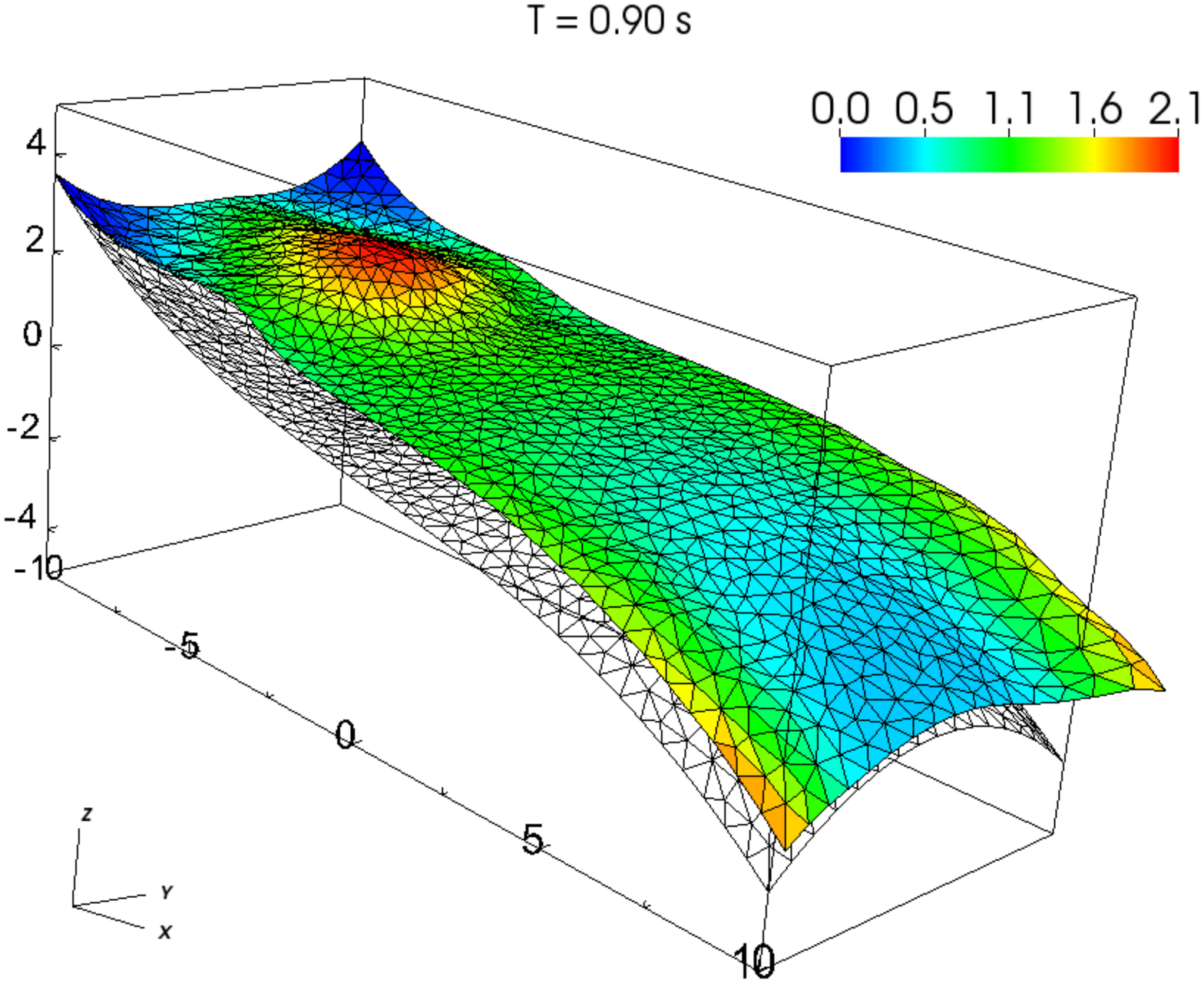}
        \caption{}
        \label{subfig:TC3t=090s}
    \end{subfigure}
    \\
    \begin{subfigure}[b]{0.45\textwidth}
        \includegraphics[width=0.9\textwidth]{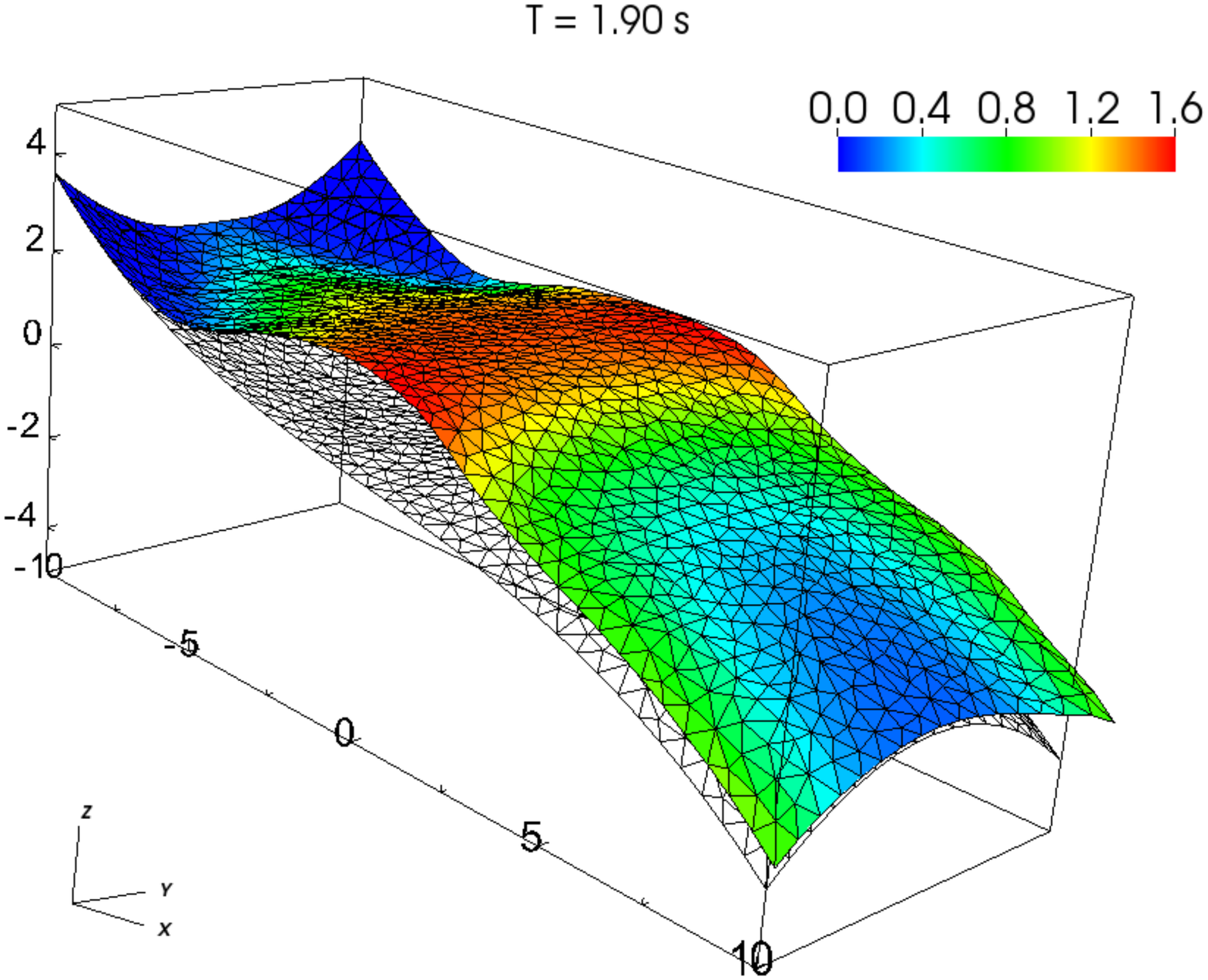}
        \caption{}
        \label{subfig:TC3t=190s}
    \end{subfigure}
    \begin{subfigure}[b]{0.45\textwidth}
        \includegraphics[width=0.9\textwidth]{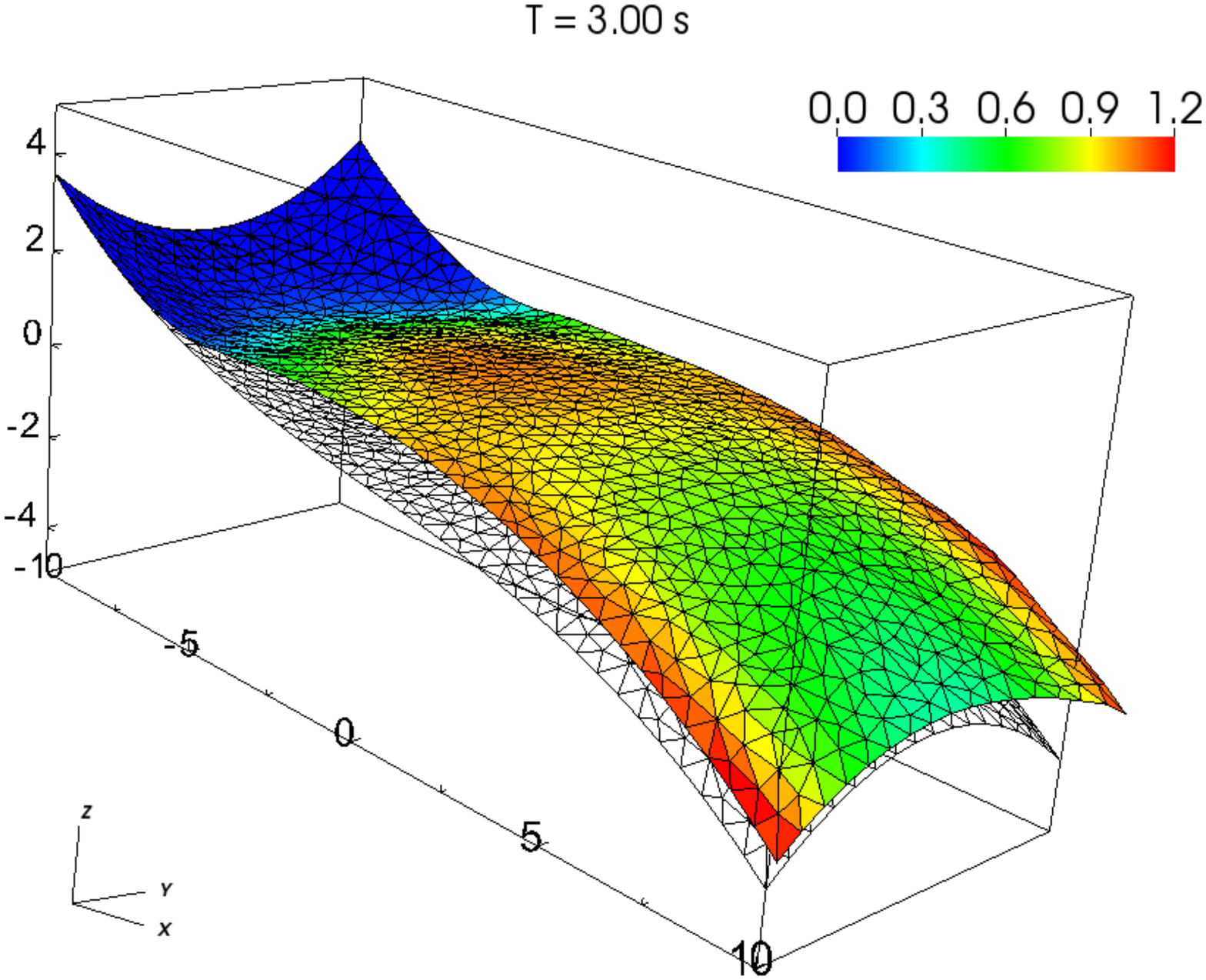}
        \caption{}
        \label{subfig:TC3t=300s}
    \end{subfigure}
    \caption{Test case 3: three-dimensional smooth curved
      bottom. Water depth [m] evolution of initial wave, shown both as
      color codes and depth elevation, the latter with a magnification
      factor of 1.3.}
    \label{fig:TC3_results}
\end{figure}

Initially, water waves propagate following the terrain
shape. In the upper portion of the domain, water accumulates
towards the center of the convex bowl forming a bell shaped wave and
emptying the upper corners of the domain (figure~\ref{subfig:TC3t=090s}).
At the same time, gravity waves form downstream, moving water
laterally from the center towards the impermeable boundary walls.
At $\tempo$=1.90~s (figure~\ref{subfig:TC3t=190s}) the upstream wave
approaches the middle and almost flat section of the bottom surface
occupying the entire width. The wave then progresses following 
the concavity of the bottom shape and leaves the domain through the
open outlet, continuing the draining of the upstream fluid.

\begin{figure}
  \centering
  \includegraphics[width=0.7\textwidth]{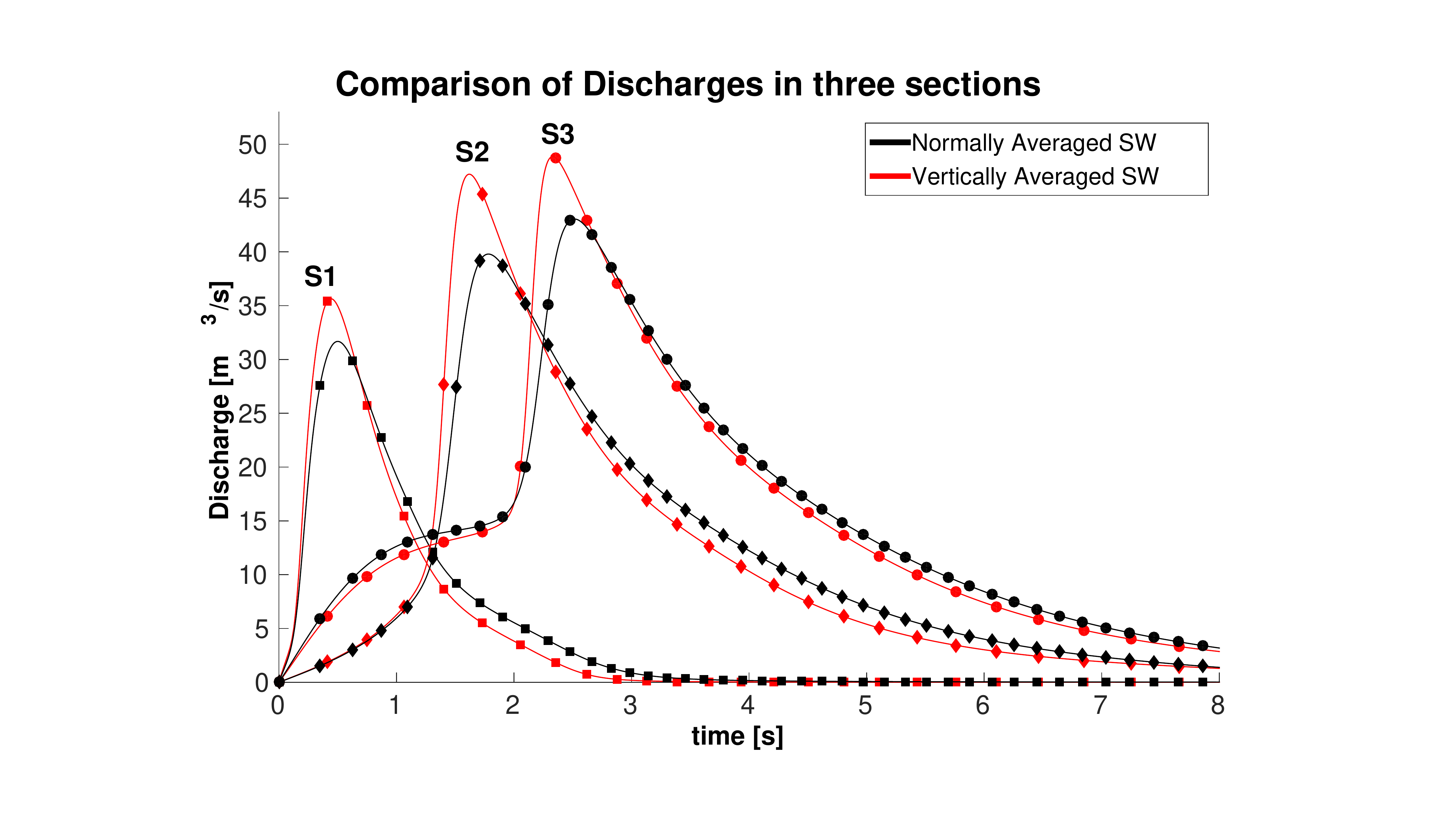}
  \caption{Test case 3: three-dimensional smooth curved
      bottom. Comparison of numerical discharges, obtained
    from the normally and vertically averaged SWE.}
  \label{fig:DischSect_TC3}
\end{figure}

The behavior of the cross-sectional discharge as function of time
differs drastically from the previous test cases as a consequence of a
more complex bottom geometry, with all sections showing varied
behavior (figure~\ref{fig:DischSect_TC3}).
We first notice that normally averaged discharges have lower peaks
that arrive at slightly later times with respect to their vertically
averaged counterparts.
Again mass conservation is satisfied, with an observable delay of the
volume arrivals.
%
  To verify the mass balance property, we
  have estimated the hydrograph volumes by evaluating the integral
  under the different curves by means of the Trapezoidal Rule. We
  obtained errors of the order of 6.4\%, 1.8\%, and 1.4\% for
  Sections~1,2, and 3, respectively. Note that the reservoir volume is
  calculated following either the normal or the vertical direction,
  depending on the assumption. Thus error for Section~1, which is
  closer to the reservoir, is influenced by this discrepancy.  The
  results, although no exact mass balance property can be proved for
  the FV scheme, seem to be sufficiently accurate and the
  differences decrease as the calculations progress.
%
Focusing on single sections, we notice that section $\Sect{1}$ already
shows important differences with a peak discharge discrepancy of
around 10\%. This is due to the influence of the metric coefficients
that are active in both directions and throughout the entire domain.
Section $\Sect{3}$ shows a peculiar behavior, with two waves clearly
perceived: the initial rising limb of the curve is associated with the
downstream gravity wave moving laterally and the following rising limb
part corresponds to the arrival of the main water wave. Notice that,
contrary to all the other curves, the normally averaged gravity wave
precedes in time the corresponding vertically averaged wave.
The 10\% difference in the peak discharge is confirmed also for the
last section.

\begin{figure}
  \centerline{
    \includegraphics[width=0.5\textwidth]{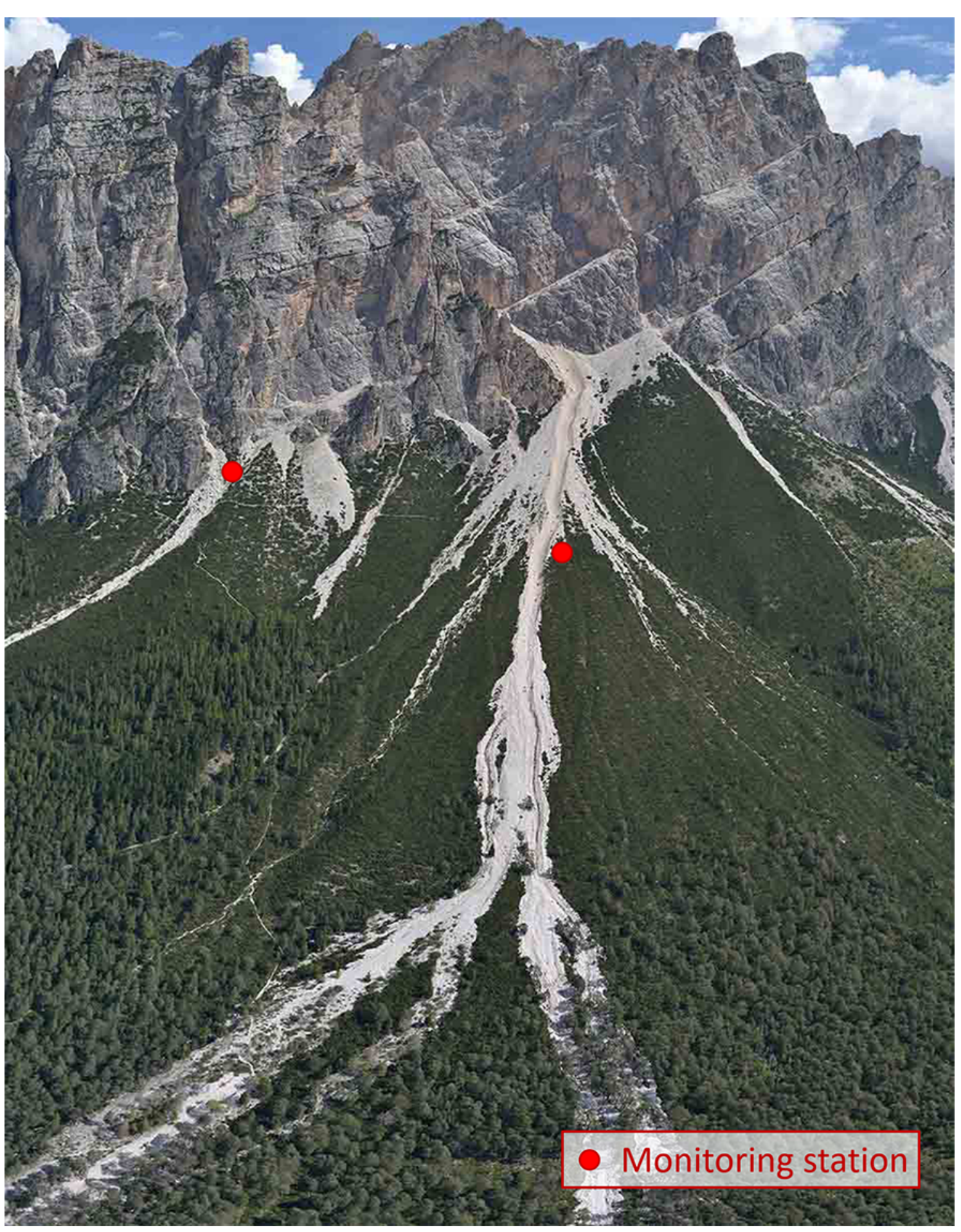}
  }
  \caption{Example of complex terrain showing the instrumented
    debris flow experimental sites located in Fiames (Dolomites),
    Northern Italy. Several other debris flow channels are clearly
    visible.  }
  \label{fig:fiames}
\end{figure}

\subsection{Discussion of the results and future research}

All these results show the importance of the effects of bottom
geometry on the dynamics of the fluid movement, as governed by shallow
water equations.  Even in the presence of relatively mild curvatures,
significant differences with respect to the corresponding standard
approach for SWE (i.e., derived by means of integration along the
vertical direction) are correlated with the magnitude of the metric
coefficients.  Differences of around 20\% are witnessed in several of
the presented test cases, notwithstanding the smoothness and
regularity of the three-dimensional surfaces employed in our
simulations. This is a confirmation of the importance of the need to
take into consideration geometrical effects of the bottom topography,
in particular for real world application, where fluids move on
generally rough terrains, as occurs, e.g., in mountainous areas.  A
typical example of the potential importance of bottom variations on
the depth-averaged flow field is embodied by debris flows. This type
of gravitational motions consist of a rapidly moving liquid-sediment
mixture usually generated in narrow steep valleys, when loose masses
of unconsolidated debris become unstable under the action of water
supplied by rainfall or snow melting~\citep{book:Takahashi2007}. Due
to inertia, the liquid-sediment mixture can travel for long distances
and eventually deposit for low enough hillslopes, or when discharging
in broad alluvial fans and in a less steep channel.  The topography
over which debris flows generate, propagate, and deposit is thus
typically characterized by both relatively high slopes and highly
variable curvatures, as can be intuitively observed from the
photograph shown in
figure~\ref{fig:fiames}~\citep{art:Gregoretti-et-al-2016}.  The
present analysis suggests that a reliable estimate of the debris flow
hydrograph requires that geometrical effects of the terrain over which
the liquid-sediment mixture moves should be introduced in the
depth-averaged models currently used to simulate this type of
phenomena.

It is important to recognize the numerous assumptions that were made
in the derivation of the normally averaged SWE on curvilinear local
reference system.  In particular, we would like to recall two
important simplifications.  The first is related to the choice of
using the normal direction to approximate the cross-flow integration
path.  The influence of this assumption needs to be quantified, in
relation to the hypothesis of
%
linear
%
pressure distribution along the different integration paths (along the
vertical or normal directions or the cross-flow path). Techniques
similar to those employed in~\citet{art:Boutounet-et-al2008}
and~\citet{art:Bresh-Noble-2011} could be used in future work to
assess these differences.  The second important simplification is
related to neglecting terms containing shear stress components
comprising the differential advective terms as well as the turbulent
and viscous stresses. These terms contain geometrical information
which could alter their importance in case of large bottom curvatures.
Disregarding these components, thus effectively incorporating their
action into the uncertainty related to the empirical resistance
coefficients, is then equivalent to a limitation on the admissible
geometry of the bottom surface.

Future research will concentrate on studying these aspects in an
attempt to contribute towards more accurate and robust simulations of
geophysical flows.  To this aim, geometrical effects of topography on
beds formed by erodible sediments and, hence, of movable nature
deserves attention in the near future.


\section{Conclusions}

The shallow water assumption, whereby horizontal variations of the
flow field occur on much larger lengths than vertical variations, is
widely adopted when deriving the depth-averaged equation used to model
many natural phenomena. Usually, this derivation is carried out by
integrating along the vertical the relevant three-dimensional mass and
momentum conservation equations. However, the terrain over which the
investigated flow fields often take place, is characterized by quite
complex geometries, exhibiting non negligible bottom slopes (thus
implying that the vertical direction does not approximate any more the
normal to the bottom) and curvatures.

The present analysis addresses the solution of this problem by
proposing the integration of the mass
and momentum conservation equations along the
local direction normal to the bottom surface. 
The geometry of this surface is defined using explicit mathematical
expressions that facilitate the enucleation of the effects of
curvatures on the flow dynamics. The results suggest that
geometrical effects (embodied by the metric coefficients appearing
into the final covariant form of the SWE equations) can significantly
affect the flow field. 
In particular, the numerical simulations carried out on both uniformly
sloping beds and synthetic but realistic smooth curved domains show
that: 
\begin{itemize}
\item
  the normally averaged SWE rather then the vertically averaged SWE
  need to be used for slopes larger than about 10\%  (as, e.g.,
  those over which hyper-concentrated and debris flows occur); 
\item
  geometrical effects can be significant also for relatively mild and
  slowly varying curvatures. They affect both the peak values and the
  shape of the hydrograph at a given cross-section;
\item
  the discretization of the proposed equations through a relatively
  simple first order FORCE-type Godunov finite volume scheme appears
  to ensure satisfactory results, achieving mass conservation and
  showing good stability properties even in the presence of flux
  functions that may be variable in space.
\end{itemize}

As stated explicitly in the introduction, our work addresses problems
related to geophysical applications, where flow is no longer laminar
and a solution in closed form for the base flow does not
exist. We are specifically interested on the dynamics of
long-waves determined by either non equilibrium of the initial
conditions (e.g., a dam break) or temporal changes in the boundary
conditions (e.g., variable input discharge or input/output water
levels) .
In addition, the aim of our work is to develop a reduced
dimensionality model that retains all the nonlinearities embodied by
non linear advection and considers curvature effects on the flow
forced by the bed topography, rather than in specifically modeling of
the long waves that arise at the water surface as a consequence of
flow instabilities (solitary pulses, roll waves) when the relevant
flow parameters (e.g., the Froude number) attain some critical
values. However, we conjecture that, in principle, these long waves
can be described by our modelling approach once the full framework of
integration along the cross-flow paths is developed.

Several improvements of the present modeling approach merit attention
in the near future.  
%
  On one hand, we would like to prove that
  our equations are an expansion of second order of a small parameter
  $\epsilon$ that takes into account the SW hypothesis of small
  vertical vs. horizontal scales and the geometric characteristics of
  the bottom surface, that the system is hyperbolic, admits a
  conserved energy in the absence of stresses, and is rotation
  invariant.  This latter property is a fundamental requirement for a
  correct definition of the associated Riemann problem, so that upwind
  schemes can be developed. Also the fact that the fluxes are
  non-autonomous requires careful analytical studies, as well as the
  relaxation of the regularity assumptions on the solution deriving by
  the use of the chain rule of differentiation and Leibnitz
  theorem~\citep{book:Ambrosio2000,art:Andreianov2011,art:Crasta2011,art:Ambrosio2012}.
  Additionally, errors arising from the use of local bed normals to
  approximate the cross-flow integration path should be analyzed to
  arrive at corrections accounting for possible deviations from the
  assumed pressure distribution.  Also well-balance properties and
  wetting and drying algorithms in the presence of a complex terrain
  should be suitably analyzed and tested.
%
Improvements and advantages associated with the
use of higher order numerical schemes also need to be considered.
Errors involved in the evaluation of the geometrical quantities of the
bottom surface need to be assessed in the case of bed geometry defined
starting from measured data such as remotely sensed digital elevation
maps.  

Finally, the modelling of a movable bed, to take into account
erosion and sedimentation, has to be properly addressed within the
context of complex terrains to obtain robust and reliable predictions
of natural phenomena as hyper-concentrated and debris flows. Future
investigations will address possibly the sudden dynamic formation of
channels incised on fans and the emergence of non-smooth bed
geometries, whose numerical treatment is still an open
issue~\citep{art:Valiani2017}. 

\section{Acknowledgments}
We thank the anonymous reviewers for their helpful comments.
This work was supported in part by the following projects: GAPDEMM ``GIS- based
integrated platform for Debris Flow Monitoring, Modeling and Hazard
Mitigation'' funded by CARIPARO foundation, 
the Multi-ITN EU-FP7 project SEDITRANS ``Sediment transport in fluvial,
estuarine and coastal environments''.
All the three-dimensional figures were prepared using the VisIt
High Performance visualization package~\cite{HPV:VisIt}.

\bibliographystyle{elsarticle-num-names}
\bibliography{Strings,bibl_SWE_MC}

\end{document}